\documentclass{article}
\usepackage{amsmath,amssymb,yhmath,bm}
\usepackage{pstricks,pst-node,pst-text,pst-3d, psfrag,graphicx}
\usepackage{slashed}
\usepackage{color}
\usepackage{array}
\usepackage{tabulary}
\usepackage{rotating}
 
 \title{Quantum Supersymmetric Bianchi IX Cosmology}

\author{  Thibault {\sc Damour}$^1$ and Philippe {\sc Spindel}$^2$
\\1 Institut des Hautes 
\'Etudes Scientifiques, 
Bures-sur-Yvette, F-91440, France\\
2 M\'ecanique et Gravitation, 
Universit\'e de Mons,
7000 Mons, Belgique}

\begin{document}

\maketitle

\begin{abstract}
We study the quantum dynamics of a supersymmetric squashed three-sphere by dimensionally reducing (to one timelike dimension) the action of $D=4$ simple supergravity for an $SU(2)$-homogeneous (Bianchi IX) cosmological model. The quantization of the homogeneous gravitino field leads to a 64-dimensional fermionic Hilbert space.
After imposition of the diffeomorphism constraints, the
 wave function of the Universe  becomes a 64-component spinor of  Spin(8,4) depending on the three squashing parameters, which satisfies Dirac-like, and  Klein-Gordon-like, wave equations describing the propagation of a ``quantum spinning particle'' reflecting off spin-dependent potential walls.
The algebra of the supersymmetry constraints and of the Hamiltonian one is found to close. One finds that the quantum Hamiltonian is built from operators that generate a 64-dimensional representation of the (infinite-dimensional) maximally compact sub-algebra of the rank-3 hyperbolic Kac--Moody algebra $AE_3$.  
The (quartic-in-fermions) squared-mass term $\widehat \mu^2$ entering the Klein-Gordon-like equation has several remarkable properties: (i) it commutes with all the other (Kac--Moody-related) building blocks of the Hamiltonian; (ii)
it is a quadratic function of the fermion number $N_F$;
and (iii) it is negative in most of the Hilbert space. The latter property leads to a possible quantum avoidance of the  singularity (``cosmological bounce"), and suggests imposing the boundary condition that the wavefunction of the Universe
vanish when the volume of space tends to zero (a type of boundary condition which looks like a final-state condition
when considering the big crunch inside a black hole).  The space of solutions is a mixture of ``discrete-spectrum states'' 
(parametrized by a few constant parameters, and known in explicit form) and  of continuous-spectrum states 
(parametrized by arbitrary functions entering some initial-value problem). The predominantly negative values of the
squared-mass term  lead to a ``bottle effect'' between small-volume-Universes and large-volume
ones, and to a possible reduction of the continuous spectrum to a discrete spectrum of  quantum states
looking like excited versions of  
the Planckian-size Universes described by the discrete  states at fermionic levels $N_F=0$ and 1.
\end{abstract}

\newpage

 \setcounter{equation}{0}\section{Introduction}

Understanding the quantum dynamics of the spacetime geometry near a spacelike (cosmological) singularity,
such as the big bang singularity that gave birth to our Universe, is one of the key problems of gravitational physics.
Since the full theory of quantum gravity is still too ill understood to allow a frontal attack on this problem, one can hope
to make progress by first studying highly symmetrical geometrical models, so that the degrees of freedom of the
gravitational, and matter, fields can be reduced to a finite number.  Among such ``minisuperspace models'', the
Bianchi IX model,  {\it i.e.}\    a spatially homogeneous ($SU(2)$-symmetric) model having spatial sections homeomorphic
to the three-sphere $S_3$, has always played a useful r\^ole.  In a classical context, the vacuum ( {\it i.e.}\   . matter-free) 
Bianchi IX model served as the paradigmatic example of the  chaotic approach towards a generic (inhomogeneous)
spatial singularity conjectured by Belinskii, Khalatnikov and Lifshitz (BKL) \cite{Belinsky:1970ew} (see also \cite{Misner:1969hg}).
The same model gave also a rich example for the quantum dynamics of space near  a big bang (or a big crunch) 
singularity \cite{Misner:1969ae}.

More recently, the Bianchi IX model has served as an important testbed for {\it supersymmetric quantum cosmology}, that
is the study of the quantum dynamics of  cosmological models, as described  within supergravity theories.
See Refs.
\cite{D'Eath:1993up, D'Eath:1993ki,  Csordas:1995kd,Csordas:1995qy, Graham:1995ni,Cheng:1994sr, Cheng:1996an,Obregon:1998hb}, as well as the books \cite{D'Eath:1996at,VargasMoniz:2010zz,Moniz:2010fea}. 
As in these references, we consider here the original ``simple'' ($\mathcal N=1$) four-dimensional supergravity theory
\cite{Freedman:1976xh,Deser:1976eh}.
Though the supersymmetric Bianchi IX model contains only a finite number of bosonic and fermionic degrees of freedom,
the previous attempts  \cite{D'Eath:1993up, D'Eath:1993ki,  Csordas:1995kd,Csordas:1995qy, Graham:1995ni,Cheng:1994sr, Cheng:1996an,Obregon:1998hb} at  studying its quantum dynamics have not succeeded in fully clarifying the structure
of its allowed states, {\it i.e.}\   the complete set of solutions of all the constraints. 

The first aim of the present work will be to remedy this situation, {\it i.e.}\   to provide a complete description of the solution space
of the quantum supersymmetric Bianchi IX model.  This will be done by using a new approach to the quantum dynamics of supersymmetric Bianchi models
that generalizes the formalism we used in \cite{Damour:2011yk}   to study the quantum dynamics of Einstein--Dirac Bianchi Universes. It differs from the formalisms used in previous works~\cite{D'Eath:1996at,VargasMoniz:2010zz,Moniz:2010fea}  in describing the gravity degrees of freedom entirely in terms of the metric components $g_{\mu\nu}$, {\it without} making use
of an arbitrary, local vielbein. We use the symmetry properties of Bianchi models to uniquely determine a specific vielbein $h_{ \ \ \mu}^{\hat\alpha}$ (with $g_{\mu\nu} = \eta_{\hat\alpha\hat\beta} \, h_{ \ \ \mu}^{\hat\alpha} \, h_{ \ \ \nu}^{\hat\beta}$) as a local function of $g_{\mu\nu}$. In other words, we gauge-fix from the start the six extra degrees of freedom contained in $h_{ \ \ \mu}^{\hat\alpha}$ that could describe arbitrary local Lorentz rotations. This gauge-fixing of the local $SO(3,1)$ gauge symmetry eliminates the need of the usual formalisms~\cite{D'Eath:1996at,VargasMoniz:2010zz,Moniz:2010fea} to impose the six local Lorentz constraints $J_{\hat\alpha\hat\beta} \approx 0$.  Another specificity of our formalism will be to describe 
the degrees of freedom of the gravitino by means of a Dirac-like gamma-matrix representation. Such a representation was
notably advocated in Refs. \cite{Macias:1987ir,Macias:1993mq,Obregon:1998hb},  and was found convenient in the 
Einstein--Dirac case  \cite{Damour:2011yk}. As we shall explicitly discussed below,  this gamma-matrix representation
of the fermionic operators is equivalent  to a representation in terms of fermionic creation and annihilation operators 
(which is, in turn, very close to the Grassmann algebra-valued  functional representation used in Refs. \cite{D'Eath:1993up, D'Eath:1993ki,  Csordas:1995kd,Csordas:1995qy, Graham:1995ni,Cheng:1994sr, Cheng:1996an}).

The second aim of the present work is to clarify the occurrence of hidden hyperbolic Kac--Moody structures in (simple, 
four-dimensional) supergravity, within a setting which goes beyond previous work both by being {\it fully quantum}, and by taking completely into account the crucial {\it nonlinearities in the fermions} that allow supergravity to exist. [Our main results on this hidden
Kac--Moody symmetry were briefly announced in  \cite{Damour:2013eua}.]  Let us recall that the existence of a
 {\it correspondence} between various supergravity theories and the dynamics of a spinning massless particle on an infinite-dimensional Kac--Moody coset space has been conjectured a few years ago~\cite{Damour:2002cu,Damour:2005zs,de Buyl:2005mt,Damour:2006xu}. Evidence for such a supergravity/Kac--Moody link emerged through the study \`a la BKL~\cite{Belinsky:1970ew} of the structure of cosmological singularities in string theory and supergravity, in spacetime dimensions $4 \leq D \leq 11$~\cite{Damour:2000hv,Damour:2001sa,Damour:2002et}.  For instance,  the well-known BKL oscillatory behavior~\cite{Belinsky:1970ew}  of the diagonal components of a generic, inhomogeneous Einsteinian metric in $D=4$ 
 (also found in the spatially homogeneous Bianchi IX model) was found to be equivalent to a billiard motion within the Weyl chamber of the rank-3 hyperbolic Kac--Moody algebra $AE_3$~\cite{Damour:2001sa}. Similarly,
the generic BKL-like dynamics of the bosonic sector of maximal supergravity (considered either in $D=11$, or, after dimensional reduction, in $4 \leq D \leq 10$) leads to a chaotic billiard motion within the Weyl chamber of the rank-10 hyperbolic Kac--Moody algebra $E_{10}$~\cite{Damour:2000hv}.  The hidden r\^ole of $E_{10}$ in the dynamics of maximal supergravity was confirmed to higher-approximations (up to the third level) in the gradient expansion $\partial_x \ll \partial_T$ of  its bosonic sector~\cite{Damour:2002cu}. In addition, the  study of the fermionic sector of supergravity theories has exhibited a related r\^ole of Kac--Moody algebras. At leading order in the gradient expansion of the gravitino field $\psi_{\mu}$,  the dynamics of $\psi_{\mu}$ at each spatial point was found to be given by  parallel transport with respect to a (bosonic-induced) connection $Q$ taking values within the ``compact'' sub-algebra of the corresponding bosonic Kac--Moody algebra: say $K(AE_3)$ for $D=4$ simple supergravity and $K(E_{10})$ for maximal supergravity~\cite{Damour:2005zs,de Buyl:2005mt,Damour:2006xu}. However,  the latter works considered only the terms {\it linear} in the gravitino, and, moreover, treated  $\psi_{\mu}$ as a ``classical'' ( {\it i.e.}\   Grassman-valued) fermionic field. By contrast, the present work will treat the
(spatially homogeneous) gravitino $\psi_{\mu}$ as a quantum fermionic operator (satisfying anti-commutation conditions),
and will keep all the nonlinearities in the fermions predicted by supergravity. This will allow us to confirm the hidden presence
of  the rank-3 hyperbolic Kac--Moody algebra   $AE_3$, notably via its ``maximal compact subalgebra''  $K(AE_3)$.

\setcounter{equation}{0}\section{Classical Lagrangian formulation}

In this work we follow the approach and notation of our previous work \cite{Damour:2013eua}. We start from the Bianchi IX metric ansatz ($i,j=1,2,3$)
\begin{equation}
ds^2 = -N(t)^2\,dt^2+g_{ij}(t)(\tau^i + N^i(t)\,dt)(\tau^j+ N^j(t)\,dt) \qquad ,\label{ds2}
\end{equation}
where, as usual, we denote by $N(t)$ and $N^i(t)$ the lapse and shift functions. The $\tau^i$ are (spatially dependent) left-invariant 1-forms on the $SU(2)$ group manifold  : 
\begin{equation}
d\tau^i = \frac12 \, C^i_{\ jk} \, \tau^j \wedge \tau^k \qquad ,
\end{equation}
where $C^i_{\ jk} = \varepsilon_{ijk}$ is the usual 3-dimensional Levi--Civita symbol ($\varepsilon_{123} = +1$).

\smallskip

This metric represents a stack of time-dependent squashed 3-spheres. Each of these deformed 3-spheres is still an homogeneous space   {\it i.e.}\    all the points on each sphere are indistinguishable from each other. However, the local geometry of each of these squashed 3-spheres is {\it anisotropic}, the anisotropy being encoded in the time-dependent quadratic form $g_{ij}(t)$. At each point the diagonalization of this quadratic form with respect to the Cartan-Killing metric
\begin{equation}
\label{CartanKilling}
k_{ij} := - \frac12 C^{r}_{\ is}  \ C^{s}_{\ jr} = \delta_{ij}  \qquad ,
\end{equation}
associated with the $SU(2)$ group symmetry, defines three special directions.

\smallskip

In order to represent the gravitino degrees of freedom, we need to introduce a vielbein (``rep\`ere mobile''). We adopt a co-frame of the form
\begin{eqnarray}
\theta^{\Hat 0} &= &N(t) \, dt \nonumber \\
\theta^{\Hat a} &= &h_i^{\Hat a} (t) \, ( \tau^i + N^i(t) \, dt ) \nonumber 
\end{eqnarray}
where $\left(h_i^{\Hat a} (t)\right)$ is a matrix squareroot of the spatial-metric matrix $\left(g_{ij} (t)\right)$  : 
\begin{equation}
\label{gab}
g_{ij} (t) = h^{\Hat a}_i (t) \,\delta_{\hat a\,\hat b}\, h_{j}^{\Hat b} (t) \qquad .
\end{equation}
An important element of our formalism is to gauge-fixed the local Lorentz co-frame $\theta^{\hat\alpha}$ by choosing as squareroot $h^{\Hat a}_i$ of $g_{ij}$ a matrix uniquely defined from the diagonalization of $g_{ij}$ with respect to the $SU(2)$ Cartan-Killing metric (\ref{CartanKilling}). The latter diagonalization is equivalent to a Gauss decomposition of $g_{ij}$,
 {\it i.e.}\   
\begin{equation}
\label{Gauss}
g_{ij} = \sum_a e^{-2\beta^a} \, S_i^{\overline a} \, S_j^{\overline a}
\end{equation}
where $S_i^{\overline a}$ is a $SO(3)$ (orthogonal) matrix, depending on three (time-dependent) Euler angles $(\varphi^a)$, and where the three eigenvalues of $g_{ij}$ with respect to $k_{ij}$ (usually denoted $a^2, b^2,c^2$ \cite{Belinsky:1970ew}) are denoted 
\begin{equation}
e^{-2\beta^1} \equiv a^2 \, , \quad e^{-2\beta^2}  \equiv b^2 \, , \quad e^{-2\beta^3} \equiv c^2 \, .
\end{equation}
In terms of the uniquely defined elements $e^{-2\beta^1}, e^{-2\beta^2}, e^{-2\beta^3}$ and $S_i^{\overline a}$ of the Gauss decomposition (\ref{Gauss}) of $g_{ij}$, we define $h_i^{\Hat a}$ as\footnote{Henceforth we will not explicitly indicate the time dependence $(t)$ of the various field components.}
\begin{equation}
\label{hki}
h_i^{\Hat a} := e^{-\beta^a} \, S_i^{\overline a}\qquad .
\end{equation} 
In addition to the co-frame $\theta^{\hat\alpha}$, it is convenient to define a non-orthonormal spatial co-frame $\theta^{\overline a}$ 
\begin{equation}
\theta^{\overline a}  : =  \tau^{\overline a}  + N^{\overline a} \, dt  : = S_i^{\overline a} ( \tau^i + N^i \, dt ) 
\end{equation}
such that 
\begin{equation}
\label{Nab}
\theta^{\Hat 0} = N dt \quad ; \quad  \theta^{\Hat a} = e^{-\beta^a} \, \theta^{\overline a} \qquad .
\end{equation}
Viewing $S_i^{\overline a} (t)$ as operating a time-dependent rotation of the spatial frame, we introduce as in ref. \cite{Damour:2011yk} the corresponding ``angular velocity'' antisymmetric tensor $w_{\overline a \overline b}$ defined as:
\begin{equation}
\label{wab}
w_{\overline a \overline b}   : = \dot S^{\overline a}_i \, S^i_{\overline b} = -w_{\overline b \overline a} 
\end{equation}
The three independent angular velocities $w_{\overline 1 \overline 2} , w_{\overline 2 \overline 3} , w_{\overline 3 \overline 1}$ are  linear combinations of $\dot\varphi^1 , \dot\varphi^2 , \dot\varphi^3$ with $\varphi^a$-dependent coefficients, as in the classical mechanics of a spinning rigid body (see, e.g., \cite{Damour:2011yk}).

\smallskip

A consistent ansatz for a homogeneous gravitino field $\psi_{\mu}^A$ in the Bianchi IX geometry is to consider that its sixteen vielbein components $\psi_{\hat\alpha}^A$, with respect to the orthonormal co-frame $\theta^{\hat\alpha}$ only depend  on time. [Here $A = 1,2,3,4$ denotes a Majorana spinor index, while $\hat\alpha =0,1,2,3$ is a Lorentz-four-vector frame index linked to $\theta^{\hat\alpha}$.]

\smallskip

In second-order form, the Lagrangian density ${\mathcal L}_{\rm tot}$ of the  ${\mathcal N} = 1$, $D=4$ supergravity action :
\begin{equation}
S=\int {\mathcal L}_{\rm tot}\,dt\wedge\tau^1\wedge\tau^2\wedge\tau^3
\end{equation}
is the sum of a gravitational Einstein--Hilbert part and a Rarita--Schwinger one:
\begin{equation}
\label{LGRS}
{\mathcal L}_{\rm tot} = {\mathcal L}_{\rm EH} (\omega) + {\mathcal L}_{\rm RS} (\ring\omega , \kappa) \qquad .
\end{equation}

\smallskip

The connection $\omega_{\Hat\alpha \Hat\beta \Hat\gamma} = \omega_{\Hat\alpha \Hat\beta \mu} \, \theta^{\mu}_{\hat\alpha}$ (note that the differentiation index is the last on $\omega$), entering the Einstein--Hilbert action [where $\theta:={\rm det}(\theta^{\hat\alpha}_\mu)$]
\begin{equation}
8\,\pi\,G\,{\mathcal L}_{\rm EH}=\frac12 \, \theta\,R(\omega)= -\frac18\,\theta \, \eta^{\mu\nu\rho\sigma} \, \eta_{\Hat \alpha\Hat \beta\Hat \gamma \Hat \delta} \, \theta^{\Hat \gamma}_{\rho} \, \theta^{\Hat \delta}_{\sigma} \, R^{\Hat \alpha\Hat \beta}_{\ \ \ \mu\nu} (\omega) \qquad ,
\end{equation}
is the sum of the Levi--Civita connection $\ring\omega_{\Hat\alpha \Hat\beta \Hat\gamma} = \ring\omega_{\Hat\alpha \Hat\beta \mu} \, \theta^{\mu}_{\hat\alpha}$ (viewed in the vielbein $\theta^{\hat\alpha}$) and of a contorsion term $\kappa_{\Hat\alpha \Hat\beta \Hat\gamma}$ quadratic in $\psi$, 
\begin{equation}
\label{wabe}
\omega_{\Hat\alpha \Hat\beta \hat\gamma}   : = \ring\omega_{\Hat\alpha \Hat\beta \hat\gamma} + \kappa_{\Hat\alpha \Hat\beta \hat\gamma}
\end{equation}
with
\begin{equation}
\kappa_{\Hat \alpha\Hat \beta\Hat \gamma}=\kappa_{\Hat \alpha\Hat \beta\mu}\,\theta^\mu_{\Hat \gamma}=\frac{1}{4}\left(\overline\psi_{\Hat \beta}\gamma_{\Hat \alpha}\psi_{\Hat\gamma}-\overline\psi_{\Hat \alpha}\gamma_{\Hat \beta}\psi_{\Hat\gamma}+\overline\psi_{\Hat \beta}\gamma_{\Hat \gamma}\psi_{\Hat\alpha}\right)
\end{equation}
corresponding to a torsion tensor equal to
\begin{equation}
T^{\Hat \alpha}_{ \ \Hat \beta\Hat \gamma} := 2 \,  \kappa^{\hat \alpha}_{ \phantom{ \Hat \alpha}  [\Hat \beta\Hat \gamma]}
= \frac12 \overline\psi_{\Hat\beta} \, \gamma^{\Hat\alpha} \, \psi_{\Hat\gamma}  =   - \frac12 \overline\psi_{\Hat\gamma} \, \gamma^{\Hat\alpha} \, \psi_{\Hat\beta}  \,\qquad .
\end{equation}
Here, we made use of the anticommuting character of the (classical) Rarita--Schwinger field which implies : 
\begin{equation}
\psi_{\Hat\alpha}^T \, M \, \psi_{\Hat\beta} = - \psi_{\Hat\beta}^T \, M^T \, \psi_{\Hat\alpha}
\end{equation}
for any even bi-spinorial matrix $M$.

By contrast, the Rarita--Schwinger action piece involves a connection ${\cal D}$ that is Levi--Civita ($\ring\omega$) with respect to the space-time vector index of $\psi_{\hat\alpha}$ (here viewed in a frame) but which is the full $\omega = \ring\omega + \kappa$ when acting on the spinor index : 
\begin{eqnarray}
8\,\pi\,G\,{\mathcal L}_{\rm RS} &= &+ \frac12 \, \theta \, \overline\psi_{\Hat\alpha} \, \gamma^{\Hat\alpha \Hat\beta \Hat\gamma} \, {\cal D}_{\hat\beta} \, \psi_{\hat\gamma}\nonumber \\
&= &+\frac12 \, \theta \, \overline\psi_{\Hat\alpha} \, \gamma^{\Hat\alpha \Hat\beta \Hat\gamma} \left(\ring\nabla_{\Hat\beta} \, \psi_{\Hat\gamma} + \frac14 \, \kappa_{\Hat\rho \Hat\sigma \Hat\beta} \, \gamma^{\Hat\rho \Hat\sigma} \, \psi_{\Hat\gamma}\right) \nonumber \qquad .
\end{eqnarray}
Here $\theta = \det (\theta_{\gamma}^{\Hat\alpha})$, and $\ring\nabla$ denotes the usual covariant derivation with respect to the Levi--Civita connection $\ring\omega_{\Hat\alpha \Hat\beta \Hat\gamma} = \omega_{\Hat\alpha \Hat\beta \mu} \, \theta^{\mu}_{\hat\alpha}$ while  
\begin{equation}
{\cal D}_{\Hat\beta} \, \psi_{\Hat\gamma} = \partial_{\Hat\beta} \, \psi_{\Hat\gamma} \, + \ring\omega_{\hat\gamma \hat\sigma \Hat\beta} \, \psi_{\Hat\sigma} + \frac14 \, \omega_{\Hat\rho \Hat\sigma \, \Hat\beta} \ \gamma^{\Hat\rho \Hat\sigma} \, \psi_{\hat\gamma}  \qquad .
\end{equation}

Let us recall that, at the classical level, the spin $3/2$ gravitino field, $\psi_{\hat\alpha}$, satisfies  the Majorana ``reality'' condition :
\begin{equation}
\overline\psi_{\hat\alpha} = \psi_{\hat\alpha}^\dagger \beta = \psi_{\hat\alpha}^T {\mathcal C} \quad .
\end{equation}
In general -- but not always -- we will not explicitly indicate the spinorial indices. The Dirac matrix $\beta$ and the charge conjugation matrix ${\mathcal C}$ obey the (representation-independent) relations $\gamma_{\mu}^\dagger = - \beta \, \gamma_{\mu} \, \beta^{-1}$ and $\gamma_{\mu}^T = - {\mathcal C} \, \gamma_{\mu} \, {\mathcal C}^{-1}  $ and may be chosen such that $\beta^\dagger = \beta$ and ${\mathcal C}^T = - {\mathcal C}$ (conditions which still leave room for some arbitrariness). The latter relations imply the (representation-independent) Dirac matrices property
\begin{equation}
{\mathcal C} \, \gamma_{\mu} = - \gamma_{\mu}^T \, {\mathcal C} = ({\mathcal C} \, \gamma_{\mu})^T \qquad .
\end{equation}

\smallskip

In a Majorana representation, where all the Dirac matrices are real, and satisfy  :  $\gamma_{\Hat 0} = - \gamma_{\Hat 0}^T$, $\gamma_{\Hat k} = + \gamma_{\Hat k}^T$, it is convenient to choose 
$$
\beta = {\mathcal C} = i \, \gamma_{\Hat 0} = -i \, \gamma^{\Hat 0} \quad .
$$
Note that the conjugation $\overline\psi = \psi^\dagger i  \, \gamma_{\Hat 0}= \psi^T  i  \, \gamma_{\Hat 0} $ defined here differs by a factor $-i$ from the convention used in \cite{Damour:2006xu}.

\smallskip

Finally, the explicit second-order form of the total Lagrangian (\ref{LGRS}) can be expressed (up to a divergence term) as  : 
\begin{equation}
\label{LTot}
8\,\pi\,G\, \ {\mathcal L}_{\rm tot} = \theta \left[\frac1{2} \ring R + \ring L_{3/2} + \frac12 \, T^{\Hat\alpha} \, T_{\Hat\alpha} - \frac14 \, T^{\Hat\alpha \Hat\beta \Hat\gamma} \, T_{\Hat\gamma \Hat\beta \hat\alpha} - \frac1{8} \, T^{\Hat\alpha \Hat\beta \Hat\gamma} \, T_{\Hat\alpha \Hat\beta \Hat\gamma} \right]
\end{equation}
where $\ring R$ is the scalar curvature associated to the $\ring\omega$ connection (the standard Einstein--Hilbert Lagrangian) and $\ring L_{3/2}$ is the Rarita--Schwinger Lagrangian part quadratic into the spinorial field
\begin{equation}
\ring L_{3/2} = \frac 12 \, \overline\psi_{\Hat\alpha} \, \gamma^{\Hat\alpha \Hat\beta \Hat\gamma} \, \ring\nabla_{\Hat\beta} \, \psi_{\Hat\gamma} \qquad .
\end{equation}
The detailed computation of the Bianchi IX reduction of the (simpler) Einstein--Dirac Lagrangian was discussed in \cite{Damour:2011yk}. Here the calculation is analogous except for the following facts: (i) the part of the Lagrangian that is quadratic in the fermions involves an extra contribution due to the vectorial part of the  $\psi_{\hat\alpha}$ field, and, (ii) there are now terms quartic  in the fermions.

\smallskip

The total Lagrangian (\ref{LTot}) consists of three kinds of terms: (i) a gravitational part $\frac\theta2 \, \ring R$; (ii) terms quadratic in $\psi : \theta \ \ring L_{3/2}$; and (iii) terms quartic in $\psi : \propto T^2$. Let us look in detail at the structure of the first two types of terms.

\smallskip

In terms of the rotating frame components $N^{\overline a}$ of the shift vector and of the angular velocity components ($w^1  : = w_{\overline 2 \overline 3}$; $w^2 : = w_{\overline 3 \overline 1}$; $w^3 : = w_{\overline 1 \overline 2}$), see Eqs. (\ref{Nab}), (\ref{wab}), the Einstein--Hilbert Lagrangian density reads (with $g:= {\rm det} g_{ij}$) : 
\begin{eqnarray}
8\pi G \, \ring{\mathcal L}_{\rm EH}&=&\frac 12 N\, \sqrt{g}\,\ring R\nonumber  \\
&=&\frac 1 {N } \, e^{-\sum_a\beta^a}\left\{-(\dot \beta^1\dot \beta^2+\dot \beta^2\dot \beta^3+\dot \beta^3\dot \beta^1)+(N^{\overline 1}+w^1)^2\sinh^2[\beta^2-\beta^3]\right.\nonumber \nonumber  \\
&&\left.+(N^{\overline 2}+w^2)^2\sinh^2[\beta^3-\beta^1]+(N^{\overline 3}+w^3)^2\sinh^2[\beta^1-\beta^2]\right\}\nonumber \nonumber  \\
&&-{N }\,\left\{ \frac14 \,e^{\sum_a\beta^a}\sum_b e^{-4\beta^b}-\frac 12 e^{-\sum_a\beta^a}\sum_b e^{2\beta^b}\right\} \qquad .\nonumber
\end{eqnarray}
This is conveniently rewritten as 
\begin{eqnarray}
\label{LEH}
8 \pi G \, \ring{\mathcal L}_{\rm EH} &=& \frac{1}{2\widetilde N} \left[ \dot\beta^a \, G_{ab} \, \dot\beta^b + (N^{\overline k} + w^k) \, K_{k\ell} (N^{\overline \ell} + w^{\ell})\right] - \widetilde N \, V_g (\beta) \nonumber \\
&\equiv &\frac1{\widetilde N} \, [T_{\beta} + T_w] - \widetilde N \, V_g(\beta) \qquad .
\end{eqnarray}
Here we defined the rescaled lapse $\widetilde N := N/\sqrt g = Ne^{\beta^1 + \beta^2 + \beta^3}$, and we introduced the 
quadratic form $G_{ab}$ defined by 
\begin{equation}
G_{ab} \, \dot\beta^a \, \dot\beta^b := \sum_a (\dot\beta^a)^2 - \left( \sum_a \dot\beta^a \right)^2 = -2 (\dot\beta^1 \, \dot\beta^2 + \dot\beta^2 \, \dot\beta^3 + \dot\beta^3 \, \dot\beta^1) \qquad ,
\end{equation}
 {\it i.e.}\   
\begin{equation}
\label{Gab}
G_{ab} = - \begin{pmatrix} 0 &1 &1 \\ 1 &0 &1 \\ 1 &1 &0 \end{pmatrix}
\end{equation}
to express the kinetic terms $T_{\beta} = \frac12 \, G_{ab} \, \dot\beta^a \, \dot\beta^b$ of the logarithmic scale factors: $\beta^1 = -\log a$, $\beta^2 = -\log b$, $\beta^3 = -\log c$, measuring the squashing of the three-geometry. The matrix $G_{ab}$ has signature $(- \, + \, +)$ and will play a crucial r\^ole below where it will appear as the metric of the Cartan subalgebra of the hyperbolic Kac--Moody algebra ${\rm AE}_3$ \cite{Damour:2002et}. 

\smallskip 

The kinetic term $T_w$ is associated with the ``rotational kinetic energy of the frame'', and involves the inertia matrix
\begin{equation}
\label{Kkld}
K_{kl}= 2
\left(
\begin{array}{ccc}
 \sinh^2[\beta^2-\beta^3] & 0  &  0 \\
0  &  \sinh^2[\beta^3-\beta^1]   & 0  \\
0  &  0 &    \sinh^2[\beta^1-\beta^2] 
\end{array}
\right)
\end{equation}
which becomes singular on the ``symmetry walls'' : $\beta^1 = \beta^2$, $\beta^2 = \beta^3$ or $\beta^3 = \beta^1$. 

\smallskip

The potential term $V_g (\beta)$ entering the gravitational action is given by
\begin{equation}\label{Vg}
V_g = \frac14 \sum_a e^{-4\beta^a}-\frac 12 e^{-2\sum_b\beta^b}\sum_a e^{2\beta^a} \qquad .
\end{equation}
It involves the ``gravitational wall forms'' $\beta^a + \beta^b$. [For the general definition of symmetry walls and gravitation walls see \cite{Damour:2002et}.]

\smallskip

Let us now consider the quadratic spinorial term $\ring{\mathcal L}_{3/2}$. 
Similarly to what was used in \cite{Damour:2011yk} and in many previous works, one can simplify the kinetic term of the gravitino $\psi$ by replacing it by the following rescaled gravitino field $\Psi$ : 
\begin{equation} \label{rescaledpsi}
\Psi_{\Hat\alpha} := g^{1/4} \, \psi_{\Hat\alpha}
\end{equation}
where $g= a^2 b^2 c^2 = e^{- 2 (\beta^1 + \beta^2 + \beta^3)}$ denotes the determinant of $g_{ij}$. This leads to :
{\small \begin{eqnarray}
\ring{\mathcal L}_{3/2} &=& \frac12 \ \overline\Psi_{\Hat p} \, \gamma^{\Hat p \Hat 0 \Hat q} \, \dot\Psi_{\hat q} \\
&& +\frac12 {e^{\sum_k\beta^k}} \left(\overline\Psi_{\hat 0}\sum_p e^{-2\beta^p}{{\widetilde\sigma}}_{\Hat p}\Psi_{\Hat p}+\sum_p e^{-2\beta^p}\overline\Psi_{\Hat p}\gamma_\star\Psi_{\Hat p}\right)\nonumber  \\
&&+\frac 1{2\,N}\sum_{\circlearrowright \{p,q,r\}}\overline\Psi_{\Hat p} ({{\widetilde\sigma}}_{\Hat p}(N^q+w^q)e^{\beta^r-\beta^p}+{{\widetilde\sigma}}_{\Hat q}(N^p+w^p)e^{\beta^r-\beta^q}+(\dot\beta^q-\dot\beta^p){{\widetilde\sigma}}_{\Hat r})\Psi_{\Hat q}\nonumber  \\
&&-\frac 1{2\,N}\sum_{\circlearrowright \{p,q,r\}}\overline\Psi_{\Hat p} ({{\widetilde\sigma}}_{\Hat q}(N^q+w^q)e^{\beta^r-\beta^p}+{{\widetilde\sigma}}_{\Hat r}(N^r+w^r)e^{\beta^q-\beta^p})\Psi_{\Hat p}\nonumber  \\
&&+\frac 1{2\,N}\overline\Psi_{\hat 0}\sum_{\circlearrowright \{p,q,r\}}\left[(\dot\beta^q+\dot\beta^r)\gamma^{\Hat p}+\sinh(\beta^p-\beta^r)(N^q+w^q)\gamma^{\Hat r}\right.\nonumber  \\
&&\phantom{+\frac 1{2\,N}\overline\Psi_{\hat 0}\sum_{\circlearrowright \{p,q,r\}}\left[(\dot\beta^q+\dot\beta^r)\gamma^{\Hat p}\right.}\left.-\sinh(\beta^p-\beta^q)(N^r+w^r)\gamma^{\Hat q}\right]\Psi_{\Hat p}\nonumber  \\
&&+\frac14 {e^{\sum_k\beta^k}}\sum_{\circlearrowright \{p,q,r\}}\overline\Psi_{\Hat p}  (e^{-2\beta^p}+e^{-2\beta^q}-e^{-2\beta^r})\gamma_{\Hat r}\Psi_{\Hat q}\nonumber  \\
&&+\frac {1}{2\,N}\sum_{\circlearrowright \{p,q,r\}}\overline\Psi_{\Hat p} \cosh(\beta^p-\beta^q)\gamma_{\hat 0}(N^r+w^r)\Psi_{\Hat q}
 \end{eqnarray}}
where we introduced the notation
\begin{equation}
\sum_{\circlearrowright \{i,j,k\}}A_iB_jC_k  : =A_1B_2C_3+A_2B_3C_1+A_3B_1C_2 \qquad ,
\end{equation}
to indicate a sum on all circular  permutations of the indices, and
\begin{eqnarray}
{{\widetilde \sigma}}_{\Hat i} := \frac 12 \varepsilon_{{\Hat i}{\Hat j}{\Hat k}} \gamma^{\hat 0}\gamma^{{\Hat j}{\Hat k}}&\qquad,\qquad& \gamma_\star := \gamma_{\hat 1}\gamma_{\hat 2}\gamma_{\hat 3}\\
 {\widetilde \sigma}_{\Hat i}^T={\cal C}{\widetilde \sigma}_{\Hat i}{\cal C}^{-1}&\qquad,\qquad& \gamma_\star^T={\cal C}\gamma_\star{\cal C}^{-1}
\end{eqnarray}

Before discussing the full structure of the gravitino action, let us focus on its kinetic term 
\begin{equation}
\label{T32}
T_{3/2} = +\frac12 \, {\overline\Psi}_{\hat a} \, \gamma^{\hat a \hat 0 \hat b} \, \dot\Psi_{\hat b} = -\frac12 \, \dot{\overline\Psi}_{\hat a} \, \gamma^{\hat a \hat 0 \hat b} \, \Psi_{\hat b}
\end{equation}
where ${\overline\Psi}_{\hat a} = \Psi_a^T \, {\mathcal C}$. The structure of this kinetic term is clarified by replacing the (rescaled) gravitino field $\Psi_{\hat a}$ by the new gravitino variables 
\begin{eqnarray}
\label{Phia}
\Phi^a & : = &\gamma^{\hat a} \, \Psi_{\hat a} \qquad \mbox{(no sum on $a$)}\\
\label{Phiab}
\overline\Phi^a &= &-\overline\Psi_{\hat a} \, \gamma^{\hat a} 
\end{eqnarray}
that proved to be convenient in the study of fermionic Kac--Moody billiards \cite{Damour:2009zc}. In terms of these new gravitino variables (and choosing ${\cal C} = i \gamma_{\hat 0}$) the kinetic term (\ref{T32}) simplifies to
\begin{equation}
T_{3/2} = + \frac i2 \, G_{ab} \, \Phi^{a T} \, \dot\Phi^b \qquad .
\end{equation}
This simple form makes more manifest the (super)symmetry between the $\beta^a$'s and the $\Phi^a$'s (and the
fact that supergravity is a ``squareroot" of General Relativity \cite{Teitelboim:1977fs,Tabensky:1977ic}).

\setcounter{equation}{0}\section{Hamiltonian formulation}

In the following we shall use units such that $c = \hbar = 1$, and such that the value of Einstein's gravitational constant  $(8\pi G)^{-1}$ (which we factored out of the total supergravity action) absorbs the spatial-volume factor $V_3 = \int_{SU(2)} \tau^1 \wedge \tau^2 \wedge \tau^3$ of the undeformed three-sphere corresponding to $a=b=c=1$. In view of the normalization $C^i_{jk} = \varepsilon_{ijk}$ of the one-forms $\tau^i$, this round three-sphere (homeomorphic to the group manifold $SU(2)$) with $a=b=c=1$ has a curvature radius equal to $R=2$ and hence a volume $V_3 = 2\pi^2 \, R^3 = 16\pi^2$. In other words, we set $8\pi G = V_3 = 16\pi^2$. 

\smallskip

With such a choice, the bosonic momenta are 
\begin{equation}
\label{pia}
\pi_a = \frac{\partial {\mathcal L}}{\partial \, \dot\beta^a} = \frac1{\widetilde N} \, G_{ab} \, \dot\beta^b + \Pi_a \qquad ,
\end{equation}
\begin{equation}
\label{pa}
p_a = \frac{\partial {\mathcal L}}{\partial \, w^a} = \frac1{\widetilde N} \, K_{ab} (N^{\overline b} + w^b)  + P_a \qquad ,
\end{equation}
where the extra terms $\Pi_a , P_a$ are quadratic in $\Psi$ and come from velocity-dependent couplings associated with the $\ring\omega$ part of the connection (\ref{wabe}). More precisely, such velocity couplings come from the connection terms in $\overline\Psi \ring\nabla \Psi$ with (without any summation on repeated indices)
\begin{equation}
\ring\omega_{\hat 0 \hat a \hat b} = \frac1N \left( \dot\beta^a \, \delta_{ab} + \sinh (\beta^a - \beta^b) (w_{\overline a \overline b} + \varepsilon_{abc} \, N^{\overline c}) \right) \qquad ,
\end{equation}
\begin{equation}
\ring\omega_{\hat a \hat b \hat 0} = -\frac1N \, \cosh (\beta^a - \beta^b) (w_{\overline a \overline b} + \varepsilon_{abc} \, N^{\overline c})  \qquad ,
\end{equation}
and give rise to an action contribution of the form 
\begin{equation}
\label{L32V}
{\mathcal L}_{(3/2 \, {\rm vel})}  \equiv  \dot\beta^a \, \Pi_a + (N^{\overline a} + w^a) \, P_a \qquad .
\end{equation}

Expressed in terms of   $\Psi_{\hat a}$ and $ \Psi^{\hat 0}$ the expressions of $\Pi_a$ and $P_a$ are rather complicated. 
They simplify when replacing the $\Psi_{\hat a}$'s by  the new gravitino variables $\Phi^a$, Eq. (\ref{Phia}), and 
$ \Psi^{\hat 0}$ by its following shifted version 
\begin{equation}
\label{psi'0}
\Psi^{\prime\hat 0}  : =  \Psi^{\hat 0} -   \gamma^{\hat 0} \sum_a  \gamma^{\hat a}   \Psi_{\hat a} =  \Psi^{\hat 0} -   \gamma^{\hat 0} \sum_a    \Phi^{ a}\qquad ,
\end{equation}
whose vanishing defines a convenient ``Kac--Moody coset gauge"  \cite{Damour:2006xu}.
In terms of these variables, the spin-dependent contributions $\Pi_a $ and $P_a$ to the momenta 
$\pi_a$, $p_a$ read
\begin{equation}
\label{Pia}
\Pi_a = \frac12  G_{ab}  \overline\Psi'^0 \Phi^b
\end{equation}
 and
\begin{equation}
\label{Pa}
P_a =\frac12 \, \sum_{k,l} \varepsilon_{\hat a \hat k \hat l} \left( \cosh (\beta^k - \beta^{l}) \, S^{[kl]} - \overline\Psi^{\prime\hat 0} \sinh (\beta^k - \beta^{l}) \, \gamma^{\hat k \hat l} \, \Phi^{l}\right) \end{equation}
where
\begin{eqnarray}
S^{[12]} &= &\frac12 \Bigl( \overline\Phi^3 \gamma^{\hat 0 \hat 1 \hat 2} (\Phi^1 + \Phi^2) + \overline\Phi^1 \gamma^{\hat 0 \hat 1 \hat 2} \, \Phi^1    + \overline\Phi^2 \gamma^{\hat 0 \hat 1 \hat 2}  \, \Phi^2 - \overline\Phi^1 \gamma^{\hat 0 \hat 1 \hat 2}  \, \Phi^2  
\Bigl)\nonumber \\
\label{S12}&=& - \, S^{[21]}  \, .
\end{eqnarray}
Similar objects $S^{[23]} $ and $S^{[31]} $ are defined by cyclic permutations of the indices. 

\smallskip

The spatial components of the Levi--Civita connection, {\it i.e.}\   (without summation on repeated indices)
\begin{equation}
\ring\omega_{\hat a \hat b \hat c} = \frac12 \, e^{\sum_d \beta^d} (e^{-2\beta^a} + e^{-2\beta^b} - e^{-2\beta^c}) \, \varepsilon_{abc} \qquad ,
\end{equation}
give rise to velocity-independent action terms coupling the $\beta$'s to quadratic terms in the fermions, namely\footnote{We define $\gamma_5 \equiv \gamma^5= \gamma^{\hat 0} \, \gamma^{\hat 1} \, \gamma^{\hat 2} \, \gamma^{\hat 3}$ and $\gamma_* = \gamma^{\hat 1} \, \gamma^{\hat 2} \, \gamma^{\hat 3}$.} 
\begin{eqnarray}
V_{s,2} &= &\frac12 \, \overline\Psi^{\hat 0} \sum_p e^{-2\beta^p} \gamma_5 \, \Phi^p + \frac12 \sum_p \overline\Phi^p e^{-2\beta^p} \gamma_* \, \Phi^p \nonumber \\
&- &\frac14 \sum_{\circlearrowright  \{ p,q,r \} } \overline\Phi^p (e^{-2\beta^p} + e^{-2\beta^q} - e^{-2\beta^r}) \, \gamma_* \, \Phi^q \qquad .  
\end{eqnarray}
There are also quartic terms in the fermions, issuing from the quadratic terms in the torsion that appears in the total Lagrangian (\ref{LTot}). They consist of terms quadratic or linear in $\Psi^{\hat 0}$ and of terms independent from $\Psi^{\hat 0}$ :  
\begin{eqnarray}
\label{Vs4}
V_{s,4} &= &\frac18 \, \left( \overline\Psi^{\hat 0} \, \Phi \right)^2 - \frac1{32} \sum_{k,\,l} \left( \overline\Psi^{\hat 0} (\gamma^{\hat k} \, \gamma^{\hat l} \, \Phi^{l} + \gamma^{\hat l} \, \gamma^{\hat k} \, \Phi^k) \right)^2 \nonumber \\
&- &\frac14 \sum_k \left( \overline\Psi^{\hat 0} \gamma_{\hat 0} \, \gamma^{\hat k} \, \Phi^k \right)\left(\overline\Phi^k \Phi \right) \nonumber \\
&- &\frac18 \sum_{k,l} \left( \overline\Psi^{\hat 0} \gamma_{\hat k} \, \gamma^{\hat l} \, \Phi^{l} \right)\left( \overline\Phi^{l} \gamma^{l} \, \gamma^{\hat 0} \, \gamma^k \, \Phi^k \right) \nonumber \\
&+ &\frac18 \sum_k \left( \overline\Phi \gamma^{\hat k} \, \Phi^k \right) \left( \overline\Phi^k \gamma^{\hat k} \, \Phi \right) \nonumber \\
&+ &\frac1{16} \sum_{k,\,l,\, n} \left( \overline\Phi^k \gamma^{\hat k} \, \gamma^{\hat l} \, \gamma^{\hat n} \, \Phi^n\right) \left( \overline\Phi^k \gamma^{\hat k}  \, \gamma^{\hat n} \, \gamma^{\hat l} \, \Phi^{l} \right) \nonumber \\ 
&- &\frac1{32} \sum_{k,\,l} \left( \overline\Phi^k \gamma^{\hat k}  \, \gamma^{\hat 0} \, \gamma^{\hat l} \, \Phi^{l} \right) \left( \overline\Phi^k \gamma^{\hat k}  \, \gamma^{\hat 0} \, \gamma^{\hat l} \, \Phi^{l} \right) \nonumber \\
&+ &\frac1{32} \sum_{k,\,l,\, n} \left( \overline\Phi^k \gamma^{\hat k}  \, \gamma^{\hat n} \, \gamma^{\hat l} \, \Phi^{\hat l} \right) \left( \overline\Phi^k \gamma^{\hat k}  \, \gamma^{\hat n} \, \gamma^{\hat l} \, \Phi^{\hat l} \right) \qquad .
\end{eqnarray}
Here, and below, we use the short-hand notation  
\begin{equation}
\label{Phi}
\Phi  : = \sum_a \Phi^a  
\end{equation}
in terms of which we  have $ \Psi'^{\hat 0}  =  \Psi^{\hat 0} - \gamma^{\hat 0} \, \Phi $.

Let us finally discuss the issue of the Hamiltonian formulation of the gravitino variables.
When dealing with classical ( {\it i.e.}\ Grassmannian) fermionic variables $\Psi_A$, the fundamental Poisson brackets ($\{\,,\,\}_{_P} $) between them and their canonical conjugate momenta $\varpi^B$ are (see ref. \cite{HenTei})
\begin{equation}
\label{PBF}
\{ \Psi_A , \varpi^B \}_{_P} = -\delta_A^B
\end{equation}
As usual for Grassmannian degrees of freedom, the Lagrangian is of   first order in the time derivative. Let us consider a general one given by : 
\begin{equation}
{\mathcal L}_F = \frac12 \, \Psi_A \, M^{(AB)} \, \dot\Psi_B = -\frac12 \, \dot\Psi_A \, M^{(AB)} \, \Psi_B \qquad .
\end{equation}
The conjugate momenta are defined by a left derivative : 
\begin{equation}
\varpi^A  : = \frac{\partial^L {\mathcal L}_F}{\partial \, \dot\Psi_A} = -\frac12 \, M^{(AB)} \, \Psi_B \qquad .
\end{equation}
As a consequence we have the constraints
\begin{equation}
\label{MC}
\chi^A  : = \varpi^A + \frac12 \, M^{(AB)} \, \Psi_B \approx 0 \qquad ,
\end{equation}
and using the Poisson brackets (\ref{PBF}) we obtain : 
\begin{equation}
\{ \chi^A , \chi^B \} _{_P}= -M^{(AB)} \qquad .
\end{equation}
Thus, assuming the kinetic matrix $M^{AB}$ to be invertible,  the constraints (\ref{MC}) are of second class. Accordingly, the Dirac brackets ($\{\,,\,\}_{_D} $) of the fermionic variables are given by :
\begin{equation}
\{ \Psi_A , \Psi_B \}_{_D} = M_{(AB)} 
\end{equation}
where $M_{AB}$ are the components of the inverse of the matrix $(M^{AB})  :  M^{AB} \, M_{BC} = \delta_C^A$. The canonical quantization corresponding to these fermionic Dirac brackets will then lead to anticommutators $\{ \, , \, \}$ equal to
\begin{equation}
\{ \widehat\Psi_A , \widehat\Psi_B \} = i \{ \Psi_A , \Psi_B \}_{_D} = i \, M_{AB} \qquad .
\end{equation}

\smallskip

In the case that interests us here, starting from the kinetic term (\ref{T32}) we obtain
\begin{equation}
\varpi_A^{\hat p} = \frac{\partial^L \, T_{3/2}}{\partial \, \dot\Psi_{\hat p}^A} = \frac12 \, (\Psi_{\hat q}^T \, {\mathcal C} \, \gamma^{\hat 0 \hat q \hat p})_A
\end{equation}
which implies linear second class constraints, from which we infer
\begin{equation}
\label{PsiAC}
\{ \Psi_{\hat p}^A , \Psi_{\hat a}^B \}_{_D} = -\frac12 \, (\gamma_{\hat a} \, \gamma_{\hat p} \, \gamma^{\hat 0} \, {\mathcal C}^{-1})_{AB} \qquad .
\end{equation}
In the Majorana representation we use this simplifies to
\begin{equation}
\label{PsiAC2}
\{ \Psi_{\hat p}^A , \Psi_{\hat q}^B \}_{_D} = -\frac i2 \, (\gamma_{\hat q} \, \gamma_{\hat p})_{AB} \qquad .
\end{equation}
When using the new gravitino variables (\ref{Phia}), this further simplifies to 
\begin{equation}
\{ \Phi^a_A , \Phi^b_B \}_{_D} = - i \, G^{ab} \, \delta_{AB} \qquad .
\end{equation}

After a tedious, but standard, calculation we obtain an Hamiltonian action ${\cal L}_H = p\, \dot q  - H_{\rm tot} (q,p)$ of the form 
\begin{equation}
\label{HTot}
{\cal L}_H = \pi_a \, \dot\beta^a + p_a \, w^a + \frac i2 \, G_{ab} \, \Phi^{aT} \, \dot\Phi^b + \widetilde N \, \overline\Psi'^A_{\hat 0} {\cal S}_A - \widetilde N H - N^i \, H_i \qquad .
\end{equation}
Here $\widetilde N = N g^{-1/2}$ as above. 
The structure of the total Hamiltonian entering (\ref{HTot}) is the one expected in a theory with local invariances. It involves eight Lagrange multipliers corresponding to eight local gauge symmetries :  the four components of $\overline\Psi'_{\hat 0}$ (local supersymmetry), the rescaled lapse function $\widetilde N$ (local temporal diffeomorphisms) and the three shift functions $N^i$ (local spatial diffeomorphisms). The variation of these Lagrange multipliers leads to the eight corresponding constraints : 
\begin{enumerate}
\item[--] the four supersymmetry constraints (henceforth often abbreviated as ``susy constraints'')
\begin{equation}
\label{CSA}
{\cal S}_A \approx 0
\end{equation}
\item[--] the four diffeomorphisms constraints, that can be split into the Hamiltonian constraint, linked to time reparametrizations
\begin{equation}
\label{CH}
H \approx 0
\end{equation}
and the momentum constraints
\begin{equation}
\label{Cpa}
H_i \approx 0
\end{equation}
reflecting spacelike coordinate reparametrisations.
\end{enumerate}
Let us note in passing the remarkable fact (already emphasized in \cite{Deser:1977ur}) that, starting from Lagrangian action that is quartic in the fermions, the Hamiltonian ends up being linear in $\Psi_{\hat 0}$. [The use of the shifted variable $\Psi'_{\hat 0}$, Eq. (\ref{psi'0}), is convenient, and linked to the Kac--Moody-coset gauge-fixing used in \cite{Damour:2006xu}.] 

\smallskip

Before giving the explicit form of ${\cal S}_A$, $H$ and $H_i$, let us note the fact that $H_i$ has a very simple link with the momentum $p_a$ conjugated to the angular velocity $w^a$. Indeed, as appears in Eqs. (\ref{LEH}), (\ref{pa}), (\ref{L32V}), the shift vector enters the Lagrangian action always in the combination
$$
w_{\overline a \overline b} + \varepsilon_{abc} \, N^{\overline c} = \varepsilon_{abc}  (w^c + N^{\overline c} )
$$
where we recall that
$$
N^{\overline a} \equiv S_i^{\overline a} \, N^i \qquad .
$$
As a consequence, one concludes that (similarly to the Einstein--Dirac case \cite{Damour:2011yk})
\begin{equation}
\label{Hipa}
H_i = -S_i^{\overline a} \, p_a \qquad .
\end{equation}
The coincidence between these Euler-angle-related momenta and the spatial diffeomorphism constraints is the result of the coincidence between  the adjoint representation of the homogeneity group of the Bianchi IX cosmological model, and the $SO(3)$ 
automorphism group of the structure constants $C^a_{\ b c}$  which was used in the Gauss decomposition Eq. (\ref{Gauss}) to parametrise the 3-beins $h_i^{\hat k}$. Let us emphasise that the Poisson brackets of the $p_a$ between themselves do not vanish\footnote{For the interpretation of the minus sign occurring on the right hand side of these Poisson brackets see ref. \cite{Damour:2011yk}, section (3.2). Notice that they are the typical Lie-Poisson brackets obtained from a reduction of the $so(3)$ algebra by a Poisson map.[See ref.\cite{JMTR}]}
\begin{equation}
\{p_a , p_b \} = -\varepsilon_{abc}  \,p_c
\end{equation}
as it is the case in general for the momenta constraint of a diffeomorphism invariant theory of gravity coupled to matter. The rotating frame components of the momenta $p_a$ are (Euler-angle-dependent) linear combinations of the momenta conjugate to the Euler angles. 

\smallskip

In addition, the dependence of the other constraints, {\it i.e.}\   ${\cal S}_A$ and $H$, on the rotational momenta $p_a$ is found to be quite simple; namely we have
$$
{\cal S}_A = {\cal S}_A^{(0)} + {\cal S}_A^{\rm rot} \qquad ,
$$
$$
H = H^{(0)} + H^{\rm rot} \qquad ,
$$
where the superscript $(0)$ indicates a reduction to zero rotational momenta, and where
\begin{equation}
\label{Srotclass}
{\cal S}_A^{\rm rot}= + \frac 14 \, \frac{p_3}{\sinh (\beta^1-\beta^2)} \left(\gamma^{\hat 1\hat 2}( \Phi^1 - \Phi^2)\right)_A+\text{cyclic}_{123} \, ,
\end{equation}
\begin{equation}
\label{Hrotclass}
H^{\rm rot} = \frac14 \, \frac1{\sinh^2  (\beta^1-\beta^2)} \left(  p_3^2 -  2 \, p_3 \,  \cosh  (\beta^1-\beta^2) \,  S_{12} \right) + {\rm cyclic}_{123} \, .
\end{equation}
With this notation the $p_a$-independent piece of ${\cal S}_A$ explicitly reads
\begin{equation}
\label{SaredCla}
{\cal S}^{(0)}_A = - \frac12 \sum_a \, \pi_a \,  \Phi_A^a + {\cal S}_A^g + {\cal S}_A^{\rm sym} +  {\cal S}_A^{\rm cubic}
\end{equation}
with
\begin{equation}
\label{SgACl}
 {\cal S}_A^g = \frac12 \sum_a e^{-2\beta^a} \left(\gamma^5 \, \Phi^a\right)_A\qquad ,
\end{equation}
\begin{equation}
\label{SsACl}
{\cal S}_A^{\rm sym} = -\frac14 \coth [\beta^1 - \beta^2] \, S_{12} \left( \gamma^{\hat1\hat2} \left( \Phi^1 - \Phi^2 \right)\right)_A +{\rm cyclic}_{123} \ ,
\end{equation}
and   
\begin{eqnarray}
\label{SigcA}
{\cal S}_A^{\rm cubic} &= &\frac1{8} \sum_{k \neq l}\left[  \left( \overline{\Phi} \, \gamma^{\hat 0\hat k \hat l} \, \Phi^k\right) \left( \gamma^{\hat k \hat l} \left( \Phi^k - \Phi^{l} \right)\right)_A- \left( \overline{\Phi}^k \gamma^{\hat 0 \hat k \hat l} \, \Phi^{l} \right) \left( \gamma^{\hat k \hat l} \,  \Phi^{l} \right)_A \right]\nonumber \\
& & + \frac14 \sum_k \left[\left(\overline{   \Phi} \gamma^{\hat 0}  \Phi^k \right)  \Phi^k_A  - \left( \overline{  \Phi} \gamma^{\hat k} \,  \Phi^k \right) \left( \gamma^{\hat 0 \hat k} \,   \Phi^k \right)_A\right] \qquad .
\end{eqnarray}
As for the $p_a$-independent piece of $H$ it has the structure
\begin{equation}
H^{(0)} = \frac12 \, G^{ab} \, \pi_a \, \pi_b + V (\beta , \Phi)
\end{equation}
where
\begin{equation}
\label{IGab}
(G^{ab}) = \frac12 \begin{pmatrix} 
1&-1&-1 \\
-1&1&-1 \\
-1&-1&1 \end{pmatrix}
\end{equation}
is the inverse of the matrix (\ref{Gab}), and where $V (\beta , \Phi)$ is
 a $\Phi$-dependent potential term of the form
\begin{equation}
\label{Vtotclass}
V(\beta , \Phi) = V_g (\beta) + V_2 (\beta , \Phi) + V_4 (\beta,\Phi) \, .
\end{equation}
Here $V_g (\beta)$ is the usual, purely bosonic (i.e. $\Phi$-independent), Bianchi IX potential (\ref{Vg}), and  the potential contribution quadratic in $\Phi$ has the structure 
\begin{equation}  \label{V2}
V_2 (\beta , \Phi) = \frac12 e^{-2\beta^1} \, J_{11} (\Phi) + \frac12 e^{-2\beta^2} \, J_{22} (\Phi) + \frac12 e^{-2\beta^3} \, J_{33} (\Phi)
\end{equation}
where the $J_{aa} (\Phi) \sim \overline \Phi \, \Phi$ are some quadratic fermionic terms that will be discussed
below [see Eq. (\ref{J11op})]. The final term in Eq. (\ref{Vtotclass}) is quartic in $\Phi$ and is made of two types of contributions: 
\begin{eqnarray}  \label{V4}
V_4 (\beta , \Phi) &= &\frac14 \coth^2 (\beta^1 - \beta^2) \, S_{12}^2 (\Phi) + \frac14 \coth^2 (\beta^2 - \beta^3) \, S_{23}^2 (\Phi) \nonumber \\
&+ &\frac14 \coth^2 (\beta^3 - \beta^1) \, S_{31}^2 (\Phi) + \Phi^4\mbox{-terms} \quad .
\end{eqnarray}

In view of the link (\ref{Hipa}) and of the structure of the rotational contribution to $H$ and ${\cal S}_A$, the eight constraints Eqs. (\ref{CSA}), (\ref{CH}), (\ref{Cpa}) are equivalent to the following eight constraints
\begin{eqnarray}
\label{newconstraints}
p_a &\approx &0 \\
{\cal S}^{(0)}_A &\approx &0 \\
H^{(0)} &\approx &0 \qquad .
\end{eqnarray}
As a consequence of the classical consistency of supergravity, and of the consistency of its Bianchi IX reduction, one can verify that this set of constraints defines an (open) algebra under the classical Dirac-Poisson brackets of the form 
\begin{eqnarray}
\{p_a , p_b \} &= &- C^c_{ \ ab} \, p_c \qquad ,  \\
\label{ClassicalSusyAlg}
\{ {\cal S}^{(0)}_A , {\cal S}^{(0)}_B \}_{_D} &= & 4 \, L^C_{AB} (\beta , \Phi) \, {\cal S}_C^{(0)} - i\,\frac12 \, H^{(0)} \, \delta_{AB} \qquad , \\
\{ {\cal S}^{(0)}_A , H^{(0)} \}_{_D} &= &M_A^B \, {\cal S}_B^{(0)} + N_A \, H^{(0)} \qquad .
\end{eqnarray}
We will discuss below the (more demanding) quantum analog of the above set of constraints.

\setcounter{equation}{0}\section{Quantization}\label{sec2}

We quantize the constrained dynamics defined by the Hamiltonian action (\ref{HTot}) \`a la Dirac, {\it i.e.}\   by: (i) replacing Poisson-Dirac brackets by appropriate (anti-) commutators; (ii) verifying that this allows one to construct operators providing a deformed version of the classical algebra of constraints; and (iii) imposing the quantum constraints $\widehat{\mathcal C}$ as conditions restricting physical states $\vert{\bm\Psi}\rangle$: $\widehat{\mathcal C} \vert{\bm\Psi}\rangle = 0$.

\smallskip

For the bosonic degrees of freedom we adopt a Schr\"odinger picture. The wave function of the Universe is seen as a function of the three exponents $\beta^1$, $\beta^2$ and $\beta^3$ of the scale factors  and of the three Euler angles $\varphi^a$ that parametrize the rotation matrix entering the diagonalization Eq. (\ref{Gauss}) of the metric tensor $g_{ij}$. Accordingly the basic conjugate quantum momenta operators are represented as ($\hbar = 1$)
\begin{eqnarray}
\widehat\pi_a &= &\frac1i \, \partial_{\beta^a} \nonumber \\
\widehat p_{\varphi^a} &= &\frac 1i \, \partial_{\varphi^a} \qquad . \nonumber
\end{eqnarray}
The rotational momenta $p_a$, Eq. (\ref{pa}), associated with the rotational velocity $w^a$ (which are linear in the $\dot\varphi^a$'s) are quantized by the natural ordering corresponding to differential operators acting on the group manifold (see, e.g., \cite{Damour:2011yk}). This ordering guarantees that these operators satisfy a $SU(2)$ algebra:
\begin{equation}
[ \widehat p_a , \widehat p_b ] = -i \, \varepsilon_{abc} \, \widehat p_c \qquad .
\end{equation}
The fermionic operators have to obey anticommutations relations dictated by the Dirac brackets (\ref{PsiAC} -- \ref{PsiAC2}) : 
\begin{equation}
\{\widehat \Psi_{\Hat a}^A , \widehat \Psi_{\Hat b}^B\} = i \{ \Psi_{\hat a}^A , \Psi_{\hat b}^B \}_{_D} = -\frac i2 \left(\gamma_{\Hat b} \, \gamma_{\Hat a} \, \gamma^{\hat 0} \, {\mathcal C}^{-1}\right)_{AB}
\end{equation}
or in terms of operators associated to the new gravitino variables (\ref{Phia}) : 
\begin{equation}
\label{GenAC}
\{\widehat \Phi^a_A , \widehat \Phi^{b}_B \} = -i \, G^{ab} \, (\gamma^{\hat 0} \, {\mathcal C}^{-1})_{AB} = +i \, G^{ab} \, (\gamma_{\hat 0} \, {\mathcal C}^{-1})_{AB}
\end{equation}
where $G^{ab}$ is the inverse of $G_{ab}$ [see Eq. (\ref{IGab})].

The anticommutator (\ref{GenAC}) is written in a way independent of the Dirac-matrices representation. In a Majorana representation where ${\cal C} = i \, \gamma_{\hat 0}$ it simplifies to : 
\begin{equation} \label{ACRPhi}
\{\widehat \Phi^a_A , \widehat \Phi^{b}_B \} = G^{ab} \, \delta_{AB} \qquad .
\end{equation}

\smallskip

This shows that the twelve quantum fermionic operators $\widehat \Phi_A^a$ have to satisfy a Clifford algebra in a 12 dimensional space with signature $(+^8 , -^4)$. Thus the gravitino operators can be represented by $64\times64$ Dirac matrices and the wave function of the Universe by a 64-dimensional spinor, depending on $\beta^a$ and $\varphi^a$~:  ${\bm \Psi} = {{\Psi}}_{\sigma} (\beta^a , \varphi^b)$, with $\sigma = 1,\ldots ,64$. The constraints (\ref{CSA}, \ref{CH}, \ref{Cpa}) have to be represented by operators $\widehat {\cal S}_A$, $\widehat H$ and $\widehat p_a$ and imposed \`a la Dirac on the state $\vert{\bm\Psi}\rangle$ : 
\begin{equation}
\label{DiracQEq}
\widehat {\cal S}_A \, \vert{\bm\Psi}\rangle = 0 \qquad , \quad \widehat H \, \vert{\bm\Psi}\rangle = 0 \qquad , \quad \widehat H_i \, \vert{\bm\Psi}\rangle = 0 \qquad .
\end{equation}
Actually, it shall be more convenient to work with the following alternative form of the constraints,
\begin{equation}
\label{newQconstraints}
\widehat {\cal S}_A^{(0)} \, \vert{\bm\Psi}\rangle = 0 \qquad , \quad \widehat H^{(0)} \, \vert{\bm\Psi}\rangle = 0 \qquad , \quad \widehat p_a \, \vert{\bm\Psi}\rangle = 0 \qquad ,
\end{equation}
in which one has separated out, as in Eq. (\ref{newconstraints}), the ``rotational'' contributions to $\widehat {\cal S}_A$ and $\widehat H$, and used the (naturally ordered) quantum version of the diffeomorphism constraint {\it i.e.} 
\begin{equation}
\label{QHipa}
\widehat H_i = - S^{\overline a}_i \, \widehat p_a
\end{equation}
We have checked that the two sets of quantum constraints (\ref{DiracQEq}), (\ref{newQconstraints}) are equivalent. This follows from the following facts. First, equation (\ref{QHipa}) shows the equivalence of the diffeomorphism constraints to the last constraint in Eq. (\ref{newQconstraints}). Second, the rotational contributions to ${\cal S}_A$ and $H$ (written, at the classical level, in Eqs. (\ref{Srotclass}, \ref{Hrotclass})) are simple and additive. At the quantum level, they do not introduce any ordering ambiguities because the $\widehat p_a$'s commute with the $\beta$'s and $\Phi$'s, and because the only
terms that are quadratic in the $\widehat p_a$'s  are their squares $\widehat p_a^2$:
\begin{equation}\label{SrotQuant}
\widehat {\cal S}_A^{\rm rot}= +\frac 1{4  \sinh (\beta^2-\beta^3)} \widehat p_1\,\left(\gamma^{\hat 2\hat 3}(\widehat \Phi^2-\widehat\Phi^3)\right)_A+\text{cyclic}_{123} \ ,
\end{equation}
\begin{equation}
\label{Hrot}
\widehat H^{\rm rot} = \frac14 \, \frac{1}{\sinh^2 (\beta^2 - \beta^3)}  \left( \widehat p_1^2   -  2 \, \widehat p_1 \,  \cosh  (\beta^2-\beta^3) \,  \widehat S_{23}\right)+ {\rm cyclic}_{123} \quad .
\end{equation}

\setcounter{equation}{0}\section{Ordering of the quantum constraints}

We have seen above that the ordering of the quantum Euler-angle momenta $\widehat p_{\varphi^a}$ is naturally solved by working with the related rotational momenta $\widehat p_a$. There is no ambiguity in the relative ordering of the $\beta^a$'s and their conjugate momenta $\pi_a$ because (after our choice of rescaled lapse $\widetilde N = N e^{\beta^1 + \beta^2 + \beta^3}$) there are no mixed terms $\propto \pi \, f(\beta)$ in the constraints. The $\pi_a$'s appear linearly with $\beta$-independent coefficients in ${\cal S}^{(0)}_A$, while they appear quadratically (again with $\beta$-independent coefficients) in $H^{(0)}$.

\smallskip

Finally, the only quantum ordering ambiguity that might   {\it a priori}   be present in our framework concerns the ordering of the gravitino variables among themselves. However, this issue is {\it uniquely} solved by imposing the following two requests: (i) that the operators $\widehat{\cal S}^{(0)}_A$ satisfy the same hermiticity condition, say 
\begin{equation} \label{hermiticitySA}
\widetilde{\widehat{\mathcal S}}_A^{(0)} = \widehat{\mathcal S}_A^{(0)} \, , 
\end{equation}
than the $\widehat\Phi$ operators they are built from ($\widetilde{\widehat\Phi}_A^{a} = {\widehat\Phi}_A^{a}$) [Here, and henceforth, we use a tilde to denote the Hermitian conjugate ---  in the sense of Eq. (\ref{Xtilde}) below ---  of an operator.]; and (ii) that the anticommutators of the $\widehat{\mathcal S}_A^{(0)}$'s close, similarly to the classical result (\ref{ClassicalSusyAlg}), on $\widehat H^{(0)} \, \delta_{AB}$ modulo a linear combination of the $\widehat{\mathcal S}_A^{(0)}$'s. The requirement (i) will define a unique ordering of the $\widehat{\mathcal S}_A^{(0)}$'s, while the requirement (ii) will then define a unique ordering for $\widehat H^{(0)}$.

\smallskip

The hermiticity conditions on the $\widehat{\cal S}^{(0)}_A$'s  can be imposed purely algebraically,  by using the basic rules: $\widetilde{AB} = \widetilde B \, \widetilde A$, $\widetilde i = -i$, $\widetilde{\widehat \pi}_a = \widehat \pi_a$, $\widetilde{\widehat p}_a = \widehat p_a$, $\widetilde{\widehat\Phi}_A^{a} = {\widehat\Phi}_A^{a}$. It is, however, important to know how it can be practically realized when explicitly representing the Clifford-algebra elements $\widehat\Phi_A^a$ as $64 \times 64$ complex matrices. Indeed, the Clifford algebra ${\rm Spin} \, (8^+ , 4^-)$ can be realized (after diagonalizing the quadratic form $G^{ab} \, \delta_{AB}$) by means of
twelve Dirac matrices that verify ($M,N = 1,\ldots ,12$)
\begin{equation}
\Gamma_M \, \Gamma_N + \Gamma_N \, \Gamma_M = 2\, \eta_{MN}
\end{equation}
where $\eta_{MN} = {\rm diag} \, ( \underbrace{+ \ldots +}_{8} , \underbrace{- \ldots -}_{4})$. They may be chosen such that
\begin{equation}
 \Gamma_M^\dagger =  \Gamma_M  \qquad M = 1,\ldots , 8
\end{equation}
\begin{equation}
\Gamma_M^\dagger = - \Gamma_M \qquad M = 9,\ldots , 12 \qquad .
\end{equation}
Here, the dagger denotes the usual matrix Hermitian conjugation $\Gamma^{\dagger} \equiv \overline\Gamma^T$.
If we introduce the product of the time-like $\Gamma$'s, namely
\begin{equation}
\label{h}
h := \Gamma_9 \, \Gamma_{10} \, \Gamma_{11} \, \Gamma_{12} \, ,
\end{equation}
which satisfies 
\begin{equation}
h = h^{\dagger} = \overline h^T \, , \qquad h^2 = 1 \ \quad ,
\end{equation}
we obtain
\begin{equation}
\Gamma_A^\dagger = h \, \Gamma_A \, h^{-1} \qquad A = 1,\ldots,12 \qquad ,
\end{equation}
{\it i.e.}  $\widetilde\Gamma_A = \Gamma_A$ with
\begin{equation}\label{Xtilde}
\widetilde X := h^{-1} \, X^{\dagger} \, h \equiv h^{-1} \, \overline X^T \, h \qquad .
\end{equation}
This definition of Hermitian conjugation of the fermionic variables is related to endowing the 64-dimensional fermionic Hilbert space ({\it i.e.}  the space of ${\rm Spin} (8,4)$ spinors) with the pseudo-hermitian inner product
\begin{equation}
\label{scalprod}
\langle u \mid v \rangle_h = \overline u^T h \, v
\end{equation}
satisfying $\overline{\langle u \mid v \rangle}_h = \langle v \mid u \rangle_h$. Indeed, the hermitian conjugate is easily checked to be such that 
\begin{equation}
\langle u \mid X \, v \rangle_h = \langle \widetilde X \, u \mid v \rangle_h \qquad .
\end{equation}
Note, however, that the sesquilinear form $\langle u \mid v \rangle_h$ is pseudo-hermitian, rather than being hermitian in the usual sense: the norm $\langle u \mid u \rangle_h$ is real but not positive definite. Actually, as a real quadratic form it has signature $+^{32} , -^{32}$.

Similarly to the usual Dirac-equation case where the hermitian properties of the $\gamma$ matrices, and the reality of the 
mass term, ensure the conservation of the Dirac current $J^{\mu} = - i \bar \psi \gamma^\mu \psi$, the hermiticity condition
(\ref{hermiticitySA}) satisfied by the susy constraints ensure the conservation (in $\beta$-space; {\it i.e.} 
$\partial_{\beta^a} J^a_A=0$) of the four currents
\begin{equation} \label{current1}
J^a_A(\beta) :=  \langle \Psi(\beta) \mid \widehat \Phi^a_A \, \Psi(\beta) \rangle_h = \Psi^\dagger h  \Phi^a_A \, \Psi
\end{equation}
for any solution $\Psi(\beta)$ of the susy constraints.
  
 When using Lorentzian coordinates in $\beta$-space, say $\xi^{\hat 0}, \xi^{\hat 1}, \xi^{\hat 2}$  [as defined below, see Eqs. (\ref{Lorxivar}) or Eqs. (\ref{Apxivar})],
 the local conservation law $\partial_{\hat a} J^{\hat a}_A=0$ implies the global conservation of the four ``charges''
 \begin{equation} Q_A= \int d \xi^{\hat 1}  d \xi^{\hat 2} J^{\hat 0}_A
 \end{equation}
Contrary to the usual Dirac charge $Q= \int d^3 x J^0=\int d^3 x \psi^\dagger \psi$, these conserved
charges are not positive-definite sesquilinear forms in the wavefunction $\Psi$. [The ``chiral'' representation
of the $\Phi^a_A$'s introduced below will also make clear that the four integrated charges $Q_A$ vanish
when considering a wavefunction having a fixed fermion number $N_F$ (because $\Phi \sim b + \widetilde b$).
On the other hand, it will not generally vanish if one considers a wavefunction that contains components
within, say, two successive fermion-number levels.]
However, one should note that the system of first-order PDE's  on the wavefunction $\Psi(\beta)$
defined by the susy constraints : 
 $$
  \widehat {\cal S}_A^{(0)} \, \vert{\bm\Psi}\rangle = 0 
  $$
 constitutes, like the usual Dirac equation  $ \gamma^\mu ( \partial_\mu - i e A_\mu) \psi + m \psi=0$,  a 
 {\it first-order symmetric-hyperbolic} system.  The definition of these systems  \cite{Choquet-Bruhat:2009xil}
 is that they admit a formulation
 in terms of real variables and real coefficients
 where the derivative terms are of the form $ (A \partial_{\hat 0}  + B^{\hat i} \partial_{\hat i} ) \psi + \cdots $,
 where the real matrices $A$  and $B^{\hat i}$ are both symmetric, and where $A$ is positive-definite. 
 When working with a complex system, it is easily seen (by decomposing into real and imaginary parts) that one can replace the conditions
 of symmetry by  conditions of hermiticity: $A^\dagger=A$,  $B^{\hat i \, \dagger} = B^{\hat i}$ for complex
 matrices. By considering one particular spinor index $A$ (say $A=1$), and by multiplying the corresponding 
 susy constraint on the left by the anti-hermitian  $64 \times 64$  matrix $\Phi^{\hat 0}_1$, we obtain a first-order
 evolution system of the type $\partial_{\hat 0} \Psi =  B^{\hat i}  \partial_{\hat i} \Psi + \cdots$ where
$B^{\hat i} =   \Phi^{\hat 0}_1 \Phi^{\hat i}_1$ is easily checked to be hermitian. Note in passing that this
ensures that the positive-definite norm $\Psi^\dagger \Psi$, though not strictly conserved, satisfies a
conservation law (involving the corresponding spatial current  $\Psi^\dagger B^{\hat i} \Psi$) modulo lower-derivative terms.
As a consequence, it is natural to assume that the wavefunction $\Psi$ is (at least) square-integrable (a fact
that we shall exploit below).

\setcounter{equation}{0}\section{Quantum (rotationally reduced) susy constraints}

The requirement of hermiticity of the $\widehat{\mathcal S}^{(0)}_A$'s determines them to be equal to
\begin{equation}\label{Sared}
\widehat {\cal S}^{(0)}_A = - \frac12 \sum_a \widehat \pi_a \, \widehat \Phi_A^a + \widehat {\cal S}_A^g + \widehat {\cal S}_A^{\rm sym} + \widehat {\cal S}_A^{\rm cubic}
\end{equation}
with
\begin{equation}
\label{SgA}
\widehat {\cal S}_A^g = \frac12 \sum_a e^{-2\beta^a} \left(\gamma^5 \, \widehat \Phi^k\right)_A\qquad ,
\end{equation}
and
\begin{eqnarray}
\label{SsA}
\widehat {\cal S}_A^{\rm sym} &= &-\frac18 \coth [\beta^1 - \beta^2] \left[ \widehat S_{12} \left( \gamma^{\hat1\hat2} \left( \widehat \Phi^1 - \widehat \Phi^2 \right)\right)_A + \left( \gamma^{\hat1\hat2} \left( \widehat \Phi^1 - \widehat \Phi^2 \right)\right)_A \widehat S_{12} \right] \nonumber \\
&+&{\rm cyclic}_{123}
\end{eqnarray}
where
\begin{eqnarray}
\label{Spin}
\widehat S_{12} &=&\frac12 \left( \overline{\widehat\Phi}^3 \, \gamma^{\hat 0\hat 1\hat 2} (\widehat\Phi^1 + \widehat\Phi^2) + \overline{\widehat\Phi}^1 \, \gamma^{\hat 0\hat 1\hat 2} \, \widehat\Phi^1
+ \overline{\widehat\Phi}^2 \, \gamma^{\hat 0\hat 1\hat 2} \, \widehat\Phi^2 - \overline{\widehat\Phi}^1 \, \gamma^{\hat 0\hat 1\hat 2} \, \widehat\Phi^2 \right)  \nonumber \\
&=& \frac12 \left( \overline{\widehat \Phi} \,  \gamma^{\hat 0\hat 1\hat 2} \left( \widehat \Phi^1 + \widehat \Phi^2 \right) - 3 \, \overline{\widehat \Phi}^1  \gamma^{\hat 0\hat 1\hat 2} \, \Phi^2 \right)\qquad .
\end{eqnarray}
The operator  $ \widehat S_{12}$, together with  similarly defined operators $\widehat S_{23},\ \widehat S_{31}$,  are spin-like operators verifying the usual $su(2)$ commutation relations : $[\widehat S_{23},\ \widehat S_{31}]=+ i\,\widehat S_{12}$, etc . [The Kac--Moody meaning of these spin operators will be further discussed below.]

\smallskip

The last contribution in Eq. (\ref{Sared}) is cubic in the $\Psi$'s and reads: 
\begin{equation}
\label{ScA}
\widehat {\cal S}_A^{\rm cubic} = \frac12 \left( \widehat\Sigma_A^{\rm cubic} + \widetilde{\widehat\Sigma}_A^{\rm cubic} \right)
\end{equation}
where
\begin{eqnarray}
\widehat{\Sigma}_A^{\rm cubic} &= &\frac14 \sum_a (\overline{\widehat\Psi'_{0}} \, \gamma^{\hat 0} \, \widehat\Psi_{\hat a}) \, (\gamma^{\hat 0} \, \widehat\Psi_{\hat a})_A - \frac18 \sum_{a,b} (\overline{\hat\Psi}_{\hat a} \, \gamma^{\hat 0} \, \widehat\Psi_{\hat b}) \, (\gamma^{\hat a} \, \widehat\Psi_{\hat b})_A \nonumber \\
&+ &\frac18 \sum_{a,b} (\overline{\widehat\Psi'_{0}} \, \gamma^{\hat a} \, \widehat \Psi_{\hat b}) ((\gamma^{\hat a} \, \widehat \Psi_{\hat b})_A + (\gamma^{\hat b} \, \widehat \Psi_{\hat a})_A) \, , \nonumber
\end{eqnarray}
with $\widehat\Psi'_{0} := \gamma_{\hat 0} \, \sum_a \gamma^{\hat a} \, \widehat\Psi_{\hat a}=  \gamma_{\hat 0}  \widehat \Phi$. 
In terms of the $\Phi$'s, it reads
\begin{eqnarray}
\label{SigcAQ}
\widehat \Sigma_A^{\rm cubic} &= &\frac1{16} \sum_{k \neq l}  \left( \overline{\widehat \Phi} \, \gamma^{\hat 0\hat k \hat l} \left( \widehat \Phi^k - \widehat \Phi^{l} \right) \right) \left( \gamma^{\hat k \hat l} \left( \widehat \Phi^k - \widehat \Phi^{l} \right)\right)_A \nonumber \\
&+&  \frac14 \sum_k \left(\overline{ \widehat \Phi} \gamma^{\hat 0} \widehat \Phi^k \right) \widehat \Phi^k_A -\frac18  \sum_{k \neq l}  \left( \overline{\widehat \Phi}^k \gamma^{\hat 0 \hat k \hat l} \, \widehat \Phi^{l} \right) \left( \gamma^{\hat k \hat l} \, \widehat \Phi^{l} \right)_A  \nonumber \\
& - &  \frac{i}{8} \widehat \Phi_A - \frac14 \sum_k \left( \overline{\widehat \Phi} \gamma^{\hat k} \, \widehat \Phi^k \right) \left( \gamma^{\hat 0 \hat k} \, \widehat \Phi^k \right)_A \qquad ,
\end{eqnarray}
where the (antihermitian) term $ - \frac{i}{8} \widehat \Phi^k_A$ will drop out of  $\widehat {\cal S}_A^{\rm cubic}$.
These operators completely determine the (reduced) Hamiltonian operator that we will now discuss.

\setcounter{equation}{0}\section{Supersymmetry algebra and ordering of the  quantum Hamiltonian operator}

We have shown, by direct computation, that the (rotationally reduced) supersymmetry operators satisfy anticommutation relations of the form : 
\begin{equation}
\label{SSLSH} 
\widehat {\cal S}^{(0)}_A \, \widehat {\cal S}^{(0)}_B + \widehat {\cal S}^{(0)}_B \, \widehat {\cal S}^{(0)}_A = 4 \, i \ \widehat L_{AB}^C (\coth \beta , \widehat \Phi) \, \widehat {\cal S}^{(0)}_C + \frac12 \, \widehat H^{(0)} \, \delta_{AB} \qquad ,
\end{equation}
where $ \widehat H^{(0)}$ reduces when $\hbar \to 0$ to the classical value (\ref{Vtotclass}) of the Hamiltonian, and where  the coefficients  $\widehat L_{AB}^C (\coth \beta , \widehat \Phi)$ are linear in the $\widehat \Phi$'s, namely
$$
\widehat L_{AB}^C (\coth \beta , \widehat \Phi) =  L_{AB;a}^{C \ \ D} (\coth \beta) \,  \widehat \Phi^a_D \qquad ,
$$
with numerical coefficients $ L_{AB;a}^{C \ \ D} (\coth \beta)$ that are linear in the three hyperbolic cotangents
$$
 L_{AB;a}^{C \ \ D} (\coth \beta) =  L_{AB;a0}^{C \ \ D} + \sum_{b < c}  L_{AB;abc}^{C \ \ D} \coth (\beta^b - \beta^c) \qquad .
$$
We give in Appendix  \ref{AppLform} the explicit values of the  $\widehat L_{AB}^C (\coth \beta , \widehat \Phi)$'s in a special
(chiral) basis for the $\Phi's$ that will be introduced below.

Note that, in Eq. (\ref{SSLSH}), the supersymmetry constraints $\widehat {\cal S}^{(0)}_C$ entering the right-hand side appear {\it on the right}. This shows that the four supersymmety constraints $\widehat {\cal S}^{(0)}_A \vert{\bm\Psi}\rangle =0$ imply the Hamiltonian constraint $\widehat H^{(0)} \vert{\bm\Psi}\rangle =0$. It is easily seen that Eq. (\ref{SSLSH}) implies further commutation relations of the form 
$$
[ \widehat{\cal S}^{(0)}_A , \widehat H^{(0)} ] = i \,  \widehat M_A^B \, \widehat{\cal S}_B^{(0)} +i \, \widehat  N_A \, \widehat H^{(0)} \qquad .
$$
As the operators $\widehat p_a$ commute both with  the $\widehat{\cal S}^{(0)}_A$'s and with $\widehat H^{(0)}$, we conclude that the three quantum constraints $\widehat {\cal S}^{(0)}_A$, $\widehat H^{(0)}$, $\widehat p_a$ 
entering (\ref{newQconstraints}) form and open (or ``soft'') algebra, and that the Dirac equations (\ref{newQconstraints}) are,    {\it a priori}, formally consistent.

In view of the above results, the set of quantum constraint equations (\ref{newQconstraints}) is equivalent to the reduced
set of $3+4$ constraints
\begin{equation}
\label{reducedQconstraints}
\widehat p_a \, \vert{\bm\Psi}\rangle = 0 \, , \quad \widehat {\cal S}_A^{(0)} \, \vert{\bm\Psi}\rangle = 0 \quad .
\end{equation}
The first three equations in (\ref{reducedQconstraints}) are equivalent to requiring that 
the wave function of the Universe ${\bm\Psi}$ does not depend on the three Euler angles, and therefore
is a 64-component spinor of Spin(8,4) that only depends on the logarithms of the scaling factors of the metric,  $\beta^1$, $\beta^2$ and $\beta^3$ :

\begin{equation}
{\bm \Psi}=\Psi_{\sigma} [\beta^a], \quad (\sigma = 1,\ldots,64) \label{Psiform}  \, .
\end{equation}

Then, the second set of equations in (\ref{reducedQconstraints}) consists, in view of the explicit form (\ref{Sared}) of the
supersymmetry operators, in imposing four simultaneous  Dirac-like equations restricting the propagation
of  the 64-component spinor $\Psi_{\sigma} [\beta^a]$ in the three-dimensional Minkowski space of the $\beta$'s.

\smallskip

Let us add two comments concerning the structure of the anticommutation relations (\ref{SSLSH}):
\begin{enumerate}
\item[(i)] There exists a version of these anticommutation relations of the form 
\begin{eqnarray}
\label{SSLSHh} 
\widehat {\cal S}^{(0)}_A \, \widehat {\cal S}^{(0)}_B + \widehat {\cal S}^{(0)}_B \, \widehat {\cal S}^{(0)}_A &= &2 \, i \ (\widehat L_{AB}^C (\coth \beta , \Phi) \widehat {\cal S}^{(0)}_C + \widehat {\cal S}^{(0)}_C \widehat L_{AB}^C (\coth \beta , \Phi) ) \nonumber \\
&+ &\frac12 \, \widehat H^{(0)}_h \, \delta_{AB} \qquad ,
\end{eqnarray}
where $\widehat H^{(0)}_h$ differs from $\widehat H^{(0)}$ by a quantum reordering. [In this form $\widehat H^{(0)}_h$ is hermitian, while, as we shall see later,  the $\widehat H^{(0)}$ defined by (\ref{SSLSH}) contains a non-hermitian piece, of order $O(\hbar)$,  that will be conveniently reabsorbed by redefining the wave function $\Psi (\beta)$.]
\item[(ii)] Contrary to the usual superalgebra appearing in supersymmetric quantum mechanics, of the form 
\begin{equation}
\label{SSLSHQM} 
\widehat {\cal S}_A \, \widehat {\cal S}_B + \widehat {\cal S}_B \, \widehat {\cal S}_A =  \frac12 \, \widehat H \, \delta_{AB} \qquad ,
\end{equation}
the presence of the supersymmetry operators $\widehat{\mathcal S}_C^{(0)}$ on the right-hand side of Eq. (\ref{SSLSH}) does not allow one to use the $\widehat{\mathcal S}_A^{(0)}$'s as ladder operators generating new solutions of the susy constraints by acting on old ones.
\end{enumerate}

\setcounter{equation}{0}\section{Explicit structure of the quantum Hamiltonian}

Similarly to the well-known fact that the (second-order) Klein-Gordon equation $(\Box  - \mu^2) \psi$ is a necessary
consequence of the (first-order) Dirac equation $(\gamma^\mu \partial_\mu - \mu) \psi=0$, 
the Hamiltonian constraint  (which is a Wheeler-DeWitt-(WDW)-type equation) 
\begin{equation} \label{WDW}
\widehat H^{(0)} \, \vert{\bm\Psi}\rangle = 0
\end{equation}
 is a necessary consequence of
the four susy constraints $ \widehat {\cal S}_A^{(0)} \, \vert{\bm\Psi}\rangle = 0 $.
However, like in the Dirac/Klein-Gordon-case, it is useful to have in hands the explicit structure of the Hamiltonian constraint because it brings out
more clearly the physical meaning of the various interaction terms predicted by supergravity.

The explicit expression of the (rotationally reduced) Hamiltonian operator $\widehat H^{(0)} $ [defined as the operator
appearing on the right-hand side of the anticommutation relations (\ref{SSLSH})] is given, 
in the $\beta$-space Schr\"odinger representation, by : 

\begin{eqnarray} \label{H0}
2 \, \widehat H^{(0)}&=& G^{ab} (\hat\pi_a + i \, A_a(\beta) )(\hat\pi_b + i \, A_b(\beta) ) + \hat\mu^2 + \widehat W (\beta)  \\ \nonumber
  &=& -G^{ab}   (\partial_a-A_a(\beta) )(\partial_b-A_b(\beta) ) +  \hat\mu^2 + \widehat W (\beta)
\end{eqnarray}
In this equation,  $\hat\pi_a = -i \, \partial_a$ (with $\partial_a := \partial / \partial \beta^a$), and the ``vector potential'' $A_a(\beta) $
 is a {\it real} vector field\footnote{This real vector field comes
from the reordering of the manifestly hermitian anticommutation relations (\ref{SSLSHh}) into the right-ordered
form (\ref{SSLSH}).} in $\beta$-space.  [We have omitted 
 an explicit identity operator $1_{64}$ in front of the differential operators].
One finds that the vector potential\footnote{Actually, in an analogy with the electromagnetically
 coupled Klein-Gordon equation, the vector potential would be the purely imaginary field $i A_a$.} $A_a(\beta)$ 
 is a pure gradient: 
 \begin{equation} \label{AgradF}
 A_a(\beta) = \partial_a \, \ln \, F  = F^{-1}   \partial_a  F 
 \end{equation}
 with 
  \begin{equation} \label{F}
 F(\beta)  = e^{\frac34 \, \beta^0} (\sinh \beta_{12} \, \sinh \beta_{23} \, \sinh \beta_{31} )^{-1/8} \, ,
  \end{equation} 
 where we introduced the convenient short-hand
 $$
 \beta^0 :=   \beta^1 +  \beta^2 +  \beta^3\qquad .
 $$
 As the vector potential $A_a$ occuring in equation   (\ref{H0}) is a pure gradient,  it can be eliminated, without changing the other terms, by working with the rescaled wave function 
 \begin{equation} \label{psi'}
 \Psi' (\beta) = F(\beta)^{-1} \Psi (\beta) ,
 \end{equation}
 in terms of which the Hamiltonian operator reads 
   \begin{equation}  \label{H'}
 2 \, \widehat H'  \Psi' :=  2 \, F^{-1} \, \hat H (F \Psi') =  (G^{ab} \hat\pi_a  \hat\pi_b  + \hat\mu^2 + \widehat W (\beta)) \Psi'
   \end{equation} 

Let us now comment the structure of the   ``spin-dependent" potential terms in the WDW-type equation   (\ref{H0}).  Both terms,
$ \hat\mu^2$ and  $ \widehat W (\beta)$, are  $64\times 64$ matrices acting in spinorial space. The separation between
these two types of term is defined so that the   ``mass-squared'' term $ \hat\mu^2$ does not depend on
the $\beta$'s, and survives as a constant, but  spin-dependent, term in the limit where all the exponential terms 
present in the potential  $ \widehat W (\beta)$ tend to zero. 

Indeed,  the remaining potential term $ \widehat W (\beta))$ can be separated into several pieces :
\begin{equation} \label{decU}
\widehat W(\beta) =  W_g^{\rm bos}(\beta) + \widehat W_g^{\rm J}(\beta) + \widehat W_{\rm sym}^{\rm spin}(\beta) \qquad .
\end{equation}
The first one, $ W^{\rm bos}_g$ is spin-independent ({\it i.e.}  diagonal in spinorial space), and is simply twice
the usual  bosonic potential, (\ref{Vg}), describing the mixmaster dynamics of Bianchi IX  models~\cite{Belinsky:1970ew,Misner:1969hg}:
\begin{equation}
\label{Wg}
W_g^{\rm bos} (\beta) = 2 \, V_g(\beta)= \frac12 \, e^{-4\beta^1} - e^{-2 (\beta^2 + \beta^3)} + {\rm cyclic}_{123} .
\end{equation}
 In the framework of supergravity this potential term is accompanied by two complementary spin-dependent pieces that decay exponentially 
 as some linear combinations of the $\beta$'s get large and positive. The first one, 
\begin{equation}\label{VJ}
\widehat W_g^{\rm J}(\beta)= e^{-2\,\beta^1} \,\widehat J\strut_{11}(\widehat\Phi)+e^{-2\,\beta^2} \,\widehat J\strut_{22}(\widehat\Phi)+e^{-2\,\beta^3} \,\widehat J\strut_{33}(\widehat\Phi)
\end{equation}
involves the products of exponentials of  $ 2 \beta^1$,  $ 2 \beta^2$,   $ 2 \beta^3$, ({\it i.e.}  {\it half} the linear combinations of the $\beta$'s that enter in the dominant potential terms in $W_g^{\rm bos} (\beta)$, and that drive the BKL oscillatory
dynamics of the $\beta$'s) by operators that are  quadratic in the gravitino field. E.g.  the linear form $ 2 \beta^1$ (gravitational wall form) is coupled to
\begin{equation}\label{J11op}
\widehat J\strut_{11}(\widehat\Phi)=\frac12\left[\,\overline{\widehat\Phi}^1\gamma^{\hat 1\hat 2\hat 3}(4\,\widehat \Phi^1+\widehat\Phi^2+\widehat\Phi^3)+\overline{\widehat\Phi}^2\gamma^{\hat 1\hat 2\hat 3}\widehat\Phi^3\right]
\end{equation}
We shall discuss in the next section the Kac--Moody meaning of the three operators  $\widehat J\strut_{11}$,
$\widehat J\strut_{22}$,, $\widehat J\strut_{33}$, defined by considering cyclic permutations of Eq. (\ref{J11op}).

The second spin-dependent, and $\beta$-dependent, contribution is quartic in the gravitino field. It reads
(with $1_{64}$ denoting the identity operator in the 64-dimensional spinor space):
\begin{equation}\label{Vsym}
\widehat W_g^{\rm sym}(\beta)=  \frac12 \ \frac{(\widehat S_{12} (\widehat\Phi))^2 - 1_{64}}{\sinh^2 (\beta^1-\beta^2)} + {\rm cyclic}_{123} \, .
\end{equation} 
The operators  $\widehat S_{12} (\widehat\Phi)$, etc.,  whose squares enter  $\widehat W_g^{\rm sym}(\beta)$, are
the quadratic-in-$\Phi$ ``spin operators''  that were introduced above in Eq. (\ref{Spin}), and that entered 
linearly in the supersymmetry operators $\widehat{\mathcal S}$'s.

Let us now discuss the squared-mass term $\widehat \mu^2$ entering the WDW equation.  This term is $\beta$-independent, but  it is spin-dependent, i.e. it is a $64 \times 64$ matrix in spinorial space. It originates from quartic fermionic contributions to the Hamiltonian.  More precisely, it comes from two types of  $\Phi^4$ contributions: (i) the original quadratic-in-torsion
(and therefore quartic-in-fermion) terms in the second-order action; and (ii)  additional terms quadratic in the spin
operators coming from Eq. (\ref{V4}) because of the identity $ \coth^2 \beta \equiv 1 + 1/\sinh^2 \beta$.

As we shall discuss in detail in the next section,  the term   $\widehat \mu^2$ plays a crucial r\^ole when considering
the quantum billliard limit where a wave packet propagates between the well--separated  Toda-like exponential walls
defined by the various terms in $\widehat W_g^{\rm sym}(\beta)$.  In this regime, the wave function far from all
the exponential walls can be approximated by a plane wave in $\beta$ space:

\begin{equation}
\Psi\propto \exp[i\,\pi_a\beta^a]\quad 
\end{equation}

Actually, as we are discussing here the ``primed'' form of the WDW equation,
 $$
 2 \, \widehat H'  \Psi' =  (G^{ab} \hat\pi_a  \hat\pi_b  + \hat\mu^2 + \widehat W (\beta)) \Psi' =0 \, ,
  $$ 
  we need to work with the rescaled wavefunction $\Psi'(\beta)$, Eq. (\ref{psi'}).
  
  In view of the form (\ref{F}) of the rescaling factor, when one is far from all the walls, this rescaling leads to
  a wave function of the form
  \begin{equation}
   \Psi'\propto \exp[i\,\pi'_a\beta^a]
  \end{equation}
  involving a primed momentum differing from the original  $\beta$ momentum  $\pi_a$ by a purely imaginary  shift :
  \begin{equation}
   \pi_a=\pi'_a-i\,\varpi_a
   \end{equation}
The components  $\varpi_a$ entering this complex shift  are given by a permutation of $\{1,\frac34,\frac12\}$ which depends
on the choice of billiard chamber (among six possibilities; see below). For instance, when using what will
be in the following our canonical billiard (or Weyl) chamber, labelled $(a)$, and corresponding to the inequalities
$ 1 \ll \beta^1 \ll \beta^2 \ll \beta^3$, the (covariant) components of  $\varpi_a$ will be  
$\{\varpi_1=1,\varpi_2=\frac34,\varpi_3=\frac12\}$. [In a Weyl chamber obtained by a permutation $\sigma$ of $(1,2,3)$,
such that $ 1 \ll \beta^{\sigma_1} \ll \beta^{\sigma_2} \ll \beta^{\sigma_3}$, they will be  $\{\varpi_{\sigma_1}=1,\varpi_{\sigma_2}=\frac34,\varpi_{\sigma_3}=\frac12\}$.]

The important point we wish to make here (anticipating on its derivation below) is that the diagonalization of
the squared-mass operator determining the mass-shell conditions for the shifted $\beta$ momentum entering
various pieces of the wave function $\Psi'$
 \begin{equation}\label{pi2m2}
 \pi'_a\pi^{\prime a}\equiv G^{ab}  \pi'_a  \pi'_b=-\mu^2
 \end{equation}
 leads to the following list of eigenvalues
 \begin{equation}\label{mu2val}
  \mu^2=\left(\left.\left.
-\frac{59}8\right\vert^{1}_{0}, -3\right\vert^{6}_{1},  \left. - \frac 38\right\vert^{15}_{2},\left. +\frac12\right\vert^{20}_{3}, \left. \left. - \frac 38\right\vert^{15}_{4},    -3\right\vert^{6}_{5}, \left.  -\frac{59}8\right\vert^{1}_{6}\right)
\end{equation}
Here we have given the different eigenvalues taken by the mass-squared operator, ordered (as indicated by the subscript
going from 0 to 6) by the value of a certain Fermion number $N_F$, which will be defined below. The superscript indicates
the dimensions of the various spaces having a given value of  $N_F$.  For instance, the  $N_F =2$ subspace is
of dimension 15, and this subspace is an eigenspace of $\widehat \mu^2$ with eigenvalue $- \frac38$.
We shall discuss in detail below the structure of the solutions of the susy constraints corresponding
to the list of eigenvalues (\ref{mu2}), but we wanted to emphasize from the start that, among the 64 dimensions of the
total spinorial space,   $\widehat \mu^2$ is {\it negative} ({\it i.e.}  tachyonic) in 44 of them !

\setcounter{equation}{0}\section{Hidden Kac--Moody structure of supersymmetric Bianchi IX cosmoloy}

One of the main results of this work concerns the Kac--Moody structures hidden in the (exact) quantum Hamiltonian (\ref{H0}). First, let us recall that the wave function of the Universe $\Psi (\beta)$ is a 64-component spinor of ${\rm Spin} (8,4)$ which depends on the three logarithmic scale factors $\beta^1 , \beta^2 , \beta^3$. In other words, supergravity describes a Bianchi IX Universe as a relativistic {\it spinning particle} moving in $\beta$-space. The spinorial wave function $\Psi (\beta)$ must satisfy four separate Dirac-like equations $\hat{\mathcal S}_A \Psi = \left( + \frac{i}2 \, \Phi_A^a \partial_a + \ldots \right) \Psi = 0$ (where the $\Phi_A^a$'s are four separate triplets of $64 \times 64$ gamma matrices). As shown above, these first-order Dirac-like equations imply that $\Psi$ necessarily satisfy the second-order, Klein-Gordon-like equation $\hat H \Psi = \left(-\frac12 \, G^{ab} \partial_a \partial_b + \ldots \right) \Psi = 0$. 

On the other hand, studies  \`a la BKL  of the structure of cosmological singularities in string theory and supergravity (in dimensions
 $4 \leq D \leq 11$) have found that the chaotic BKL oscillations could be interpreted as a billiard motion
 in the Weyl chamber of an hyperbolic Kac--Moody algebra~\cite{Damour:2000hv,Damour:2001sa,Damour:2002et}.
 This interpretation was extended by including the dynamics of the gravitino, and led to the conjecture of a
 {\it correspondence} between various supergravity theories and the dynamics of a spinning massless particle on an infinite-dimensional Kac--Moody coset space~\cite{Damour:2002cu,Damour:2005zs,de Buyl:2005mt,Damour:2006xu}.
 In the particular case of  pure vacuum gravity in $D=4$,  the conjectured Kac--Moody algebra corresponding to the
 gravity dynamics is  $AE_3$~\cite{Damour:2001sa}. In this section, we shall study in detail the structure of the
 quantum dynamics of the 64-component supergravity spinorial wave function $\Psi (\beta)$ in $\beta$-space, to
 exhibit to what extent it contains Kac--Moody related elements. This will contribute to showing to what extent
 the conjectured  Kac--Moody-coset/gravity correspondence holds.

\smallskip

The first basic Kac--Moody feature hidden in this dynamics of the Universe is the fact that the (Lorentzian-signature) metric $G_{ab}$ defining the kinetic term of the ``$\beta$-particle'' is the metric in the Cartan subalgebra of the hyperbolic Kac--Moody algebra $AE_3$~\cite{Damour:2001sa}. Second, the potential term $\hat W (\beta)$ in Eq.~(\ref{H0}) is naturally decomposed (see Eq. (\ref{decU})) into three different pieces which all carry a deep Kac--Moody meaning. The first term, $\widehat W_g (\beta)$, given by Eq. (\ref{Wg}),
is the well-known bosonic potential describing the usual dynamics of  Bianchi IX oscillations~\cite{Belinsky:1970ew,Misner:1969hg}. Its Kac--Moody meaning is that it is constructed from Toda-like exponential potentials $\sim e^{-2\alpha_{ab}(\beta)}$ involving the following six linear forms in the $\beta$'s: 
\begin{equation}\label{gravroot}
\alpha_{ab}^g (\beta) := \beta^a + \beta^b \quad , \qquad a,b = 1,2,3 \quad .
\end{equation} 
These six linear forms coincide with the six roots of $AE_3$ located at level $\ell = 1$ (``gravitational walls'', linked to the level-$1$ $AE_3$ ``dual-graviton'' coset field $\phi_{ab} = \phi_{ba}$ of Ref.~\cite{Damour:2002et}). 

\smallskip

Third, the purely bosonic (spin-independent) potential $W_g^{\rm bos} (\beta)$ is accompanied, in supergravity, by a spin-dependent complementary piece given by Eq. (\ref{VJ}). This spin-dependent potential $\widehat W_g^{\rm spin} (\beta , \hat\Phi) = e^{-\alpha_{11}^g (\beta)}  \hat J_{11} (\hat\Phi)  + \ldots$
 involves the three dominant (gravitational) Kac--Moody roots $\alpha_{11}^g (\beta) = 2\beta^1$, etc. each one being coupled to an operator that is {\it quadratic} in the gravitino variables, see Eq. (\ref{J11op}).
 
\smallskip

The third contribution to $\widehat W(\beta)$ involves the three level-$0$ Kac--Moody roots 
\begin{equation}\label{}
\alpha_{12}^{\rm sym} (\beta) := \beta^2 - \beta^1  \quad , \quad \alpha_{23}^{\rm sym} (\beta) := \beta^3 - \beta^2 \quad , \quad \alpha_{13}^{\rm sym} := \beta^3 - \beta^1 \quad .
\end{equation}
These three linear forms are called ``symmetry wall forms''; each one of them is coupled to an operator that is {\it quartic} in the $\hat\Phi$'s.   See Eq. (\ref{Vsym}) which involves the squares of the three spin operators $\widehat S_{12} (\hat\Phi)$, $\widehat S_{23} (\hat\Phi)$, $\widehat S_{31} (\hat\Phi)$   defined in Eq. (\ref{Spin}) (modulo cyclic permutations).

\smallskip

A truly remarkable fact, which clearly shows the hidden r\^ole of Kac--Moody structures in supergravity, is that the operators entering $\widehat H$ as (spin-dependent) basic blocks, $\hat S_{12}, \hat S_{23} , \hat S_{31}, \hat J_{11} , \hat J_{22} , \hat J_{33}$ generate (via commutators) a Lie-algebra which is a 64-dimensional representation of the (infinite-dimensional) ``maximally compact'' sub-algebra, $K(AE_3)$, of $AE_3$.

Let us first indicate why such a structure is related to the conjectured Kac--Moody/supergravity correspondence \cite{Damour:2002cu,Damour:2005zs,de Buyl:2005mt,Damour:2006xu}. 

\smallskip

According to the latter conjecture, the dynamics of the bosonic degrees of freedom is equivalent to geodesic motion on a coset space $G/K$, where $G$ is a hyperbolic Kac--Moody group (over the reals) and $K$ its maximal compact subgroup. When considering $D=11$, ${\mathcal N} = 1$ supergravity, it is conjectured that $G$ is the group associated with $E_{10}$. In the case we are considering here of $D=4$, ${\mathcal N} = 1$ supergravity, $G$ gets reduced to $AE_3$, and $K$ to the corresponding maximal compact subgroup of $AE_3$, say $K(AE_3)$. A geodesic on $G/K$ is described by a one-parameter family of group elements $g(t) \in G$, considered modulo right multiplication by an arbitrary element $k(t)$ in $K$. Decomposing the Lie-algebra valued ``velocity'' of $g(t)$ in $P \in {\rm Lie} (G) \ominus {\rm Lie} (K)$ and $Q\in {\rm Lie} (K)$ pieces,
\begin{equation}\label{cosetvel}
\partial_t g \, g^{-1} = P(t) + Q(t) \, ,
\end{equation}
the coset Lagrangian describing a geodesic on $G/K$ is simply
\begin{equation}\label{cosetL}
{\mathcal L} = \frac1{2n(t)} \, (P \mid P)
\end{equation}
where $(\cdot \mid \cdot)$ denotes the (unique) invariant bilinear form on ${\rm Lie} (G)$.

\smallskip

The coset ``lapse'' function $n(t)$ is a Lagrange multiplier enforcing the constraint that the considered geodesic is null: $(P \mid P) = 0$. The equation of motion of $g(t)$ can be written (in the coset gauge $n(t) = 1$) as
\begin{equation}\label{cosetpdot}
\partial_t \, P(t) = [Q(t) , P(t)]
\end{equation}
where $[\cdot , \cdot]$ denotes a Lie-algebra bracket. Eq. (\ref{cosetpdot}) shows that the $Q$-piece of the velocity ({\it i.e.}  the piece within the compact algebra ${\rm Lie} (K)$) can be viewed as the connection describing (via its Lie-bracket action) how the (bosonic) coset velocity $P$ rotates along the geodesic.

\smallskip

According to the coset/supergravity conjecture, the same ${\rm Lie} (K)$-valued piece of the velocity plays also the role of the connection describing how the {\it fermionic} degrees of freedom rotate as some one-parameter coset fermion $\Psi^{\rm coset} (t)$ propagates along the considered bosonic geodesic of the supersymmetric space $G/K$:
\begin{equation}\label{cosetpsidot}
\partial_t \, \Psi^{\rm coset} = Q^{\rm vs} \cdot \Psi^{\rm coset} \, .
\end{equation}
Here $Q^{\rm vs} \cdot \Psi^{\rm coset}$ denotes the linear action of the abstract Lie-algebra element $Q \in {\rm Lie} (K)$ on a member $\Psi^{\rm coset}$ of a vector space, on which $Q^{\rm vs}$ defines a {\it representation} of ${\rm Lie} (K)$. In Refs. \cite{Damour:2005zs,de Buyl:2005mt,Damour:2006xu,Damour:2009zc} the coset fermion $\Psi^{\rm coset}$ was taken as a classical, Grassmannian object living in a finite-dimensional vector space (of dimension 12 for the $K(AE_3)$ case \cite{Damour:2009zc}), and $Q^{\rm vs}$ was, accordingly, a $12 \times 12$ ``vector-spinor'' representation of $K(AE_3)$.

\smallskip

Let us indicate here the Kac--Moody structures hidden within our {\it quantum} supergravity framework which, indeed, lead to a gravitino of motion resembling the conjectured one, Eq. (\ref{cosetpsidot}). At the quantum level, the equations of motion of the gravitino operators $\widehat\Phi_A^a$ derive, according to the general Heisenberg rule, from the commutator of the Hamiltonian operator $\widehat{\mathcal H}$ with the $\widehat\Phi_A^a$'s. In the gauge where $\Psi'_{\hat 0} = 0$, $\widetilde N = 1$ and $N^a = 0$, the Hamiltonian operator following from Eq. (\ref{HTot}) is simply $\widehat H$. The Heisenberg equation of motion for the gravitino operators are
$$
\partial_t \, \widehat\Phi_A^a = i \left[ \widehat H , \widehat\Phi_A^a \right] \, .
$$
For these equations to resemble the classical, coset-expected equations of evolution (\ref{cosetpsidot}), the quantum Hamiltonian $\widehat H$ should parallel the classical structure of the $K(AE_3)$-connection $Q$, which was found in previous works  \cite{Damour:2005zs,de Buyl:2005mt,Damour:2006xu,Damour:2009zc}  to be of the form,
$$
Q = \sum_{\alpha} Q_{\alpha} \, J_{\alpha} \, ,
$$
where $\alpha$ labels the positive roots of $AE_3$, and where
\begin{equation}\label{Jalpha}
J_{\alpha} = E_{\alpha} - E_{-\alpha} \equiv E_{\alpha} + \omega (E_{\alpha})
\end{equation}
 is the generator of $K(AE_3)$ associated with the positive root $\alpha$. [Here, $E_{\alpha}$ denotes a generator of $AE_3$ associated with the root $\alpha$, and $\omega$ denotes the Chevalley involution, which, by definition, fixes the set $K(AE_3)$.] In addition, the numerical coefficients $Q_{\alpha}$ are, roughly ({\it i.e.}  when separately considering the effect of each root in the coset Hamiltonian) of the form $Q_{\alpha} \sim e^{-\alpha (\beta)} \, p_{\alpha} \, J_{\alpha}$, where $p_{\alpha}$ is the momentum conjugated to the variable $\nu_{\alpha}$ parametrizing the $E_{\alpha}$-dependent piece in the velocity $\partial_t g \, g^{-1}$ (see, e.g., section 2.4 of \cite{Damour:2009zc}). Such a Kac--Moody-related structure is present in our quantum Hamiltonian $\widehat H$, especially if we consider it {\it before} its reduction to zero rotational momenta. 
 
 First, $\widehat H$ contains the following contributions that are quadratic in the $\widehat\Phi$'s and that are
 related to the three dominant gravitational roots : 
 \begin{equation}
 \widehat H_g^J = \frac12 \, C_{ \ 23}^1 \, e^{-\alpha_{11}^g(\beta)} \, \widehat J_{11} + \frac12 \, C_{ \ 31}^2 \, e^{-\alpha_{22}^g (\beta)} \, \widehat J_{22} + \frac12 \, C_{ \ 12}^3 \, e^{-\alpha_{33}^g (\beta)} \, \widehat J_{33}\qquad .
\end{equation}
 
In addition, the terms linear and quadratic in the rotational momenta $p_a$ conjugate to the angular velocities $w_a$ (see Eq. (\ref{pa})) contribute to the Hamiltonian terms of the form : 
\begin{equation}\label{HS2}
\widehat H_{\rm sym}^S = \frac14 \, \frac1{\sinh^2 \, \alpha_{12}^{\rm sym} (\beta)} \left( \widehat p_3 - \cosh \alpha_{12}^{\rm sym} (\beta) \, \widehat S_{12} \right)^2 + {\rm cyclic}_{123} \qquad .
\end{equation}
The terms quadratic in the $\widehat\Phi$'s in the latter expression are
\begin{equation}
-\frac12 \, \widehat p_3 \, \frac{\cosh \alpha_{12}^{\rm sym} (\beta)}{\sinh^2 \alpha_{12}^{\rm sym} (\beta)} \, \widehat S_{12} + {\rm cyclic}_{123} \qquad .
\end{equation}
When inserting these contributions in the Heisenberg equations of motion, one will have contributions to $\partial_t \, \widehat\Phi_A^a$ of the respective form
$$
\sim C_{ \ 23}^1 \, e^{-\alpha_{11}^g (\beta)} \ i \left[ \widehat J_{11} , \widehat\Phi_A^a \right] + {\rm cyclic}_{123}
$$
and
$$
\sim - \widehat p_3 \, \frac{\cosh \alpha_{12}^{\rm sym} (\beta)}{\sinh^2 \alpha_{12}^{\rm sym} (\beta)} \ i \left[ \widehat S_{12} , \widehat\Phi_A^a \right] + {\rm cyclic}_{123} \, .
$$
These terms will be of the expected form
$$
Q_{\alpha} \cdot \Phi_A^a \sim e^{-\alpha (\beta)} \, p_{\alpha} \, J_{\alpha}^{\rm vs} \cdot \Phi_A^a
$$
if the commutators $i \left[ \widehat J_{11} , \widehat\Phi_A^a \right]$, $- i \left[ \widehat S_{12} , \widehat\Phi_A^a \right]$ (respectively associated with the roots $\alpha_{11}^g$ and $\alpha_{12}^{\rm sym}$) correctly reproduce the corresponding actions $J_{\alpha}^{\rm vs} \cdot \Phi_A^a$, within the vector-spinor representation of $K(AE_3)$.

\smallskip

That this is indeed the case follows from the {\it functorial} property of the second quantization of the gravitino. Indeed, similarly to what was noticed in the spin-$\frac12$ case \cite{Damour:2011yk}, the quantization conditions (\ref{ACRPhi}) ensure that if we are given a {\it first-quantized} operation ${\mathcal O}^{1q}$, acting as a $12 \times 12$ matrix on the combined vector-spinor index $(a,A)$ of $\Phi_A^a$, the corresponding {\it second-quantized} operator $\widehat{\mathcal O}^{2q}$ defined as
\begin{equation}
\widehat{\mathcal O}^{2q} := \frac12 \sum_{a,b,A} G_{ab} \, \Phi_A^{a} (\widehat{\mathcal O^{1q} \, \Phi } )_A^b
\end{equation}
will generate, by commutators, the action of ${\mathcal O}^{1q}$ on $\Phi_A^a$, {\it i.e.} 
\begin{equation}
\left[ \widehat{\mathcal O}^{2q} , \widehat\Phi_A^a \right] = ( \widehat{{\mathcal O}^{1q} \, \Phi} )_A^a \, ,
\end{equation}
and will also satisfy quantum commutation relations that exactly parallel the matrix commutation relations satisfied by the first-quantized matrices ${\mathcal O}^{1q}$, {\it i.e.} 
\begin{equation}
\left[ \widehat{\mathcal O}^{2q}_1 , \widehat{\mathcal O}^{2q}_2 \right] = \left[ \widehat{{\mathcal O}^{1q}_1 , {\mathcal O}^{1q}_2} \right] \, .
\end{equation}

We have checked that, modulo a conventional factor $\pm i$ needed to pass from the anti-hermitian generators\footnote{Note also that the gravitational-root generator $J_{11}$ was denoted $J_{1;23}$ in  \cite{Damour:2009zc}.} used in Refs. \cite{Damour:2006xu,Damour:2009zc} to the (formally) hermitian ones used in the present work, we had indeed such a first-quantized $\to$ second-quantized mapping between the vector-spinor representation generators $J_{\alpha}^{\rm vs}$ of previous works \cite{Damour:2006xu,Damour:2009zc}, and our quantum operators $ \widehat S_{12}, \ldots , \widehat J_{11} , \ldots$ entering the Hamiltonian $\widehat H$, namely
$$
\widehat S_{12}^{ \ {\rm here}} = \frac12 \, G_{ab} \, \widehat\Phi^{a \, T} \left( \widehat{-i \, J_{\alpha_{12}^{\rm sym}}^{\rm vs} \, \Phi} \right)^b
$$
$$
\widehat J_{11}^{ \ {\rm here}} = \frac12 \, G_{ab} \, \widehat\Phi^{a \, T} \left( \widehat{+i \, J_{\alpha_{11}^g}^{\rm vs} \Phi} \right)^b \, .
$$
As a consequence of the structure of the Lie algebra $K(AE_3)$, we can conclude from this result that the basic blocks 
 $\hat S_{12}, \hat S_{23} , \hat S_{31}, \hat J_{11} , \hat J_{22} , \hat J_{33}$ generate (via commutators) a Lie-algebra which is a 64-dimensional representation of the (infinite-dimensional) ``maximally compact'' sub-algebra, $K(AE_3)$, of $AE_3$.  First, we note that
 the $\hat S$'s generate the $(\ell = 0)$ sub-algebra $SO(3)$ of $K(AE_3)$:
 $$
  [\widehat S_{12} , \widehat S_{23}] = + i \, \widehat S_{31} \, , \, {\rm etc.}
  $$
  Second, though the quantum Hamiltonian  explicitly features only the three gravitational-wall generators
  $\widehat J_{11}$,   $\widehat J_{22}$,  $\widehat J_{33}$, associated with the real-roots $\alpha_{11}^g$, $\alpha_{22}^g$, $\alpha_{33}^g$,  the ones associated with the subdominant  gravitational-wall roots  $\alpha_{12}^g$, $\alpha_{23}^g$, $\alpha_{31}^g$ are generated by acting with the $\widehat S_{ab}$'s on the dominant  $\widehat J_{aa}$'s. For instance,
  $$
  \widehat J_{12} := -\frac{i}2 \, [\widehat S_{12} , \widehat J_{11}]
  $$
  Then, having so constructed quantum generators for  $K(AE_3)$ at levels $0$ and $1$, the commutators of
  level-1 generators among themselves will generate (modulo level-0 generators) the level-2 generators.
  By induction, all generators can be obtained, and the consistency of the (12-dimensional)  vector-spinor representation guarantees
  than one so generates a consistent (though unfaithful) representation of the full $K(AE_3)$  Lie algebra
  by $64 \times 64$ matrices

\smallskip

Above, we focussed on the Kac--Moody meaning of the terms in $\widehat H$ that are quadratic in fermions. On the other hand, we see on Eq. (\ref{HS2}) an analog of a well-known fact: a Lagrangian containing a linear coupling to velocities, say $L = \frac12 \, \dot q^2 + A \, \dot q$ (so that $p = \partial L / \partial \dot q = \dot q + A$), leads to the Hamiltonian $H = \frac12 \, (p-A)^2$, which contains, besides the linear coupling $-Ap$, an extra term quadratic in $A = p - \dot q$. It was argued in Ref. \cite{deBuyl:2005zy}, in the context of a coset model including spin-$\frac12$ fermions $\chi$, rather than the spin-$\frac32$ fermions $\Psi$ of supergravity that this mechanism will generate a squared-mass term $\mu^2$ formally given by the quadratic Casimir of the compact Lie-algebra $K$, {\it i.e.} 
$$
\mu^2_{\rm coset} = \frac12 \sum_{\alpha} (i \, J_{\alpha}^s)^2
$$
where the superscript $s$ refers to a spinor representation of $K$. (See also the discussion in \cite{Damour:2011yk}.)

\smallskip

The extension of this result to a second quantized spin-$\frac32$ coset model would suggest that an operatorial squared-mass term of the form
$$
\widehat\mu^2_{\rm coset} = \frac12 \sum_{\alpha} \left(i \, \widehat{J_{\alpha}^{\rm vs}}\right)^2 \, ,
$$
{\it i.e.}  the quantum version of the formal definition of the (hermitian) Casimir of $K$. If that were the case, we would expect the operator $\widehat\mu^2$ to commute with all the generators of the compact Lie algebra $K$ ($K(AE_3)$ in our case).

\smallskip

It is remarkable that our (uniquely defined) result for the squared-mass generator $\widehat\mu^2$ happens indeed to belong to the {\it center} of the algebra generated by the quantum $K(AE_3)$ generators $\widehat S_{ab}$, $\widehat J_{ab}$ ({\it i.e.}  it commutes with all of them). 
This term gathers many complicated, quartic-in-fermions contributions: not only contributions quadratic in the spin operators
$\widehat S_{ab}$ [via Eq. (\ref{HS2})] , but also all  the infamous $\psi^4$ terms present in the original, second-order supergravity action. In spite of this mixed origin, at the end of the day, the structure of the operator  $\widehat\mu^2$
is remarkable simple. Not only does it 
 belong to the {\it center} of the algebra generated by the $K(AE_3)$ generators $\hat S_{ab} , \hat J_{ab}$,
 but the {\it quartic} in fermions operator  $\hat\mu^2$ can finally be expressed in terms of the square of a very simple operator (which also commutes with $\hat S_{ab} , \hat J_{ab}$), namely, we find
\begin{equation}
\label{mu2}  
\widehat\mu^2 = \frac12 - \frac78 \, \widehat C_F^2
\end{equation}
where 
\begin{equation}\widehat C_F := \frac12 \, G_{ab} \overline{\widehat  \Phi}^a \, \gamma^{\hat 1\hat 2\hat 3} \, \widehat\Phi^b \, .\label{CFGPhi}
\end{equation}
 As we shall discuss next, $\widehat C_F$ is related to the fermion number operator $\widehat N_F$ by : 
 \begin{equation}\label{CFNF}
 \widehat C_F \equiv \widehat N_F - 3\qquad .
\end{equation}

 Let us now recall  the definition of the Weyl chamber of
 $AE_3$ and show its connection with various elements of the Bianchi IX  dynamics.
  In Kac--Moody theory a Weyl chamber is defined as  a polyhedron of $\beta$-space (identified with the
 space parametrizing a Cartan subalgebra of  $AE_3$) which is bounded by  $r$ hyperplanes  $\alpha_i(\beta)=0$
 corresponding to
 a set of ``simple" roots of  $AE_3$,
 {\it i.e.}  a set of linear forms $\alpha_i(\beta)$, $i= 1, \ldots, r$ (where $r$ denotes the rank; equal to 3 in the present case)
 such that all the other roots $\alpha(\beta)$ can be written  as a linear combination of the simple  roots with integer
 coefficients which can  be taken to be either all positive (for ``positive'' roots) or all negative (for ``negative'' roots).
 In the case of  $AE_3$  one can take as simple roots $\alpha_1(\beta) = \beta^2 - \beta^1$, 
 $\alpha_2(\beta) = \beta^3 - \beta^2$, and $\alpha_3(\beta) = 2 \beta^1$. The first two roots are symmetry-wall forms
 $\alpha_{ab}^{\rm sym}(\beta)$ (modulo some choice of signs), while the last root is a gravitational-wall form   $\alpha_{11}^{\rm g}(\beta)$. The corresponding  $AE_3$ Weyl chamber is, by definition,
 the polyhedron of $\beta$ space where   $\alpha_1(\beta) \geq 0$,   $\alpha_2(\beta) \geq 0$ and  $\alpha_3(\beta) \geq 0$. In other words, it is such that $ 0 \leq \beta^1 \leq   \beta^2 \leq  \beta^3$. We shall refer to it as being the
 ``canonical Weyl chamber'' in $\beta$-space. 
Its boundaries are the two symmetry walls   $\beta^1 = \beta^2$, 
 and $\beta^2 = \beta^3$, as well as the gravitational wall  $2 \beta^1 = 0$.
 The canonical Weyl chamber in $\beta$ space (as well as some of the equivalent Weyl chambers, see below)
 is illustrated in Fig.  \ref{betaspace}.
 
{\begin{figure}[h]
\begin{center}
\includegraphics[height=75mm]{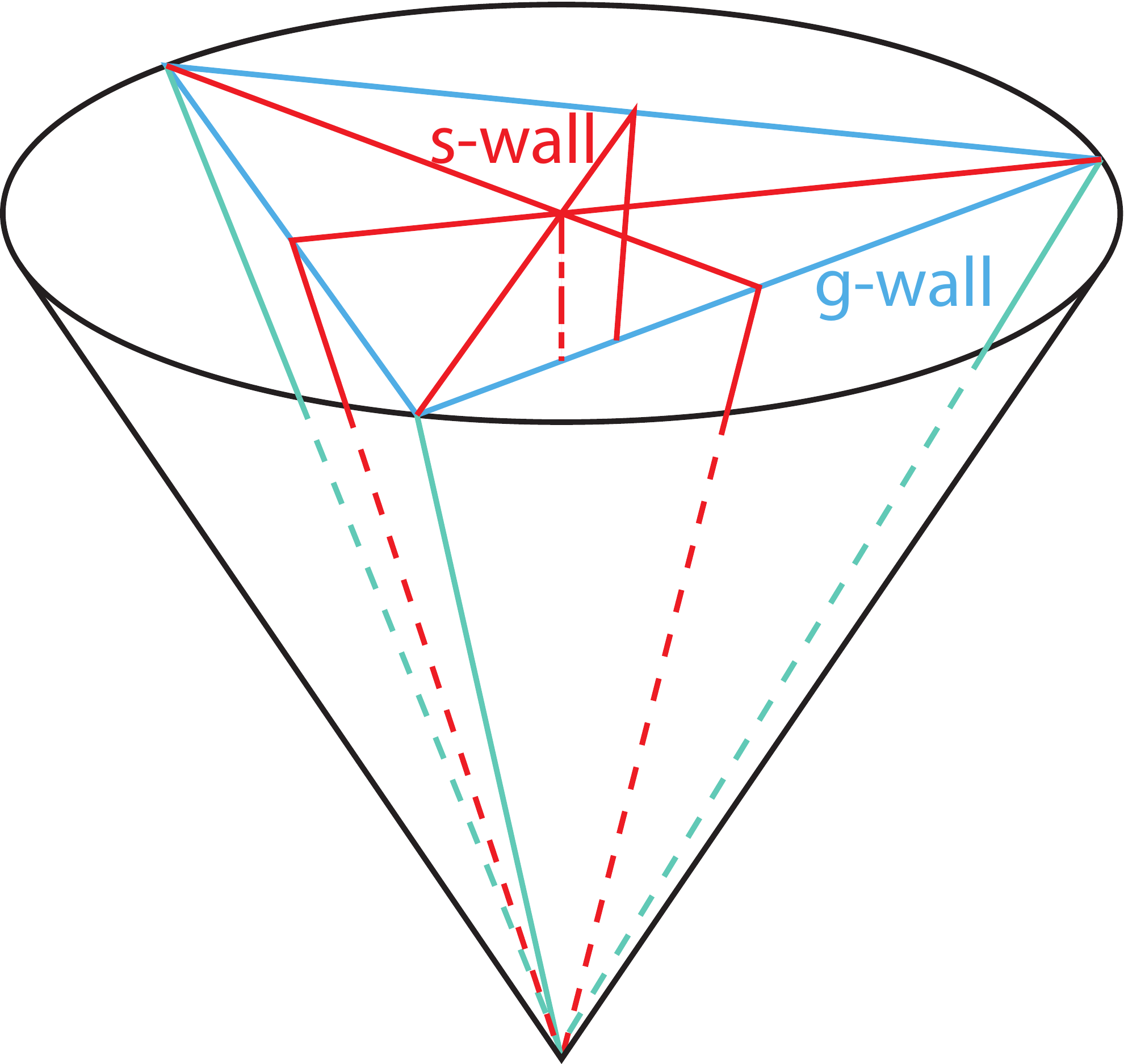}
\end{center}
\caption{\label{betaspace}\footnotesize $\beta$-space light-cone and its decomposition in  Weyl chambers  separated either by symmetry (s) of gravitational (g) walls. Six Weyl chambers are shown. One of them will be used as our canonical
Weyl chamber.}
\end{figure}}

 The role, in our Hamiltonian,  of the boundaries of the Weyl chambers  is somewhat dissymetric.  The three symmetry walls
 $\beta^1 = \beta^2$, $\beta^2 = \beta^3$ and $\beta^3 = \beta^1$
 are such that the terms containing hyperbolic cotangents of the corresponding symmetry-wall forms  $\alpha_{ab}^{\rm sym}(\beta)$  (associated with corresponding spin operators $S_{ab}$)  become singular
 on them (see e.g.  (\ref{SsA}) ). By contrast,  the terms containing the gravitational wall forms   $\alpha_{ab}^{\rm g}(\beta)$ (either in the susy constraints or in the Hamiltonian) 
 do not become singular on the gravitational walls . Rather,  the corresponding gravitational-wall potential
 terms are ``soft'' potential walls which start being repulsive as the $\beta$ particle representing the
 dynamics of the geometry starts penetrating within, say, the  $\alpha_{11}^{\rm g}(\beta)$ gravitational wall 
 defining one of the boundaries of the canonical Weyl chamber. It is only  when considering the near-singularity billiard limit,
 where all the $\beta$'s tend to large, positive values, that the gravitational wall tends to define a sharp limit
 similar to the sharp walls associated with the symmetry-wall forms. [This will be seen explicitly below when discussing
 the effect of $\mu^2$ on the approach to the cosmological singularity.]
 
 The Kac--Moody/gravity conjecture  assumes that the symmetry between symmetry walls and gravitational walls
 (and thereby between all possible choices of Weyl chambers) will be somehow restored when considering
 the quantum dynamics of the unifying theory behind supergravity. In the present paper, we shall stay at the
 level of the supergravity description. At this level, though there will be a dissymetry between symmetry roots
 and gravitational roots, there will still be a (nearly) manifest permutation symmetry between the three
 (or six, if we include their   sign-reversed versions)
 different symmetry roots   $\alpha_{ab}^{\rm sym}(\beta)$. This symmetry is simply the group
 of permutation of three objects $S_3$ (say of the three $\beta^a$'s). 
 This is illustrated in  Fig. \ref{WeylChfig}. This figure is obtained by intersecting the polyhedral 
 Weyl chambers of Fig.  \ref{betaspace} by a hyperplane $ \beta^1 + \beta^2 + \beta^3= $ constant.
 Our canonical Weyl chamber  where $ 0 \leq \beta^1 \leq   \beta^2 \leq  \beta^3$ is labelled as $(a)$
 in this figure.  The action of the permutation group   $S_3$ maps this canonical chamber into six 
 equivalent chambers (labelled $a,\ b,\ c,\ d,\  e,\ f$).  In the following, because of the soft, penetrable nature
 of the gravitational walls, we shall have to distinguish between the usual Kac--Moody definition of
 a Weyl chamber (which, e.g., in the case of the chamber labelled a would stop at the gravitational
 wall $\beta^1=0$), and the definition of the corresponding chamber of $\beta$-space in which we shall solve
 the susy constraints (which will actually be the full dihedron  between the two symmetry walls 
  $\beta^1 = \beta^2$, and $\beta^2 = \beta^3$, {\it i.e.}  the domain  $  \beta^1 \leq   \beta^2 \leq  \beta^3$,
  without restriction on the value of $\beta^1$).   The permutation symmetry  $S_3$ between the six
  chambers  $a,\ b,\ c,\ d,\  e,\ f$ in  Fig. \ref{WeylChfig} is rooted in the basic diffeomorphism symmetry
  of supergravity. More precisely,   $S_3$ can be considered as a group of ``large diffeomorphisms".
  The constraint linked to small diffeomorphisms, {\it i.e.}  $H_i \, \vert{\bm\Psi}\rangle = 0$,
  or equivalently,  $\widehat p_a \, \vert{\bm\Psi}\rangle = 0$, was saying that $\Psi$ does not depend
  on the Euler angles. It is natural to think that the gauge invariance under large diffeomorphisms
  is furthermore saying that the wavefunction $\Psi(\beta)$ ``lives'' only in one of the six equivalent chambers;
  the other ones being just gauge-equivalent description of the same physics. 
  In the following,
  we shall therefore often restrict our study of the wavefunction to the canonical chamber (a), {\it i.e.}  
  $  \beta^1 \leq   \beta^2 \leq  \beta^3$.

{\begin{figure}[h]
\begin{center}
\includegraphics[height=75mm]{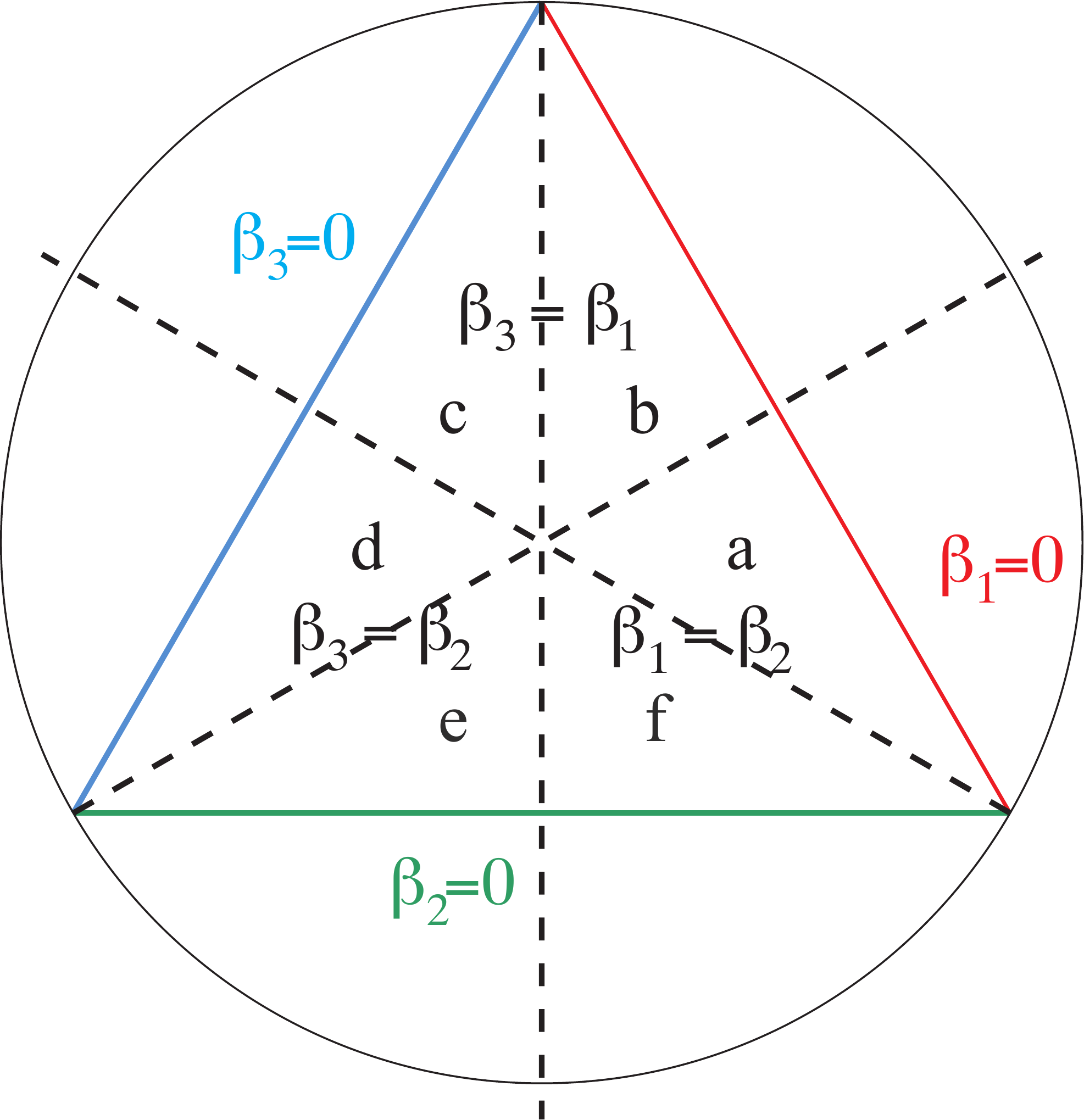}
\end{center}
\caption{\label{WeylChfig}\footnotesize The various Weyl chambers. Chamber $(a)$, where 
$ 0 \leq \beta^1 \leq   \beta^2 \leq  \beta^3$ will generally be taken as our canonical Weyl chamber.}
\end{figure}}

\setcounter{equation}{0}\section{Fermion number operators in spinorial space}

\smallskip

To be able to describe in detail the set of solutions of the supersymmetry constraints 
$$\widehat {\cal S}^{(0)}_A \vert {\bm \Psi}\rangle = 0 \, ,
$$
it will be convenient to replace the $3 \times 4$   ``real''  ({\it i.e.}  Majorana)  operators $\widehat \Phi^a_A$,  that enter both the $\beta$-derivative terms in the supersymmetry operators, and the subsequent potential (and mass-type) terms
\begin{equation}
\widehat {\cal S}^{(0)}_A = + \frac{i}{2} \Phi^a_A \partial_{\beta^a} + \widehat V_A(\beta,\Phi)  \qquad .
\end{equation}
by   $3 \times 2$ complex
``annihilation operators''    $b_\epsilon^a$ (where the index $\epsilon$ takes two values, say $ +, -$), and the
corresponding  hermitian-conjugated ``creation operators''   $\widetilde b_\epsilon^a$.  [Henceforth, to ease the notation we do not put hats on the $b_{\epsilon}^a$, and $\widetilde b_\epsilon^a$ operators.]
The definition of the $b$'s
and $\widetilde b$'s we shall use is:
\begin{eqnarray}
b_+^a &= &\widehat \Phi_1^a + i \, \widehat \Phi_2^a \nonumber \\
b_-^a &= &\widehat \Phi_3^a - i \, \widehat \Phi_4^a \nonumber \\
\widetilde b_+^a &= &\widehat \Phi_1^a - i \, \widehat \Phi_2^a \nonumber \\
\widetilde b_-^a &= &\widehat \Phi_3^a + i \, \widehat \Phi_4^a \qquad . \label{bbd} 
\end{eqnarray}
The various signs appearing in these combinations are related to our convention for the value of the matrix
$\gamma^5$ in the Majorana representation we use. We use the following real $\gamma$ matrices:
\begin{eqnarray}
\hspace{-6mm}  \gamma^{\hat 1}=
\left(
\begin{array}{cccc}
0  &   1&0&0   \\
1  &  0 & 0  &0\\
 0 & 0  & 0  &1\\
 0 & 0  &1   & 0
\end{array}
\right)\hspace{6mm}\quad& , &\quad
\gamma^{\hat 2}=
\left(
\begin{array}{cccc}
-1  &  0&0&0   \\
0  &  1 & 0  &0\\
 0 & 0  & -1  &0\\
 0 & 0  &0  & 1
\end{array}
\right)\quad,\\
\hspace{-6mm} \gamma^{\hat 3}=
\left(
\begin{array}{cccc}
0  &  0&0&-1   \\
0  & 0 & 1  &0\\
 0 &1  & 0  &0\\
 -1 & 0  &0  & 0
\end{array}
\right)\quad &, &\quad\gamma^{\hat 0}=
\left(
\begin{array}{cccc}
0  &  0&0&1   \\
0  & 0 &-1  &0\\
 0 &1  & 0  &0\\
 -1 & 0  &0  & 0
\end{array}
\right)\quad, 
\end{eqnarray}
leading to
\begin{equation} \label{gamma5}
\gamma^5 =\gamma^{\Hat 0}\gamma^{\Hat 1}\gamma^{\Hat 2}\gamma^{\Hat 3}=
\left(
\begin{array}{cccc}
0  &  -1&0&0   \\
1  & 0 &0  &0\\
 0 &0  & 0  &1\\
 0& 0  &-1  & 0
\end{array}
\right) \qquad .
\end{equation}

The above definition is such that the $b$'s correspond to a ``chiral'' projection of the $\Phi's$ in the sense that 
$b_+^a$ and $ b_-^a$
are proportional to the two independent  spinor components of
$$
( 1 - i \, \gamma^5) \widehat \Phi^a 
$$
while  $\widetilde b_+^a $ and   $\widetilde b_-^a $ are proportional to those of $ ( 1 + i \, \gamma^5) \widehat \Phi^a $.

The definitions above are such that the   $ b$'s and  $\widetilde b$'s satisfy  usual-type 
anti-commutation relations  for annihilation/creation operators, modulo the replacement of the
expected Euclidean metric, by the Lorentzian-signature $\beta$-space metric $G^{ab}$
metric 
\begin{eqnarray}
\label{bbCR}
&&\left\{ b_{\epsilon}^k , \tilde b_{\epsilon'}^{l} \right\} = 2 \, G^{kl} \delta_{\epsilon\, \epsilon'}\qquad ,
\\
&&\left\{ b_{\epsilon}^k , b_{\epsilon'}^{l} \right\} = \left\{ \tilde b_{\epsilon}^k , \tilde b_{\epsilon'}^{l} \right\} = 0 \qquad .
\end{eqnarray}
As indicated here, we shall henceforth indicate the $\beta$-space indices (either on $\beta$, $\Phi$, $b$, $G$, etc) by
arbitrary latin indices, $ a, b, c, \ldots, k, l, m, \ldots$, without limiting ourselves (as we did up to now) to the
first part of the Latin alphabet. Note that the position (up or down) of these  $\beta$-space indices is meaningful, and
should be respected. For instance, the indices on the $\Phi$'s, and therefore on the $b$'s are contravariant.
This is why it is the inverse metric $G^{kl}$ which appears in the anti-commutation relations (\ref{bbCR}).

The operators   $ b$'s and  $\widetilde b$'s are useful because they allow one to decompose 
the 64-dimensional spinorial space ${\mathbb H}$ on which they act into ``slices'' corresponding to the usual
Fock-type construction of a fermionic Hilbert space. More precisely, there exists a unique ``vacuum state''
such that
\begin{equation}
b_{\epsilon}^k   \, {{\vert\, 0\,\rangle\hspace{-0.5mm}\strut_-}} = 0
\end{equation}
or, equivalently (in terms of the real Clifford algebra $\Phi$'s)
\begin{equation}
 ( 1 - i \, \gamma^5) \widehat \Phi^k \, {{\vert\, 0\,\rangle\hspace{-0.5mm}\strut_-}} = 0
\end{equation}

We then obtain a basis of the whole space ${\mathbb H}$ by acting on ${{\vert\, 0\,\rangle\hspace{-0.5mm}\strut_-}}$ with the 64 possible products of all different $\tilde b_{\epsilon}^k$ operators. This construction defines a bi-grading 
($N_{+}^F$ , $N_{-}^F$) 
on ${\mathbb H}$, defined by  (separately) counting the number of operators $\tilde b_{+}^k$ 
and  $\tilde b_{-}^k$ that act on ${{\vert\, 0\,\rangle\hspace{-0.5mm}\strut_-}}$. In other words we obtain two fermionic number operators $\widehat N_{\pm}^F$, that can be represented as
\begin{equation}
\widehat N_{\epsilon}^F = \frac12 G_{kl} \, \tilde b_{\epsilon}^k \, b_{\epsilon}^{l}
\end{equation}
These operators satisfy the commutation relations
\begin{equation}
[\widehat N_{\epsilon}^F , b_{\epsilon'}^k] = -\delta_{\epsilon\, \epsilon'} \, b_{\epsilon'}^k \, ; \quad [\widehat N_{\epsilon}^F , \tilde b_{\epsilon'}^k] = + \delta_{\epsilon\,\epsilon'} \, \tilde b_{\epsilon'}^k
\end{equation}
and their eigenvalues run from $0$ to $3$.

\smallskip

We also consider the total fermionic number operator
\begin{equation}\label{NF}
\widehat N_F = \widehat N_+^F + \widehat N_-^F
\end{equation}
whose eigenvalues vary from $0$ to $6$. This operator will play an important r\^ole in the structure of the solution space
because, as we shall soon see, it has nice commutation relations with the chiral components of the supersymmetry
operators. 

As was already mentioned above,  the fermion-number operator $\widehat N_F$  is very simply
related to a remarkably simple quadratic fermion operator $\widehat C_F$ that crucially enters in the 
 ``squared-mass term''  $\widehat\mu^2$ occuring in the Hamiltonian $\widehat H' $. Namely,
\begin{equation}\label{CF}
\widehat C_F = \widehat N_F - 3 = \frac12 \, G_{kl} \, \overline{\widehat \Phi}^k \gamma^{\hat1\hat2\hat3} \, \widehat \Phi^{l}
\end{equation}

It is worthwhile to notice that the ladder compact generators $\widehat J_{11}$, $\widehat J_{22}$ and $\widehat J_{33}$ that occur in the $\widehat V^J$ part (\ref{VJ}) of the potential (\ref{decU}) commute both with $\widehat N_F$ and the $\widehat N^F_\pm$ operators while the spin operators (\ref{Spin}) only commute with $\widehat N_F$, except for $\widehat S_{12}$ that commutes with $\widehat N^F_\pm$.

As we shall use it systematically in the following, let us describe in detail the decomposition of the 64-dimensional space ${\mathbb H}$ first into eigenspaces  ${\mathbb H}_{[N_F]}$ of the total fermion number operator  $\widehat N_F$,
and then  into common eigenspaces ${\mathbb H}_{(N^F_+ , N^F_-)}$ of the two separate fermion number operators
$ \widehat N_+^F ,  \widehat N_-^F $ (with $N_F= N^F_+  +  N^F_-$). To do that, one must take  into account both the $\widehat N_{\epsilon}^F$ eigenvalues and the symmetry (or lack of symmetry) of the Lorentzian indices $k,l,\ldots$ of the products of the $\widetilde b_{\epsilon}^k$ operators acting on ${{\vert\, 0\,\rangle\hspace{-0.5mm}\strut_-}}$.

\smallskip

The $N_F = 0$ space is the one-dimensional space generated by ${{\vert\, 0\,\rangle\hspace{-0.5mm}\strut_-}}$:
\begin{equation}
{\mathbb H}_{[0]} = {\mathbb H}_{(0,0)}=  {\rm span}_{\mathbb C}  {{\vert\, 0\,\rangle\hspace{-0.5mm}\strut_-}}  \, .
\end{equation}
Here  ${\rm span}_{\mathbb C} \, \{{\cal B}\}$ denotes the vector space generated by all complex  linear combinations of elements of the set $\{{\cal B}\}$. 

 The $N_F=1$ subspace ${\mathbb H}_{[1]}$ is 6-dimensional, and splits into two 3-dimensional subspaces 
 $ {\mathbb H}_{[1]} = {\mathbb H}_{(1,0)} \oplus {\mathbb H}_{(0,1)} $ with : 
\begin{eqnarray}
{\mathbb H}_{(1,0)} &= &{\rm span}_{\mathbb C} \, \{\tilde b_+^k \, {{\vert\, 0\,\rangle\hspace{-0.5mm}\strut_-}} \} \nonumber \\
{\mathbb H}_{(0,1)} &= &{\rm span}_{\mathbb C} \, \{\tilde b_-^k \, {{\vert\, 0\,\rangle\hspace{-0.5mm}\strut_-}} \} \nonumber
\end{eqnarray}

The  $N_F = 2$ eigenspace  ${\mathbb H}_{[2]}$ is 15-dimensional. It naturally decomposes itself into $3+3+3 + 6$ dimensional subspaces:
\begin{eqnarray} \label{decompH2}
{\mathbb H}_{(2,0)} &= &{\rm span}_{\mathbb C} \, \{\tilde b_+^{k} \, \tilde b_+^{l} \, {{\vert\, 0\,\rangle\hspace{-0.5mm}\strut_-}} \} \nonumber \\
{\mathbb H}_{(0,2)} &= &{\rm span}_{\mathbb C} \, \{\tilde b_-^{k} \, \tilde b_-^{l} \, {{\vert\, 0\,\rangle\hspace{-0.5mm}\strut_-}} \} \nonumber \\
{\mathbb H}_{(1,1)_A} &= &{\rm span}_{\mathbb C} \, \{\tilde b_+^{[k} \, \tilde b_-^{l]} \, {{\vert\, 0\,\rangle\hspace{-0.5mm}\strut_-}} \} \nonumber \\
{\mathbb H}_{(1,1)_S} &= &{\rm span}_{\mathbb C} \, \{\tilde b_+^{(k} \, \tilde b_-^{l)} \, {{\vert\, 0\,\rangle\hspace{-0.5mm}\strut_-}} \}  \qquad .\nonumber 
\end{eqnarray}
In the first three spaces we have (either naturally, or by explicit projection\footnote{with $T_{[kl]} := \frac12 (T_{kl} - T_{lk})$
and  $T_{(kl)} := \frac12 (T_{kl} + T_{lk})$.}) 
antisymmetry over the two indices $k l$, corresponding to three independent possibilities. By contrast, the symmetry over $ k l$ in ${\mathbb H}_{(1,1)_S}$
leads to a 6-dimensional space.

The next level, $N_F=3$,  ${\mathbb H}_{[3]}$, is 20-dimensional. It splits into two 10-dimensional subspaces that themselves decompose into $1$, $3$ and $6$-dimensional subspaces : 
\begin{eqnarray}
{\mathbb H}_{(3,0)} &= &{\rm span}_{\mathbb C} \left\{\tilde b_+^1 \, \tilde b_+^2 \, \tilde b_+^3 \, {{\vert\, 0\,\rangle\hspace{-0.5mm}\strut_-}} \right\} \nonumber \\
{\mathbb H}_{(2,1)_A} &= &{\rm span}_{\mathbb C} \left\{ \frac12 \, \eta^{[k}_{\ \ pq} \, \tilde b_-^{l]} \, \tilde b_+^p \, \tilde b_+^q \, {{\vert\, 0\,\rangle\hspace{-0.5mm}\strut_-}} \right\} \nonumber \\
{\mathbb H}_{(2,1)_S} &= &{\rm span}_{\mathbb C} \left\{ \frac12 \, \eta^{(k}_{ \ \ pq} \, \tilde b_-^{l)} \, \tilde b_+^p \, \tilde b_+^q \, {{\vert\, 0\,\rangle\hspace{-0.5mm}\strut_-}} \right\} \qquad , \nonumber
\end{eqnarray}
and similarly for ${\mathbb H}_{(0,3)}$,  ${\mathbb H}_{(1,2)_A} $, and ${\mathbb H}_{(1,2)_S} $. Above, we have used
the Levi-Civita tensor $\eta_{lpq}$ in $\beta$-space (with one index raised by $G^{kl}$).

At this stage we have described half of the ${\mathbb H}$ space. The second half can be obtained in two equivalent ways: 
either (i) by  continuing to act on the ``minus'' vacuum state $ {{\vert\, 0\,\rangle\hspace{-0.5mm}\strut_-}}  $ by means of
creation operators  $\widetilde b_{\pm}^k$, or, (ii) by
exchanging the roles of the $\tilde b_{\epsilon}^k$ operators and $b_{\epsilon}^k$ operators and by starting from the ``filled''  fermionic  state 
$$
{{\vert\, 0\,\rangle\hspace{-0.5mm}\strut_+}} = \frac 14\prod_{\epsilon} \, \prod_k \, \tilde b_{\epsilon}^k \, {{\vert\, 0\,\rangle\hspace{-0.5mm}\strut_-}}
$$
  {\it i.e.}\    the (unique) state\footnote{Here normalized so that $ b_+^1 \,  b_+^2 \,  b_+^3 \, {{\vert\, 0\,\rangle\hspace{-0.5mm}\strut_+}}$  coincides with $\tilde b_+^1 \, \tilde b_+^2 \, \tilde b_+^3 \, {{\vert\, 0\,\rangle\hspace{-0.5mm}\strut_-}}$.}
 that is  annihilated by all $\tilde b_{\epsilon}^k$ operators~: 
\begin{equation}
\label{Psi64}
\tilde b_{\epsilon}^k \, {{\vert\, 0\,\rangle\hspace{-0.5mm}\strut_+}} = 0
\end{equation}
In the second construction (from the filled state), we have:
$\mathbb H_{(3,3)} = {\rm span}_{\mathbb C} \, \{{{\vert\, 0\,\rangle\hspace{-0.5mm}\strut_+}}\}$, ${\mathbb H}_{(2,3)} = {\rm span}_{\mathbb C} \, \{b_+^k \, {{\vert\, 0\,\rangle\hspace{-0.5mm}\strut_+}} \}$, etc$\ldots$.
Note that the filled state is also uniquely fixed (modulo an arbitrary factor) by the opposite-chirality condition that
fixed the empty state, namely
\begin{equation}
 ( 1 + i \, \gamma^5) \widehat \Phi^k \, {{\vert\, 0\,\rangle\hspace{-0.5mm}\strut_+}} = 0
\end{equation}

In many developments in the rest of this paper, it will be useful to have in mind the main characteristics of
each one of the  subspaces of $ {\mathbb H}_{(N^+_F,N^-_F)} $ that we have just considered,
 notably their dimensions, the corresponding
eigenvalue of $\mu^2$,  as well as the spectrum of the Kac--Moody-related operators $J_{ab}$
and $S_{ab}$ in these spaces.  Actually, it happens that while the  $J_{ab}$'s  are block diagonal with respect to (w.r.t.)
the above defined subspaces, this is not generally true for the  $S_{ab}$'s (which are only block diagonal
in larger subspaces of  $ {\mathbb H}_{[N_F]} $).  However, the squared-spin operators  $S_{ab}^2$,
which crucially enter the symmetry walls of the Hamiltonian operator turn out to be simpler,
and to be block diagonal w.r.t. the above defined subspaces of each fermion level. 
For the convenience of the reader, we shall gather this information in Appendix \ref{fermionicspaces}.

\smallskip

\setcounter{equation}{0}\section{Explicit structure of the supersymmetry operators in the chiral basis}

The main point established in the previous sections is that, in the minisuperspace framework in which we consider the quantization of ${\mathcal N} = 1$, $D=4$ Bianchi IX cosmological supergravity model, the relevant equations to be solved are
\begin{equation}
\label{SAEq}
\widehat {\cal S}^{(0)}_A \vert {\bm \Psi}\rangle = 0 \qquad .
\end{equation}
These equations consitute a system of four simultaneous Dirac equations in a 3--dimensional ``space-time" (the $\beta$-space) for a 64-component spinorial wave function $\Psi_{\sigma}(\beta)$.  The number and structure of the solutions of
this heavily overconstrained system of  partial differential equations is   {\it a priori}   unclear (and was left in great part undecided
by previous work on quantum supersymmetric Bianchi IX cosmology \cite{D'Eath:1993up,Csordas:1995kd,Csordas:1995qy,Cheng:1996an}). Here, we shall bring a rather complete answer to this issue by using the simplifications
obtained by projecting the supersymmetry operators in the chiral basis of the $b$'s and $\widetilde b$'s introduced above.

\smallskip

Similarly to the definition of the $b$ operators [see Eqs (\ref{bbd})], we define (omitting the operatorial hats) the (annihilation-type) chiral components of the supersymmetry operators as : 
\begin{eqnarray}
\label{Schiral}
{\cal S}^{(0)}_+ &=& {\cal S}^{(0)}_1 + i \,  {\cal S}^{(0)}_2, \\
{\cal S}^{(0)}_- &=& {\cal S}^{(0)}_3 - i \,  {\cal S}^{(0)}_4 \, .
\end{eqnarray}
The two non-hermitian operators  ${\cal S}^{(0)}_{\pm}$ represent half of the content of the four original hermitian 
$ {\cal S}^{(0)}_A$'s. The other half is described by the hermitian-conjugated operators   $\widetilde {\cal S}^{(0)}_{\pm}$.

With respect to such a chiral basis the supersymmetry operators have a rather simple structure. They read:
\begin{equation}
\label{Seps}
{\cal S}^{(0)}_\epsilon = \frac i2\, b
^k_\epsilon \, \partial_{\beta^k}  + \alpha_k(\beta)  \, b^k_\epsilon + \frac 12 \, \mu_{[kl] m}(\beta) \, B^{[kl] m}_\epsilon + \rho_{kl m}(\beta) \, C^{kl m}_\epsilon + \frac 12 \, \nu_{[kl] m}(\beta) \, D^{[kl] m}_\epsilon
\end{equation}
where the $B$'s, $C$'s and $D$'s are {\it cubic} in the fermion operators, and, more precisely, are of the
$ \widetilde b \, b \, b$ type, with always an annihilation operator on the right (so that   $B$, $C$, $D$, and therefore
${\cal S}$, acting on ${{\vert\, 0\,\rangle\hspace{-0.5mm}\strut_-}} $ yield zero). In addition,  the $B$'s and the $D$'s
are antisymmetric in the first two upper indices $k l$ (while the  the $C$'s do not have such a symmetry property).
Their explicit expressions are
\begin{eqnarray}\label{Bklm}
B^{klm}_\epsilon&=&\tilde b_\epsilon^m\,b_\epsilon^k\, b_\epsilon^l +G^{lm}\,b^k_\epsilon-G^{km}b^l_\epsilon \nonumber \\
&=& b_\epsilon^k\, b_\epsilon^l\,\tilde b_\epsilon^m-G^{lm}\,b^k_\epsilon+G^{km}b^l_\epsilon \qquad ,\label{Bklmbis}\\
C^{klm}_\epsilon &=&\tilde b_{-\epsilon}^m\,b_\epsilon^k\, b_{-\epsilon}^l +G^{lm}\,b^k_\epsilon \nonumber \\
&=& b_{\epsilon}^k\, b_{-\epsilon}^l\,\tilde b_{-\epsilon}^m-G^{lm}\,b^k_\epsilon
\qquad ,\label{Cklmbis}\\
D^{klm}_\epsilon&=&\tilde b_\epsilon^m\,b_{-\epsilon}^k\, b_{-\epsilon}^l\ \nonumber \\
&=& b_{-\epsilon}^k\, b_{-\epsilon}^l\,\tilde b_\epsilon^m 
\qquad .\label{Dklmbis}
\end{eqnarray}
As for the ($\epsilon$-independent) $\beta$-dependent coefficients $\alpha(\beta)$, $\mu(\beta)$, $\rho(\beta)$ and $\nu(\beta)$
entering  ${\cal S}^{(0)}_\epsilon$, they can be written as rational functions of the new variables
\begin{equation}
x := e^{2\beta^1} = \frac1{a^2}  \qquad , \qquad y := e^{2\beta^2} = \frac1{b^2} \qquad , \quad z := e^{2\beta^3}=\frac 1{c^2} \qquad .
\end{equation}
Namely (denoting the derivatives $\partial_{\beta^k}$ by $\partial_k$; note that $\partial_1= 2 x \partial_x$, etc.),
 \begin{eqnarray}\label{alphak}
\alpha_k &= &\frac i2 \left( \frac1x , \frac1y , \frac1z \right) =i\,\partial_k\alpha \qquad {\rm with}\qquad  \alpha =- \frac 14 \,  \left( \frac1x + \frac1y + \frac1z \right)   \qquad ,  \\
\mu_{klm} &= &\mu_{[k} \, G_{l]m} 
\qquad {\rm
with}\qquad
\mu_k =i \, \partial_k \mu \label{muk}\ ,\ \mu=\frac1 8   \ln\left \vert\frac{(x-y)}{x^2\,y^2\,z^2}\right\vert  \qquad , \label{mukpot} \\
\nu_{klm} &=&\nu_{[k} \, G_{l]m}  \qquad
{\rm with}\qquad
\nu_k =i \, \partial_k \nu\label{nuk}\ , \ \nu= \frac 18 \,   \ln \left\vert \frac{x- z}{y-z} \right\vert\qquad , \label{nukpot}\\
\rho_{klm}&=& \frac 1{10}\left(\phantom{\frac\strut\strut}\hspace{-1.7mm} (4\,{\overset{\text{\tiny{(1)}}}\rho}_k-{\overset{\text{\tiny{(2)}}}\rho}_k-{\overset{\text{\tiny{(3)}}}\rho}_k)G_{lm}+(4\,{\overset{\text{\tiny{(2)}}}\rho}_l-{\overset{\text{\tiny{(3)}}}\rho}_l-{\overset{\text{\tiny{(1)}}}\rho}_l)G_{km}+(4\,{\overset{\text{\tiny{(3)}}}\rho}_m-{\overset{\text{\tiny{(1)}}}\rho}_m-{\overset{\text{\tiny{(2)}}}\rho}_m)G_{kl}\right)\nonumber\\
&&+\tau_{(klm)}\label{rhoklm}\end{eqnarray}
where $\tau_{(klm)}$ is a completely symmetric traceless tensor, whose explicit form is displayed in Appendix \ref{Apptauform}, and  
\begin{eqnarray}\rho_{kl}^{\ \ l}&= : &{\overset{\text{\tiny{(1)}}}\rho}_k=i\,\partial_{\beta^k}r_1\qquad \text{where}\qquad r_1=  \frac 1{16} \ln  \left[\frac{(x-y)(x-z)^3(y-z)^3}{(x\,y\,z)^6}\right]\nonumber\\&& \label{rho1}\\
 \rho_{lk}^{\ \ l}&= :& {\overset{\text{\tiny{(2)}}}\rho}_k=i\,\partial_{\beta^k}r_2\qquad \text{where}\qquad r_2=  \frac 1{16} \ln\left [\frac{(x-y)^3(x-z)(y-z)}{(x\,y\,z)^2}\right]\nonumber\\&& \label{rho2}\\
 \rho_{l\ k}^{\ l}&= :& {\overset{\text{\tiny{(3)}}}\rho}_k=i\,\partial_{\beta^k}r_3\qquad \text{where\footnotemark}\qquad r_3= \frac 3{16} \ln\left [\frac{(x-y)(x-z)(y-z)}{(x\,y\,z)^2}\right]\nonumber\\&&\label{rho3}\end{eqnarray}\setcounter{footnote}{7}\footnotetext{Let us notice that  :  $r_1+r_2=\frac 43 r_3$ .}
 
 Note that all the coefficient functions  $\alpha_k(x,y,z)$,  $\mu_{klm}(x,y,z)$, $  \nu_{klm}(x,y,z)$, $  \rho_{klm}(x,y,z)$
 are {\it pure imaginary}, {\it i.e.}  they are of the form: $i$ times some real (rational) functions of $x,y,z$. As a consequence,
 the hermitian-conjugate of the chiral supersymmetry constraints read
 \begin{equation}\label{wtSeps}
\widetilde {\cal S}^{(0)}_\epsilon =+ \frac i2\, \widetilde b^k_\epsilon \, \partial_{\beta^k}  
- \alpha_k(\beta)  \, \widetilde b^k_\epsilon - \frac 12 \, \mu_{[kl] m}(\beta) \, \widetilde B^{[kl] m}_\epsilon - \rho_{kl m}(\beta) \, \widetilde C^{kl m}_\epsilon 
- \frac 12 \, \nu_{[kl] m}(\beta) \, \widetilde D^{[kl] m}_\epsilon
\end{equation}
Here, all operators have been tildeed, and all coefficients have changed sign, {\it except} the first which originally read
$- \frac12 b^k_\epsilon \widehat \pi_k$, and for which we used the fact that $\widetilde {\widehat \pi}_k = + \widehat \pi_k $

\par\noindent 
Globally, because of the structure $ {\cal S}^{(0)} \sim b + \widetilde b \, b \,  b$, ${\cal S}^{(0)}$ decreases the total fermion number $N_F$ by one unit, while $\widetilde {\cal S}^{(0)}$ increases  $N_F$ by one unit. But there are also some
similar conservation laws (modulo 2) when considering the finer decomposition of   ${\mathbb H}_{[N_F]}$  into
sums of ${\mathbb H}_{(N^F_+ , N^F_-)}$'s with $N_F= N^F_+  +  N^F_-$. Indeed, 
because of the specific values of the $\epsilon$ indices entering  the $B$'s, $C$'s and $D$'s  above,
the various terms appearing in (\ref{Seps}) act differently on the subspaces ${\mathbb H}_{(N^F_+ , N^F_-)}$, labelled by the separate $N^F_\pm=0,\dots,3$ eigenvalues. For instance we have
\begin{equation}
b_+^k  :  {\mathbb H}_{(N^F_+ , N^F_-)} \to {\mathbb H}_{(N^F_+ - 1 , N^F_-)} \qquad , \quad \tilde b_+^k  :  {\mathbb H}_{(N^F_+ , N^F_-)} \to {\mathbb H}_{(N^F_+ + 1 , N^F_-)}
\end{equation}
\begin{equation}
B_+^{klm}  :  {\mathbb H}_{(N^F_+ , N^F_-)} \to {\mathbb H}_{(N^F_+ - 1 , N^F_-)} \qquad , \quad \tilde B_+^{klm}  :  {\mathbb H}_{(N^F_+ , N^F_-)} \to {\mathbb H}_{(N^F_+ + 1 , N^F_-)}
\end{equation}
\begin{equation}
C_+^{klm}  :  {\mathbb H}_{(N^F_+ , N^F_-)} \to {\mathbb H}_{(N^F_+ - 1 , N^F_-)} \qquad , \quad \tilde C_+^{klm}  :  {\mathbb H}_{(N^F_+ , N^F_-)} \to {\mathbb H}_{(N^F_+ + 1 , N^F_-)}
\end{equation}
but
\begin{equation}
D_+^{klm}  :  {\mathbb H}_{(N^F_+ , N^F_-)} \to {\mathbb H}_{(N^F_+ + 1 , N^F_- -2)} \qquad , \quad \tilde D_+^{klm}  :  {\mathbb H}_{(N^F_+ , N^F_-)} \to {\mathbb H}_{(N^F_+ - 1 , N^F_- + 2)} \qquad .
\end{equation}
and similarly for the minus chirality operators, by exchanging the r\^ole of the labels $N^F_+$ and $N^F_-$.

The supersymmetry operators ${\cal S}^{(0)}_{\epsilon}$ (resp. $\widetilde {\cal S}^{(0)}_{\epsilon}$) 
satisfy the following commutation relations with the total fermionic number $\widehat N_F$ 
\begin{equation}\label{NScom}
[\widehat N_F , {\cal S}^{(0)}_{\epsilon}] = -{\cal S}^{(0)}_{\epsilon} \qquad , \quad [\widehat N_F , \widetilde {\cal S}^{(0)}_{\epsilon} ] = \widetilde {\cal S}^{(0)}_{\epsilon} 
\end{equation}
Moreover, apart for their $D_{\epsilon}^{klm}$ contribution, the various terms in ${\cal S}^{(0)}_{\epsilon}$ (resp. $\widetilde {\cal S}^{(0)}_{\epsilon}$) 
act separately on each $\epsilon$-species of fermions. 

As as been previously noticed \cite{D'Eath:1993up,Csordas:1995kd,Csordas:1995qy,Cheng:1996an}, the fact that the ${\cal S}^{(0)}_{\epsilon}$ and $\widetilde {\cal S}^{(0)}_{\epsilon}$ change the fermionic number by one unit, 
allow one to look for solutions of the supersymmetry constraints at each fixed total fermion level $N_F$.
A simple proof of this fact reads as follows. The commutation relations (\ref{NScom}) show that if  $\bm \Psi$
is a solution of  ${\cal S}^{(0)}_{\epsilon} \bm \Psi =0 $ and $\widetilde {\cal S}^{(0)}_{\epsilon} \bm \Psi =0$, then
$\widehat N_F\, \bm \Psi$ is also a solution. By iterating the action of  $\widehat N_F $,  $\widehat N_F^n \, \bm \Psi$
will be a solution for any integer $n$. If we then decompose $\bm \Psi$ in $N_F$ levels, {\it i.e.}   
 $\bm \Psi=\sum_{N_F}{{\Psi}}_{N_F}$, we see that, for any $n$
$$
\widehat N_F^n \, \bm \Psi=\sum_{N_F}  {N_F}^n \,\ {{\Psi}}_{N_F} \, ,
$$
will be a solution.  Because of the non vanishing of  a corresponding Vandermonde determinant, we see that each
separate state $\Psi_{N_F}$ must be a solution.
This remark facilitates the study of the solution space. It is enough to look for solutions of the supersymmetry constraints having a fixed fermion level $N_F$.

In addition, though the $ {\cal S}^{(0)}_{\epsilon}$and $ \widetilde{\cal S}^{(0)}_{\epsilon}$ operators do not commute with the separate fermionic numbers $N^F_\pm$,   these operators and the parity indicators $(-)^{N^F_{\pm}}$ are found
to verify the relations
\begin{equation}\label{SEIpiNP}
\{{\cal S}^{(0)}_{\pm},(-)^{N^F_{\pm}}\}=\{\widetilde{\cal S}^{(0)}_{\pm},(-)^{N^F_{\pm}}\}=[{\cal S}^{(0)}_{\pm},(-)^{N^F_{\mp}}]=[\widetilde{\cal S}^{(0)}_{\pm},(-)^{N^F_{\mp}}]=0\qquad .
 \end{equation}
Accordingly, at a given level, decomposing $\Psi_{N^F_+ + N^F_-}=\sum_{p}\Psi_{(N^F_+-p,N^F_-+p)}$ (setting to zero components with negative $N^F_\pm$ index, or index greater than 3), we obtain that 
\begin{eqnarray}
&&{\cal S}^{(0)}_{\pm}\sum_{p=0}^3\Psi_{(N^F_+-p,N^F_-+p)}=0\quad \Rightarrow \quad{\cal S}^{(0)}_{\pm}\sum_{p} (-)^p\Psi_{(N^F_+-p,N^F_-+p)}=0\nonumber \\ \,
&&\widetilde{\cal S}^{(0)}_{\pm}\sum_{p}\Psi_{(N^F_+-p,N^F_-+p)}=0\quad \Rightarrow \quad \widetilde{\cal S}^{(0)}_{\pm}\sum_{p}(-)^p\Psi_{(N^F_+-p,N^F_-+p)}=0\nonumber
\end{eqnarray}
As a consequence if $\Psi_{N^F_++N^F_-}=\sum_{p}\Psi_{(N^F_+-p,N^F_-+p)}$ is a solution of the four supersymmetry constraints equations, so are the partial sums $\sum'_{p}\Psi_{(N^F_+-p,N^F_-+p)}$ where $p$ is restricted
to even or odd values. 
In other words we may, without loss of generality, look for solutions in the subspaces (when considering $N_F \leq3$):
$$ {\mathbb H}_{(0,0)},\  {\mathbb H}_{(1 , 0)},\  {\mathbb H}_{(0 , 1)},\   {\mathbb H}_{(2 ,0)} \oplus  {\mathbb H}_{(0 ,2)}  ,  {\mathbb H}_{(1 ,1)},\ {\mathbb H}_{(3 ,0)} \oplus  {\mathbb H}_{(1 ,2)}  ,\ {\mathbb H}_{(0 ,3)} \oplus  {\mathbb H}_{(2 ,1)} .$$
In addition solutions belonging to the subspace ${\mathbb H}_{(1 ,1)}$ may be decomposed into symmetric and antisymmetric ones. To summarise we obtain 8 different classes of  possible solutions. We shall consider them in turn.

It is to be noted that, when looking for a solution at some fixed fermionic level $N_F=N$, say
\begin{equation}
\Psi_{(N)} =   f^{\epsilon_1 \epsilon_2 \cdots \epsilon_N}_{a_1 a_2 \cdots a_N} (\beta) \widetilde b^{a_1}_{\epsilon_1} \cdots    \widetilde b^{a_N}_{\epsilon_N}  {\vert\, 0\,\rangle\hspace{-0.5mm}\strut_-}
\end{equation}
the components $ f^{\epsilon_1 \epsilon_2 \cdots \epsilon_N}_{a_1 a_2 \cdots a_N} (\beta)$ of the wave function satisfy
[because of (\ref{SAEq})] a set of partial differential equations  whose explicit expression is equivalent to
\begin{equation}-i {\cal S}^{(0)}_\epsilon \Psi_{(N)} =0 , \,  -i \widetilde {\cal S}^{(0)}_\epsilon \Psi_{(N)} =0 
\end{equation}
We have written these equations with an extra factor $-i$, so that, in view of the explicit expressions of the
chiral  ${\cal S}^{(0)}_\epsilon$'s given above, {\it all} the coefficients appearing in these equations become {\it real}.
In addition, as the commutation relations of the $b$'s and $\widetilde b$'s  are also real, we see that the set of 
partial differential equations satisfied by the wave-function components   $ f^{\epsilon_1 \epsilon_2 \cdots \epsilon_N}_{a_1 a_2 \cdots a_N} (\beta)$ will be real. One can therefore construct a basis of solutions of the set of supersymmetric solutions at level $N_F$ made of {\it real} wavefunctions   $ f^{\epsilon_1 \epsilon_2 \cdots \epsilon_N}_{a_1 a_2 \cdots a_N} (\beta)$.

\bigskip

\setcounter{equation}{0}\section{Up-down symmetry in fermionic space}

Before discussing explicit solutions in detail, let us note in what sense there is a symmetry between the lower 
($N_F \leq3$) and the upper ($N_F \geq3$) parts of fermionic space. At the kinematical level, there is, as we have seen
above, the usual symmetry in the Fock construction of the state space, under which 
$$
 {{\vert\, 0\,\rangle\hspace{-0.5mm}\strut_-}}   \to \  {{\vert\, 0\,\rangle\hspace{-0.5mm}\strut_+}} 
 $$
 and
 $$
 b_\epsilon^a  \to \  \widetilde  b_\epsilon^a
 $$
But the issue is to know whether this kinematical symmetry extends to the dynamics, {\it i.e.}  whether there is a one-to-one
map between {\it solutions} of the supersymmetry constraints at the levels $N_F$ and $6-N_F$. A (positive) answer
to this question is obtained by first recalling that the difference between  $ {{\vert\, 0\,\rangle\hspace{-0.5mm}\strut_-}}$  ,
$ b_\epsilon^a$ and   $ {{\vert\, 0\,\rangle\hspace{-0.5mm}\strut_+}}$  ,
$\widetilde b_\epsilon^a$ is connected to a choice in the chiral projection
$$
 b_\epsilon^a \propto ( 1 - i \, \gamma^5) \widehat \Phi^a 
$$
versus
$$
 \widetilde b_\epsilon^a \propto ( 1 + i \, \gamma^5) \widehat \Phi^a 
$$
We need therefore to see whether there is a symmetry of the constraint equations (\ref{SAEq})  which involves a flip
in the sign of $\gamma^5 = \gamma^{\hat 0} \gamma^{\hat 1} \gamma^{\hat 2} \gamma^{\hat 3}$.  We note that  the appearance
of the $\gamma$ matrices in $\widehat {\cal S}^{(0)}_A(\beta, \Phi)$  has a special structure. In particular, after
the choice of a Majorana representation with $\beta= \mathcal C = i \gamma_{\hat 0}$, all the cubic terms in   $\widehat {\cal S}^{(0)}_A(\beta, \Phi)$ involve only the spatial gamma matrices $\gamma^{\hat a}$. As a consequence, $ \gamma^{\hat 0}$ only appears in
the gravitational-wall term
\begin{equation}
\label{SgAbis}
 {\cal S}_A^g = \frac12 \sum_k e^{-2\beta^k} \left( \gamma^{\hat 0} \gamma^{\hat 1} \gamma^{\hat 2} \gamma^{\hat 3} \, \Phi^k\right)_A\qquad .
\end{equation}
Given an initial Majorana representation for  $(\gamma^{\hat 0},  \gamma^{\hat 1}, \gamma^{\hat 2}, \gamma^{\hat 3})$, the new matrices
$(\gamma^{\prime\hat 0},  \gamma^{\prime\hat 1}, \gamma^{\prime\hat 2}, \gamma^{\prime\hat 3}) = (-\gamma^{\hat 0},  \gamma^1, \gamma^{\hat 2}, \gamma^{\hat 3})$, form a second
Majorana representation (which differs by a conjugation with $ \gamma^{\hat 1} \gamma^{\hat 2} \gamma^{\hat 3}$). This change of
representation will leave the expressions of the   $\widehat {\cal S}^{(0)}_A(\beta, \Phi)$ invariant if we additionally
perform the following complex shift of the $\beta$ variables:
\begin{equation}
\label{betashift}
\beta^a  \to \ \beta^a + i  \frac{\pi}{2}
\end{equation}
Indeed, this shift changes the sign of the gravitational potentials $e^{-2\beta^a} $, while leaving invariant
all the terms related to the symmetry walls (which are $ \propto \coth( \beta^a - \beta^b)$. In terms of the variables
$ x=e^{2\beta^1}=1/a^2, y=e^{2\beta^2}=1/b^2, z=e^{2\beta^3}=1/c^2$, the above complex shift of the $\beta$'s means
\begin{equation} \label{xshift}
x \to -x, \, y \to -y, \, z \to -z .
\end{equation}
Summarizing:  the usual (kinematical) up-down fermionic symmetry (mapping $N_F$ to $6-N_F$)
extends to the dynamical level ({\it i.e.}  maps a solution on a solution), at the cost,
however, of  the change (\ref{betashift}), {\it i.e.}   (\ref{xshift}), of the bosonic coordinates.
We note in passing that, when $N_F=3$,  we have a map between solutions at the same level.


\setcounter{equation}{0}\section{ Solutions at the fermionic level $N_F = 0$}

\smallskip

It is particularly easy to obtain the general solution at this level. The subspace is one-dimensional; thus any putative
solution must be described by a single (scalar) amplitude $f(\beta)$ with
\begin{equation}
{{\Psi}}_{(0)} = f(\beta) \, {{\vert\, 0\,\rangle\hspace{-0.5mm}\strut_-}} \qquad .
\end{equation}
As there is no subspace of level $N_F = -1$, the susy constraints of the annihilation-type ($ b + \widetilde b bb$)
are identically satisfied: 
\begin{equation}
{\cal S}^{(0)}_{\epsilon} \,\Psi_{(0)} \equiv 0 \qquad .
\end{equation}
On the other hand the conditions linked to the creation-type susy constraints ($ \widetilde b + \widetilde b \widetilde b b$)
\begin{equation}
\widetilde {\cal S}^{(0)}_+ \, {{\Psi}}_{(0)} = 0 \qquad \mbox{and} \qquad \widetilde {\cal S}^{(0)}_- \, {{\Psi}}_{(0)} = 0 
\end{equation}
lead to twice the same three equations : 
\begin{equation}
\frac i2 \, \partial_k \, f - (\alpha_k + \mu_k + \rho^{(1)}_k) \, f = 0 .
\end{equation}
Eqs (\ref{alphak}), (\ref{mukpot}) and (\ref{rho1}) showed that  each of the factors $\alpha_k$,  $ \mu_k$, and $ \rho^{(1)}_k$ 
is $i$ times the  gradient of a real function.  Therefore the equations for $f(\beta)$  are (locally) trivially integrable. The general $N_F = 0$ solution is  then found to be of the form: 
\begin{equation}
\label{Sol00}
f=C_{(0)} \left[ (y-x) (z-x) (z-y) \right]^{3/8} \, \frac{e^{- \frac12 \left( \frac1x + \frac1y + \frac1z \right)}}{(x \, y \, z)^{5/4}} \quad ,
\end{equation}
In terms of the $\beta$'s it reads
\begin{equation}
f \propto \textstyle \exp \left( - \frac74 \, \beta^0 \right) (\sinh \beta^{12} \sinh \beta^{23} \sinh \beta^{31})^{3/8} 
\textstyle\exp \left( -\frac12 \sum_a \exp (-2 \, \beta^a) \right) 
\end{equation}
where $\beta^0 \equiv \beta^1 + \beta^2 + \beta^3$, $\beta^{12} \equiv  \beta^1 - \beta^2$, etc.

This solution, which depends on a single multiplicative constant, deserves some comments.  First, if  $C_{(0)} $ is taken
to be real, the solution is real. More precisely, we have written it so that it is real in our canonical Weyl chamber (a)
where $ x \leq y \leq z$.  As was argued above, it is natural to interpret the symmetry of supergravity under large
diffeomorphisms as implying that we can restrict the moduli space ({\it i.e.}  the space of the $\beta$'s) to only one Weyl
chamber. With this interpretation, the expression (\ref{Sol00}), considered only for  $ x \leq y \leq z$, would be a full
description of the $N_F =0$ solution space.  If, on the other hand, one wanted to extend the wave function
to the six different Weyl chambers (represented in  Fig. \ref{WeylChfig}, it might be natural to continue it analytically
by passing through the successive symmetry walls where either $x=y$, $y=z$ or $z=x$. This would lead to a global
wave function of the form
\begin{equation}
\label{SolG00}
f(x,y,z)=C_{(0)} \left[ \vert (x-y) (y-z) (z-x) \vert \, e^{in\pi} \right]^{3/8} \, \frac{e^{- \frac12 \left( \frac1x + \frac1y + \frac1z \right)}}{(x \, y \, z)^{5/4}} 
\end{equation}
where the index $n$ counts modulo $6$ the number of symmetry walls crossed when turning around the $\beta^1 = \beta^2 = \beta^3$ axis (see Fig. \ref{WeylChfig}).

Independently of the way we wish to view this solution, let us note that it {\it vanishes} on the symmetry walls, 
and {\it decays} under the gravitational walls , {\it i.e.}  when  $\beta^a \to - \infty$, for each given index $a$. We recall that 
$e^{-2 \beta^1}=a^2=1/x$, so that going under the $2 \beta^1$ gravitational wall ($e^{-2 \beta^1} \to + \infty$)
means $a^2 \to + \infty$ or $ x \to  0^+$.  The exponential factor by which the $N_F=0$ solution decays under
the gravitational walls ({\it i.e.}  for large, anisotropic Universes) is
\begin{equation} \label{exp-}
e^{ - \frac12 ( a^2 + b^2 + c^2)} \equiv e^{- \frac12 \left( \frac1x + \frac1y + \frac1z \right)}
\end{equation}
Ground-state solutions, incorporating such a (real)  exponential factor, of either the ordinary bosonic 
Bianchi IX WDW equation \cite{Moncrief:1991je}, or 
its supersymmetric extension \cite{D'Eath:1993up,Csordas:1995qy,Cheng:1996an},  have been discussed in previous works. However, our new (unique) ground-state  solution further incorporates the non-trivial extra factor
\begin{equation}\frac{ \left[ (y-x) (z-x) (z-y) \right]^{3/8}} {(x \, y \, z)^{5/4}} \propto  \exp \left( - \frac74 \, \beta^0 \right) (\sinh \beta^{12} \sinh \beta^{23} \sinh \beta^{31})^{3/8} 
 \end{equation}
 which necessary follows from the presence of the symmetry-wall contributions (\ref{SsA}) in the susy constraints.

\bigskip

\setcounter{equation}{0}\section{Solutions at the level $N_F = 6$}

\smallskip

The subspace ${\mathbb H}_{(3,3)}$  is also one dimensional: 
\begin{equation}
{{\Psi}}_{(6)} = \tilde f (\beta) \,  {{\vert\, 0\,\rangle\hspace{-0.5mm}\strut_+}}
\end{equation}
where $ {{\vert\, 0\,\rangle\hspace{-0.5mm}\strut_+}}$ is annihilated by all the $\tilde b_{\epsilon}^k$ operators, defined in Eq. (\ref{Psi64}).  When imposing the susy constraints  Eqs. (\ref{SAEq}) (in chiral form), the creation-type constraints
\begin{equation}
\widetilde {\cal S}^{(0)}_{\epsilon} \, {{\Psi}}_{(6)} \equiv 0
\end{equation}
are identically satisfied, while the annihilation-type ones
\begin{equation}
{\cal S}^{(0)}_{\epsilon} \, {{\Psi}}_{(6)} = 0
\end{equation}
yield twice the equations
\begin{equation}
\frac i2 \, \partial_k \, \tilde f + (\alpha_k - \mu_k - \rho^{(1)}_k) \, \tilde f = 0\qquad .
\end{equation}
As in the $N_F=0$ case, the (imaginary) gradient nature of the vectors $\alpha_k $, $ \mu_k $, $\rho^{(1)}_k$
implies the existence of a unique solution (modulo an arbitrary  multiplicative factor $C_{(6)}$)
\begin{equation}\label{Sol33}
\tilde f = C_{(6)} \left[ (y-x) (z-x) (z-y) \right]^{3/8} \, \frac{e^{+\frac12 \left( \frac1x + \frac1y + \frac1z \right)}}{(x \, y \, z)^{5/4}} \qquad .
\end{equation}
Here, we have an explicit example of the general property we explained above. One maps a solution at level
$N_F$ to a solution at level $6-N_F$ by exchanging $b \to \widetilde b$, $ {\vert\, 0\,\rangle\hspace{-0.5mm}\strut_-} \to {\vert\, 0\,\rangle\hspace{-0.5mm}\strut_+}$ and $ (x,y,z) \to (-x,-y,-z)$.
[Here, we need to absorb a phase factor $\exp (i \pi) ^{(3/8 - 5/4)}$ in the multiplicative constants.]

Note that the transformation rule  $ (x,y,z) \to (-x,-y,-z)$ (which was seen above to be connected with the
nature of the gravitational-wall contributions (\ref{SgA}) to the susy constraints, and especially their proportionality
to $\gamma^5 \Phi^a$)  ``explains'' why the (unique)  $N_F = 6$ solution grows exponentially  under the gravitational walls,
proportionally to
\begin{equation} \label{exp+}
e^{ + \frac12 ( a^2 + b^2 + c^2)} \equiv e^{+ \frac12 \left( \frac1x + \frac1y + \frac1z \right)}
\end{equation}
while the  $N_F = 0$ solution was exponentially decaying under the gravitational walls.

Though we are not   {\it a priori}   sure of what kind of physical requirements should be imposed on the
wave function of the Universe, we shall tentatively assume in the following that one should only
retain wavefunctions that do not exhibit a growth for large values of $a^2, b^2, c^2$ as violent as  Eq. (\ref{exp+}).

\bigskip

\setcounter{equation}{0}\section{ Solutions at the level $N_F=1$}

\smallskip

A general ${\bm\Psi}$ in ${\mathbb H}_{[1]} ={\mathbb H}_{(1,0)} \oplus {\mathbb H}_{(0,1)}$ is given by a superposition : 
\begin{equation}
{{\Psi}}_{(1)} = \sum_{\epsilon = \pm} f_k^{\epsilon} \, \tilde b_{\epsilon}^k \, {{\vert\, 0\,\rangle\hspace{-0.5mm}\strut_-}} \qquad .
\end{equation}
The ${\cal S}^{(0)}_{\epsilon}$ operators project ${{\Psi}}_{(1)}$ onto ${\mathbb H}_{(0,0)}$. The image of this projection vanishes if the divergence conditions  :
\begin{equation}
\label{eq100}
\frac i2 \, \partial_k \, f_{\epsilon}^k + \varphi_k \, f_{\epsilon}^k = 0
\end{equation}
are satisfied. Here $f_{\epsilon}^k := G^{kl} \, f_l^{\epsilon}$ and $\varphi_k$ is defined by
\begin{equation}\label{phik}
\varphi_k  : = \alpha_k + \mu_k + \rho_k^{(1)}= \frac i2 \, \partial_k \, \varphi\qquad ,
\end{equation}
where $\varphi$ is defined as the logarithm of the $N_F=0$ solution $f$, Eq. (\ref{Sol00}) (with $C_{(0)}=1$).
 In what follows $\beta$-indices are raised or lowered with the metric (\ref{Gab}); the positions of the $N^F_\pm$ indices
 ($\epsilon, \epsilon', \ldots = \pm$)  are indifferent, and will be dictated by writing facilities. 
The $\widetilde {\cal S}^{(0)}_{\epsilon}$ operators lead to two (similar) sets of three equations. Indeed $\tilde S$ maps ${\mathbb H}_{(1,0)}$ (and ${\mathbb H}_{(0,1)}$) into ${\mathbb H}_{(2,0)}$, ${\mathbb H}_{(1,1)}$ and ${\mathbb H}_{(0,2)}$. Explicitly we obtain
\begin{equation}
\label{eq102}
\nu_{[k} \, f_{l]}^{\epsilon} = 0
\end{equation}
\begin{equation}
\label{eq111}
\frac i2 \, \partial_k \, f_l^{\epsilon} - \varphi_k \, f_l^{\epsilon} + 2 \, \rho_{kl}^m \, f_m^{\epsilon} = 0
\end{equation}
\begin{equation}
\label{eq120}
\frac i2 \, \partial_{[k} \, f_{l]}^{\epsilon} - \varphi_{[k} \, f_{l]}^{\epsilon} + \mu_{[k} \, f_{l]}^{\epsilon} = 0 \qquad .
\end{equation}
We see explicitly here the consequence of the commutation relations Eq. (\ref{SEIpiNP}) that was anticipated above:
because of  parity properties at the $N_F=1$ level, there is a complete decoupling of the
modes of different partial fermionic number  $N^F_+$,   $N^F_-$.  

It is not   {\it a priori}   clear that the overconstrained set of  equations (\ref{eq100}) -- (\ref{eq120}) admit any non-zero solutions. 
Because we have shown above that our way of quantizing supergravity led to a consistent algebra of constraints, we,
however, expect that the structure of the above equations will be special enough to admit non trivial solutions. 
We have explicitly verified this for all the levels that will be discussed here in full detail. 

In the present  $N_F=1$ case, the use of  the algebraic constraint Eq. (\ref{eq102}) immediately reduces the 
degrees of freedom of the ``vectorial'' wavefunctions $ f_k^{\pm}$ to scalar ones : 
\begin{equation}
f_k^{\pm} = f^{\pm} \nu_k
\end{equation}
Inserting this factorized form in the remaining equations (\ref{eq111}), (\ref{eq120})  leads to three integrable equations (plus some identities). The general solution at level $N_F =1$ is then found to be 
\begin{equation}\label{Sol10}
f_k^{\pm} = C_{(1)}^{\pm} \{ x(y-z) , y(z-x), z(x-y)\} \, \frac{e^{-\frac12 \left( \frac1x + \frac1y + \frac1z \right)}}{(x \, y \, z)^{3/4} ((x-y)(y-z)(z-x))^{3/8}} 
\end{equation}
where $C_{(1)}^{\pm}$ are two arbitrary constants.  Each constant parametrizes the unique solution having either
$N^F_+ = 1$ or  $N^F_- = 1$.

Note  that each one of the basic solutions (which have the same amplitude $f_k$, but correspond to different
quantum states) can be taken as being real. Like at level $N_F=0$ the solutions
decay exponentially under the gravitational wall, with the same (WKB) exponential decay  (\ref{exp-}).
By contrast to the  $N_F=0$ case where the solution vanished on the symmetry walls, these $N_F=1$
 solutions become singular on the symmetry walls, but in a rather mild (square-integrable) way. [More about this below.]
 Let us finally remark that all previous work on supersymmetric Bianchi IX (and other minisuperspace) models
 \cite{D'Eath:1993up, D'Eath:1993ki,  Csordas:1995kd,Csordas:1995qy, Graham:1995ni,Cheng:1994sr, Cheng:1996an,Obregon:1998hb,D'Eath:1996at}  
 have stated that it was impossible to construct solutions of the susy constraints at odd fermion levels. This difference might
 be due to a difference in the quantization scheme used. However, we rather think that it is due to the fact that
 previous work considered a too restrictive
 class of ans\"atze when trying to construct putative odd-level states. In our construction, the odd fermion levels
 do not introduce any special difficulty.

\bigskip

\setcounter{equation}{0}\section{ Solutions at level $N_F=5$}

\smallskip

The general solution at this level can be either built in analogy with the one just obtained, or simply by using
the $N_F \to 6-N_F$ rules given above. We have checked that this yields the same solutions.
Writing $\bm\Psi$ as :
\begin{equation}
{{\Psi}}_{(5)} = \sum_{\epsilon=\pm} f_k^{\epsilon} \, b_{\epsilon}^k \,  {{\vert\, 0\,\rangle\hspace{-0.5mm}\strut_+}}
\end{equation}
we obtain  (consistently with changing $ x \to -x$ etc in the $N_F=1$ solutions), 
\begin{equation}\label{Sol32}
f_k^{\pm} = C_{(5)}^{\pm} \{ x(y-z) , y(z-x), z(x-y)\} \, \frac{e^{+\frac12 \left( \frac1x + \frac1y + \frac1z \right)}}{(x \, y \, z)^{3/4} ((x-y)(y-z)(z-x))^{3/8}} 
\end{equation}
depending on the two constants $C_{(5)}^{\pm}$ parametrizing the separate unique states with $N^F_{\pm} =5$. 

Like the solutions at level $N_F=6$, these solutions grow exponentially under the gravitational walls. 
We shall therefore tentatively reject them.

\bigskip

\setcounter{equation}{0}\section{ Solutions at level $N_F=2$}

\smallskip
So far, {\it i.e.}  for $N_F=0,1,5,6$,  the solutions we obtained,  which were the most general
at these levels, only consisted of ``discrete solutions'', 
 containing  arbitrary multiplicative factors, but having
 fixed shapes as   functions of the $\beta$'s. The situation will change in the middle of the fermionic Fock space,
{\it i.e.}  for $N_F= 2,3,4$, where we will find solutions depending also on arbitrary ``initial'' functional data. Our findings are qualitatively consistent with the finding of Refs. \cite{Csordas:1995kd,Csordas:1995qy}
that there exist supersymmetric Bianchi IX solutions at fermion levels 2 and 4 depending on as many data as a solution
of the usual bosonic WDW equation. However, as we shall comment below, our results differ also significantly 
(both qualitatively and quantitatively) from previous results. Most notably, we shall construct ``continuous'' solutions at the odd fermionic level $N_F=3$, which
was considered as being impossible in previous works.

We study the solution space at level  $N_F=2$ by extending the procedure used at lower levels.  The dimension of 
${\mathbb H}_{[2]}$ is 15, so that we are   {\it a priori}   dealing with a fifteen-component wave function, say
\begin{equation}
\label{Psi2}
{{\Psi}}_{(2)} = \frac12 \sum_{  \epsilon , \epsilon' = \pm \atop k,k' = 1,\,2 ,\, 3} f_{kk'}^{\epsilon\epsilon'}(\beta)  \tilde b_{\epsilon}^k \, \tilde b_{\epsilon'}^{k'} \, {{\vert\, 0\,\rangle\hspace{-0.5mm}\strut_-}}
\end{equation}
The  wave function $ f_{kk'}^{\epsilon\epsilon'}(\beta) $ must verify the symmetry relation: 
\begin{equation}
\label{f2sym}
f_{kk'}^{\epsilon\epsilon'} = -f_{k'k}^{\epsilon'\epsilon} \qquad .
\end{equation}
which indeed implies that it contains fifteen independent components.  Note in passing that while (\ref{f2sym}) imposes
an antisymmetry on the $ k, k'$ ``tensorial''  indices when $\epsilon = \epsilon'$, it does not restrict the tensorial 
symmetry of the wave function in the opposite case where  $\epsilon \neq \epsilon'$. In the latter case, it only says that
$f_{pq}^{+ -}$ and $  f_{pq}^{- +}$ are not independent ($f_{pq}^{+ -} \equiv - f_{qp}^{- +}$).

By projecting the equations ${\cal S}^{(0)}_{\epsilon} \, {\Psi_{(2)}} = 0$ on the subspace of level $N_F=1$, we obtain two sets of equations
\begin{equation}
\label{eq21pm}
\frac i2\partial_{\beta^k}G^{kp}f^{\epsilon,-\epsilon}_{p\,n}+ \varphi^k\,f^{\epsilon,-\epsilon}_{k\,n}-  2\,\rho^{k\,l}_{\phantom{k\,l}n}f^{\epsilon,-\epsilon}_{k\,l}=0\qquad ,
\end{equation}
and   
\begin{equation}
\label{eq21pp}
\frac i2\partial_{\beta^k}G^{kp}f^{\epsilon,\epsilon}_{p\,n}+(\varphi^k-\mu^k-\nu^k)\,f^{\epsilon,\epsilon}_{k\,n} =0\qquad .
\end{equation}
The projection on the level $N_F = 3$ of the equations $\widetilde {\cal S}^{(0)}_{\epsilon} \, {\bm \Psi} = 0$ leads to four 
additional sets of equations
\begin{equation}
\label{eq23pmnu}
\nu_{[p} \, f_{qr]}^{\epsilon,-\epsilon} = 0\qquad ,
\end{equation}
\begin{equation}
\label{eq23pm}
\frac i2 \, \partial_{[p} \, f_{q]r}^{\epsilon,-\epsilon} - \varphi_{[p} \, f_{q]r}^{\epsilon,-\epsilon} + \mu_{[p} \, f_{q]r}^{\epsilon,-\epsilon} + 2  \, \rho_{[p \vert r \vert} \, ^sf_{q]s}^{\epsilon,-\epsilon} = 0\qquad,
\end{equation}
\begin{equation}
\label{eq23pp}
\frac i2 \, \partial_{p} \, f_{qr}^{\epsilon,\epsilon} - \varphi_{p} \, f_{qr}^{\epsilon,\epsilon} - 4   \, \rho_{p[q} \, ^sf_{r]s}^{\epsilon,\epsilon} + 2 \, \nu_{[q} \, f_{r]p}^{-\epsilon,-\epsilon} = 0\qquad,
\end{equation}
\begin{equation}
\label{eq23ppa}
\frac i2 \, \partial_{[p} \, f_{qr]}^{\epsilon,\epsilon} - \varphi_{[p} \, f_{qr]}^{\epsilon,\epsilon} + 2 \, \mu_{[p} \, f_{qr]}^{\epsilon,\epsilon} = 0 \qquad .
\end{equation}

It is not   {\it a priori}   evident how to deal with this complicated, redundant set of (partial differential, and algebraic) equations.
A first simplification comes from the fact (mentioned above) that, under the decomposition  Eq. (\ref{decompH2}) of
${\mathbb H}_{[2]}$ into its  $(N^F_+, N^F_-)$ subspaces, there should be a decoupling  between
 ${\mathbb H}_{(2,0)} \oplus \, {\mathbb H}_{(2,0)} $  and ${\mathbb H}_{(1,1)}$.  In terms of the components 
 $f_{pq}^{\epsilon,\epsilon'}$ this means a decoupling between ( $f_{pq}^{+ +}$, $f_{pq}^{- -}$) on one side,
 and  $f_{pq}^{+ -} \equiv - f_{qp}^{- +}$ on the other side. And, indeed one easily sees that
 Eqs. (\ref{eq21pp}, \ref{eq23pp}, \ref{eq23ppa}) contain only  the $f_{pq}^{\epsilon,\epsilon}$ components,
 while Eqs. (\ref{eq21pm}, \ref{eq23pm}, \ref{eq23pmnu}) involve only the $f_{pq}^{\epsilon,-\epsilon}$ components.
 Actually, there is even a further simplification, in that, among the  $f_{pq}^{+ -} $ components (parametrizing 
 ${\mathbb H}_{(1,1)}$) the three components   $f_{[pq]}^{+ -} $ components (parametrizing 
 ${\mathbb H}_{(1,1)_A}$) decouple from  the six $f_{(pq)}^{+ -} $ components (parametrizing 
 ${\mathbb H}_{(1,1)_S}$).
 
 Summarizing: one can {\it separately} look for solutions in the subspaces  
 $\left[{\mathbb H}_{(2,0)} \oplus \, {\mathbb H}_{(2,0)} \right]_{3+3} $,
  $\left[ {\mathbb H}_{(1,1)_A} \right]_3$ and  $\left[ {\mathbb H}_{(1,1)_S} \right]_6$, where the subscripts indicate the
  dimensions ({\it i.e.}  the number of components of the wave function).  Let us also recall that all the equations we are
  dealing are {\it real} after multiplying them by a common $i$. We can therefore look for real solutions in each subspace
  (even, if we later build general complex combinations of basic solutions). In the following we consider in turn each one
  of the above separated problems.

\smallskip
\subsection{ Level $N_F=2$: solutions in the  ${\mathbb H}_{(2,0)} \oplus {\mathbb H}_{(0,2)}$ subspace}
\smallskip

By subtracting the trace of Eq. (\ref{eq23pp}) from Eq. (\ref{eq21pp}) we obtain an extra algebraic equation that can be written as
\begin{equation}
\label{Algf2pp}
\left( 2 \, \alpha^k + \mu^k + \rho^{(1)k} - \rho^{(3)k} \right) f_{kl}^{\epsilon\epsilon} = 0\qquad .
\end{equation}
As $f_{kl}^{\epsilon\epsilon}=f_{[kl]}^{\epsilon\epsilon}$, its explicit solution is immediate. It is given by 
($\varepsilon_{klp} = \varepsilon_{[klp]}$ with $\varepsilon_{123}= +1$)
\begin{equation}
\label{f2pp}
f_{kl}^{\epsilon\epsilon} = f^{\epsilon\epsilon} \, \varepsilon_{klp} \left( 2 \, \alpha^p + \mu^p + \rho^{(1)p} - \rho^{(3)p} \right) \qquad .
\end{equation}
Inserting these components in Eqs. (\ref{eq23pp}) we obtain six coupled equations, one for each partial derivative of the two unknown functions $f^{++}$ and $f^{--}$. These equations are integrable and provide the general expression of the solution of the Eqs. (\ref{SAEq}) restricted to the subspace ${\mathbb H}_{(2,0)} \oplus {\mathbb H}_{(0,2)}$.
This general solution depends on two arbitrary constants and can be explicitly written as: 
{\small \begin{equation}
\label{f2ppa}
\left\{ f_{12}^{\epsilon\epsilon} , f_{23}^{\epsilon\epsilon} , f_{31}^{\epsilon\epsilon} \right\} = \biggl\{ x(y-z) - yz + \frac{xyz}2 , \ y(z-x) - zx + \frac{xyz}2 ,\ z(x-y) - xy + \frac{xyz}2 \biggl\}   f^{\epsilon\epsilon}
\end{equation}}
where the two independent scalar  functions $f^{\epsilon\epsilon}= (f^{++}, f^{--})$ are given by
\begin{eqnarray}
\label{pre2ppa}
f^{\epsilon\epsilon} &= &e^{-\frac12 \left(\frac1x + \frac1y + \frac1z\right)} (xyz)^{-3/4} \, \left[(x-y)(y-z)(x-z)\right]^{-1/8} \nonumber \\
&& \left[C_1 (x-z)^{-1/2} + \epsilon \, C_2 (y-z)^{-1/2}\right] 
\end{eqnarray}
with two arbitrary constants $C_1$ and $ C_2$. Note that both constants appear in  $f^{++}$ and $f^{--}$,
though in a different way [because of the sign $\epsilon$ in front of $C_2$ in Eq. (\ref{pre2ppa})].

\bigskip

\subsection{ Level $N_F=2$: solutions in the  ${\mathbb H}_{(1,1)_A}$ subspace}

\smallskip

Solutions living in ${\mathbb H}_{(1,1)_A}$ are similar to the ones just discussed, and are even easier to obtain. They   {\it a priori}   involve three arbitrary components, say
\begin{equation}\label{FA2}
{{\Psi}}^A_{(2 )}= \frac12 \, f_{[pq]}^{+-} \, \tilde b_+^p \, \tilde b_-^q \,  {\vert\, 0\,\rangle\hspace{-0.5mm}\strut_-} \qquad .
\end{equation}
From the general equations at level $N_F=2$ above, one finds that the antisymmetric tensor $ f_{[pq]}^{+-}$
has  to satisfy two sets of algebraic equations (besides some differential equations). The first one is Eq. (\ref{eq23pmnu}), the second one, similar to Eq. (\ref{Algf2pp}), is : 
\begin{equation}
\left( 2 \, \alpha^k + \mu^k + \rho^{(1)k} - \rho^{(2)k} \right) f_{[kl]}^{+-} = 0 \, ,
\end{equation}
as results from the difference between Eq. (\ref{eq23pm}) evaluated with $\epsilon = +$ and $\epsilon = -$ (taking into account the symmetry relation (\ref{f2sym})). The linear system  constituted of these four equations is found to
be  of rank 2. Accordingly we conclude that the tensor  $ f_{[pq]}^{+-}$ is parametrized by a single
independent function: 
\begin{eqnarray}
\label{f2pma}
\left\{ f_{[12]}^{+-} , f_{[23]}^{+-} , f_{[31]}^{+-} \right\} &= & \biggl\{ x(y-z) - yz + \frac{xyz}2 , y(z-x) - zx + \frac{xyz}2 , \nonumber \\
&&z(x-y) - xy + \frac{xyz}2 \biggl\}  f^{+-}(x,y,z)  \qquad . 
\end{eqnarray}
The $\beta$-space dependence of the function $f^{+-}(x,y,z)$ is then determined by using the
differential Eq. (\ref{eq23pm}). The general solution of the latter differential equation reads
\begin{equation}\label{pre2fpm}
f^{+-} = C_3 \, e^{-\frac12 \left(\frac1x + \frac1y + \frac1z\right)} (xyz)^{-3/4} \left[(x-y)(y-z)(x-z)\right]^{-1/8} (x-y)^{-1/2} \qquad .
\end{equation}
The result (\ref{f2pma})  is found to also  satisfy Eq. (\ref{eq21pp}), for an arbitrary value of the  constant $C_3$.
It therefore describes the general solution within the  ${\mathbb H}_{(1,1)_A}$ subspace.

It is interesting to note that the three-dimensional set of solutions obtained by combining the solutions in the subspaces
 ${\mathbb H}_{(2,0)} \oplus {\mathbb H}_{(0,2)}$  and ${\mathbb H}_{(1,1)_A}$ have a precisely similar
 structure as functions of $x,y,z$. Actually, they define a three-dimensional representation of the permutation group
 of  the three variables  $x,y,z$.
\par\noindent Similarly to the solutions found at levels $N_F=0$ and $N_F=1$, 
all these solutions exponentially decay under the gravitational walls, with the basic WKB behavior (\ref{exp-}).
However,  contrary to what happened at lower $N_F$ levels, 
the solutions (\ref{f2ppa}, \ref{f2pma}) exhibit now a more singular ({\it non square-integrable})
behavior when they approach the symmetry walls, say $\sim (x-y)^{-5/8} \sim (\beta^1-\beta^2)^{-5/8}$.  We would tentatively conclude that such solutions cannot be physically retained.

\bigskip

\subsection{ Level $N_F=2$: solutions in the ${\mathbb H}_{(1,1)_S}$ subspace}

\smallskip
We now turn to the more involved, and physically richer, case of solutions belonging to the subspace ${\mathbb H}_{(1,1)_S}$. On the one hand, contrary to the previous cases, 
here we have to satisfy less (namely, eleven) equations  than the number (eighteen)
of partial derivatives $\partial_k \, f_{(pq)}^{+-}$ of the corresponding tensorial wave function.  On the other hand, 
we  have more differential equations to satisfy than the number (six) of unknowns: $11 > 6$. The number of
solutions of such an overconstrained system is   {\it a priori}   unclear, and depends on its precise structure.
We shall be able to give precise answers by mixing various approaches: (i) a precise  study of
the set of partial differential equations satisfied by the wave function; (ii) a detailed mathematical discussion of 
the corresponding ``initial value problem"; and (iii) complementary studies of the general solution of our system
in various asymptotic regimes.  

We are interested in states of the form
\begin{equation}\label{2S}
{{\Psi}}^S_{(1,1)} =  \, f_{(pq)}^{+-}(\beta)  \ \tilde b_+^p \, \tilde b_-^q \,  {\vert\, 0\,\rangle\hspace{-0.5mm}\strut_-} \qquad ,
\end{equation}
parametrized by a {\it symmetric} $\beta$-space tensorial wave function $ f_{(pq)}^{+-}(\beta)$. In the following,
we shall ease the notation by denoting the latter symmetric tensor as:
\begin{equation}\label{kpqdef}
k_{pq}  : = f_{(pq)}^{+-} \qquad .
\end{equation}
This tensor wave function has to satisfy Eqs. (\ref{eq21pm}) and (\ref{eq23pm}). By taking the difference of these equations for $\epsilon = +$ and $\epsilon = -$, we obtain the complete set of differential equations that $k_{pq}(\beta)$ has to satisfy:
\begin{equation}
\label{divk}
\frac i2 \, \partial^s \, k_{sq} + \varphi^s \, k_{sq} - 2 \, \rho^{rs} \, _q \, k_{rs} = 0\qquad ,
\end{equation}
\begin{equation}
\label{rotk}
\frac i2 \, \partial_{[p} \, k_{q]r} - \varphi_{[p} \, k_{q]r} + \mu_{pq} \, ^s \, k_{sr} + 2 \, \rho_{[p\vert r \vert} \, ^s \, k_{q]s} = 0 \qquad .
\end{equation}
These equations are similar to the Maxwell equations; the first one being of the ``div'' type and the second of the ``curl'' type. From a more formal point of view,  they generalize the PDE systems linked to the $\mathcal {N} =2$ supersymmetric
quantum mechanics of a particle in external potentials.  Witten \cite{Witten:1981nf,Witten:1982im} (see also \cite{Claudson:1984th}) has shown how such supersymmetric quantum mechanical systems yield generalizations
of  the De Rham-Hodge theory
of $p$-forms on manifolds, satisfying the first-order (div and curl) equations  $\delta \omega_p =0$ and $ d \omega_p=0$.
Our  supersymmetric Bianchi IX system can be viewed as a special  $\mathcal {N} =4$ (rather than $\mathcal {N} =2$)
supersymmetric quantum mechanical system. This explains why our $N^F_+=1$, $N^F_-=1$ equations (\ref{divk}),  (\ref{rotk}) generalize 
the  1-form $\delta \omega_1 =0$ and $ d \omega_1=0$ system. [Our symmetric wavefunction
$k_{pq}$ can be roughly viewed as being 
separately 1-form-like on each index.] This raises the issue of the analogues of the well-known compatibility condition
for De Rham-Hodge theory encoded in the Cartan identities $d^2 \equiv 0$, $\delta^2\equiv 0$. We expect to have similar
identities in our context, as a consequence of the basic identity (\ref{SSLSH}) that we have proven to hold within
our quantization scheme (and which generalizes the simpler identity   (\ref{SSLSHQM}) holding in ordinary
supersymmetric quantum mechanics). To display these identities, let us rewrite  the equations of our system  (\ref{divk}),  (\ref{rotk}) as
\begin{equation}
\label{Deq}
\mathcal E_p := \partial^s  k_{sp} - \Delta_p [x,y,z;k_{ab}] \qquad ,
\end{equation}
\begin{equation}
\label{Req}
\mathcal E_{rsp}  := \partial_r \, k_{sp} - \partial_s \, k_{rp} - R_{rsp} [x,y,z;k_{ab}] \qquad .
\end{equation}
We recall in passing that, in this form, all those equations have real coefficients.

We have explicitly checked that the system of equations (\ref{Deq}),  (\ref{Req}), satisfy a certain number of 
 Bianchi-like\footnote{Here, the name Bianchi alludes to the (contracted) Bianchi identities that underlie the consistency 
 of the Einstein equations, and is disconnected from the denomination ``Bianchi IX".} identities that guarantee their compatibility. The first such identity is an algebraic one. Indeed, 
 because, on the one hand, of the symmetry of $k_{ab}$, and, on the other hand, of the specific structure of the 
  $\mu_{klm}$ and $\rho_{klm}$ tensors [see Eqs. (\ref{alphak} -- \ref{rhoklm})], we have
  \begin{equation}
\varepsilon^{pqr} \mathcal E_{pqr}  \equiv 0 \qquad .
\end{equation}
It is because of this identity that we said above that our system contained 11 equations, rather than the $3+3 \times 3=12$ it seems
to contain.
 We have also checked that our equations verify identities of the form
\begin{equation}\label{Ep}
\varepsilon^{trs} \, \partial_t \, \mathcal E_{rsp}  = O(  \mathcal E_{abc},  \mathcal E_d) \qquad ,
\end{equation}
\begin{equation}\label{Ersp}
\partial_p \, \mathcal E_q - \partial_q \, \mathcal E_p - \partial^s  \mathcal E_{pqs} = O(  \mathcal E_{abc},  \mathcal E_d) \qquad .
\end{equation}
where,  the right-hand sides  (r.h.s.'s)  are  (linear) combinations of  the equations  $\mathcal E_p$,   $\mathcal E_{rsp}$  of the system.

These Bianchi-like identities, like their general-relativistic analogs, allow one to show the consistency of separating
our system of equations  $\mathcal E_p=0$,   $\mathcal E_{rsp}=0$ into ``evolution equations'' and ``constraint equations".
To discuss such a $(2+1)$-split of our system, it is convenient to replace the original $\beta$-space coordinates $\beta^a$
by the following Lorentzian-type combinations: 
\begin{eqnarray}\label{Lorxivar}
\label{Lor0}
\xi^{\hat 0} & : = &\frac{\sqrt 6}2 \, (\beta^1 + \beta^2 + \beta^3) \\
\label{Lor1}
\xi^{\hat 1} & : = &\frac{\sqrt 2}2 \, (\beta^2 - \beta^3) \\
\label{Lor2}
\xi^{\hat 2} & : = &\frac{\sqrt 6}6 \, (2\beta^1 - \beta^2 - \beta^3) \qquad .
\end{eqnarray}
In these coordinates, the $\beta$-space metric $G_{ab}$ takes the usual Lorentz-Poincar\'e-Minkowski form
$ {\rm diag} (-1,1,1)$ . Using such  coordinates, our system of equations (which was written in a $\beta$-space
covariant way) implies the following system of first-order in $\xi^{\hat 0}$-time  evolution equations
\begin{eqnarray}
\label{Deqp0}
&&\partial_{\hat 0} \, k_{\hat 0 \hat 0} = \partial_{\hat 1} \, k_{\hat 1 \hat 0} + \partial_{\hat 2} \, k_{\hat 2 \hat 0} + \Delta_{\hat 0} \qquad ,\\
&&
\label{Reqi}
\partial_{\hat 0} \, k_{\hat 0 \hat i} = \partial_{\hat i} \, k_{\hat 0 \hat 0} + R_{\hat 0 \hat i \hat 0} \qquad (\hat i = \hat 1 , \hat 2) 
\qquad ,\\
&&
\label{Reqj}
\partial_{\hat 0} \, k_{\hat i \hat j} = \partial_{\hat i} \, k_{\hat 0 \hat j} + R_{\hat 0 \hat i \hat j} \qquad (\hat i , \hat j= \hat 1 , \hat 2) \qquad .
\end{eqnarray}
This system of $(2+1)$ evolution equations must be supplemented by a system of {\it initial constraints}.
Indeed, the following combinations of our equations do not contain any ``time'' derivatives of the $k$'s [$(\hat p=\hat 1, \hat 2)$]
\begin{eqnarray} \label{constraints}
\mathcal C_{\hat 0}&:=&\partial_{\hat 1} \, k_{\hat 0 \hat 2} - \partial_{\hat 2} \, k_{\hat 0 \hat 1} - R_{\hat 1 \hat 2 \hat 0} = 0 \qquad  \nonumber \\
\mathcal C_{\hat p}&:=&\partial_{\hat 1} \, k_{\hat p \hat 2} - \partial_{\hat 2} \, k_{\hat p \hat 1} -R_{\hat 1 \hat 2 \hat p}=  0 \qquad  \nonumber \\
\mathcal C'_{\hat p}&:=& \partial_{\hat 1} \, k_{\hat p \hat 1} + \partial_{\hat 2} \, k_{\hat p \hat 2} - \partial_{\hat p} \, k_{\hat 0 \hat 0} - \Delta_{\hat p} - R_{\hat 0 \hat p \hat 0} = 0 \qquad 
\end{eqnarray}

Summarizing: a general solution for $k_{(ab)}$ at level  ${\mathbb H}_{(1,1)_S}$ is obtained by:
(i) finding the most general solution of the five equations of constraints  (\ref{constraints}) for the six initial data $k_{(ab)}$
considered on a spacelike hypersurface $\xi^{\hat 0} = t=C^{\text{st}}$ in $\beta$ space ; and, then (ii) evolving these
initial data in  $\xi^{\hat 0}$-time by integrating the six evolution equations  (\ref{Deqp0} --  \ref{Reqj}).

We have checked that,  as in Maxwell or Einstein theories, the Bianchi-like identities given above 
insure that if the constraints are satisfied on an initial spacelike $\beta$-space section they will remain verified for all values of $\xi^{\hat 0}$. To express the above results in a proper mathematical way one should prove that the evolution
system for $k$ is well-posed (as well as the evolution system for the constraints).
 However, as we know that the full system governing the   $\xi^{\hat 0}$-time evolution of the
complete (64-component) spinorial wave function $\Psi_{\sigma}(\beta)$ is well-posed\footnote{Indeed, given consistent
initial data for  $\Psi_{\sigma}(\beta)$, any of the four simultaneous  Dirac-like equations (\ref{SAEq}) yields a well-posed
symmetric-hyperbolic evolution system for its $\beta^0$-time evolution.}, it is clear that there is a way to rewrite our
evolution system  (\ref{Deqp0} --  \ref{Reqj}) in a well-posed way. [The evolution system for the constraints should also be
a consequence of our general consistency result (\ref{SSLSH}), which shows that all the constraints are ``in involution'',
in the sense of Cartan.]

At this stage, we have reduced the problem of parametrizing the set of solutions at level ${\mathbb H}_{(1,1)_S}$
to the problem of parametrizing the set of solutions of the initial constraint equations  (\ref{constraints}).  Though this
is a {\it linear} problem, it is a highly non trivial one, notably because of the complicated 
(and singular) $\beta$-dependence of the 
coefficients $\varphi, \rho, \mu$ entering the basic system (\ref{divk}),  (\ref{rotk}).  We have succeeded in showing,
by a detailed analysis, 
that the general solution of the five real  PDE's  (\ref{constraints}) (in any initial two-plane $\xi^{\hat 1} ,\xi^{\hat 2} $)
for the six real unknowns $k_{ab}(\xi^{\hat 1} ,\xi^{\hat 2} )$ is parametrized by {\it two} arbitrary real functions of
the two variables  ( $\xi^{\hat 1} ,\xi^{\hat 2} $), together with an arbitrary constant $C_4$ (entering the initial value of 
a certain projected component  $k_{\hat 0 \hat 0}$ of $k_{ab}(\xi^{\hat 1} ,\xi^{\hat 2} )$, see below) . In order not to interrupt the logical flow of this paper, we relegate
our proof of this result (as well as the boundary conditions we imposed in looking for solutions)
to Appendix  \ref{solvingconstraints}. Let us, however, give here some brief indications about the counting
of free functions in the general solution. First, it would seem that having five constraints for six unknowns
will only leave one free function in the general solution. The reason why it is not so, is that there is actually
one {\it identity} satisfied by the constraints.  It is of the form
\begin{equation} \label{identityconstr}
\partial_{\hat 1}  \mathcal C_{\hat 1} +  \partial_{\hat 2} \mathcal C_{\hat 2}  +  \partial_{\hat 2}  \mathcal C'_{\hat 1} -  \partial_{\hat 1} \mathcal C'_{\hat 2} \equiv O( \mathcal C_{\hat 0}, \mathcal C_{\hat p},\mathcal C'_{\hat p}) \, .
\end{equation}
Second, let us make plausible our result by considering the trivial case where one keeps only the derivative
terms in the constraints, neglecting the effect of the $\beta$-dependent coefficients  $\varphi, \rho, \mu$.
In that case, one immediately sees that the $\mathcal C_{\hat 0}$ constraint implies that $k_{\hat 0 \hat p}$
is (at least locally) a gradient:  $k_{\hat 0 \hat p} =  \partial_{\hat p}  \psi$. This accounts for one free function.
Then, the two  $\mathcal C_{\hat p}$ constraints imply that  $k_{\hat p \hat q}$ is a gradient w.r.t the second index:
$  k_{\hat p \hat q} =  \partial_{\hat q}  \phi_{\hat p}$. Using now the symmetry $  k_{\hat p \hat q}  =  k_{\hat q \hat p} $,
one sees that the vector potential  $\phi_{\hat p}$ must also be (at least locally) a gradient $\phi_{\hat p} =  \partial_{\hat p}  \phi$. Finally, we have  $  k_{\hat p \hat q} =  \partial_{\hat p}   \partial_{\hat q}  \phi $ which accounts for the second
free function. [One then checks that the remaining constraints  $\mathcal C'_{\hat p}=0$ can be solved for   
$  k_{\hat 0 \hat 0}$.] 

Note that an equivalent result would follow from analyzing the system of equations (\ref{divk}), (\ref{rotk}), directly in 
$2 + 1$ dimensions. Considering only the symbols (the derivative terms) of Eqs. (\ref{divk}), (\ref{rotk}), we obtain from the latter equation  that 
$k_{pq}=\partial_p \partial_q\Phi$. The former equation then yields $\partial_p \Box\Phi=0$ {\it i.e.} $\Box\Phi=C$. Accordingly the general solution will depend on the constant $C$ and on the two arbitrary functions defining  Cauchy data for $\Box \bar \Phi =0$, where $\bar \Phi := \Phi - \frac{C}{2} G_{ab} \beta^a \beta^b$.
\bigskip

In summary, the present section has shown that, at level $N_F=2$ the full set of solutions of the
supersymmetry constraints (\ref{SAEq}) was parametrized by:
\begin{enumerate} 

\item[(i)]  three arbitrary constants $C_1, C_2, C_3$ parametrizing  three ``discrete-spectrum states" 
belonging to the subspaces
 ${\mathbb H}_{(2,0)} \oplus {\mathbb H}_{(0,2)}$ and  ${\mathbb H}_{(1,1)_A}$; 

\item[(ii)]  two arbitrary (real) functions of two variables (and one real constant $C_4$) parametrizing a general ``continuous-spectrum state"
living in the  ${\mathbb H}_{(1,1)_S}$ subspace [{\it i.e.}  having a symmetric-tensor wave function  $k_{pq}(\beta)  : = f_{(pq)}^{+-}(\beta)$].

\end{enumerate}

In view of the boundary conditions we incorporated in the analysis of the initial-value problem in Appendix  \ref{solvingconstraints}, one can check that,
by appropriately choosing the two arbitrary functions  parametrizing the initial data (e.g. with compact support, or, at least,
with fast enough decay in the  spacelike directions spanned by $\xi^{\hat 1} ,\xi^{\hat 2} $), one can ensure that
all the components of  $k_{pq}(\beta)  : = f_{(pq)}^{+-}(\beta)$ initially decay under the gravitational walls 
(or, simply, under the gravitational wall $ 2 \beta^1$ when working within our canonical chamber).  As the evolution
of these initial data in $\beta$-space (in both directions of $\beta^0$ off the initial Cauchy slice) is given (when
considering any of the Dirac-like susy constraints) by a (well-posed) first-order symmetric-hyperbolic system, 
the property of fast decay under the gravitational walls will be preserved by the  $\beta^0$-time evolution. Our construction
therefore leads to solutions of the $N_F=2$ susy constraints which decay (rather than grow) under the gravitational walls
(and which are square-integrable at the symmetry walls). [As in the usual Dirac-equation case, the property of conservation
of the current(s) $J^a_A$ ensures a preservation of the integrability of any of its $\beta^0$-time component.]

As already explained, one can deduce from these results what are the solutions at the up-down symmetric level
$N_F=4$. This is straightforward for the discrete-spectrum states which are given by explicit analytic functions
of $x, y, z$. [One then sees that the transformation $ x \to -x$ etc. will induce an exponentially growing behavior of these
modes under the gravitational walls, and will leave their behavior under the symmetry walls as singular as
it is at level 2.]  This is less straightforward for the continuous-spectrum states. One should carefully redo the
analysis given in Appendix  \ref{solvingconstraints} with the system of equations obtained by the changes
 $( x,y,z)  \to (-x,-y-z)$.  Clearly the counting of free functions will be the same, but one may have to modify our reasoning
 by choosing appropriately modified Green's functions in the proof of  Appendix \ref{solvingconstraints}.
 We, however, expect that this is possible, and that, by choosing initial data which appropriately decay
 under the initial location of the gravitational walls, they will continue to do so under the (well-posed)
$\beta^0$-time evolution.

\bigskip

\setcounter{equation}{0}\section { Solutions at level $N_F = 3$}

\smallskip

The set of equations at level $N_F=3$ is  similar to the one at level $N_F=2$. It, however,  involves more degrees of freedom, and extra complications. The most general 
$N_F = 3$ state is given by
\begin{equation}
\label{Psi3}
{{\Psi}}_{(3)} \equiv {{\Psi}}_{(3)}^+ + {{\Psi}}_{(3)}^-
=\sum_{\epsilon = \pm} \left( \frac1{3!} \, f_{[klm]}^{\epsilon} \, \tilde b_{\epsilon}^k \, \tilde b_{\epsilon}^l \, \tilde b_{\epsilon}^m + \frac12 \, h_{[kl],m}^{\epsilon} \, \tilde b_{-\epsilon}^k \, \tilde b_{-\epsilon}^l \, \tilde b_{\epsilon}^m \right)  {\vert\, 0\,\rangle\hspace{-0.5mm}\strut_-} \qquad .
\end{equation}
Here the decomposition in $+$ and $-$ is done  according to the values indicated by the $\epsilon$'s.
Note that there is a multiplicative conservation law for them: $+ \times +$ counts like $- \times -$; [see Eqs (\ref{SEIpiNP})].

As already mentioned above,  there is a complete decoupling between the dynamics of  ${{\Psi}}_{(3)}^+$ (belonging to  ${\mathbb H}_{(3,0)} \oplus {\mathbb H}_{(1,2)}$)  and that of   ${{\Psi}}_{(3)}^-$  (belonging to  ${\mathbb H}_{(0,3)} \oplus {\mathbb H}_{(2,1)}$).
Because of the $ +  \leftrightarrow -$ symmetry, we shall henceforth only consider the $\epsilon=+$ case, and drop the
$\epsilon$ index on the wave functions.  
 The fully antisymmetric tensor $f_{klm}$ contains only one independent component, say $ f_{123}=\sqrt{2}\,f$ such that
 \begin{equation} f_{[klm]} = f \, \eta_{klm}
 \end{equation} 
On the other hand, the nine independent components of $h_{[kl],q}$ can be conveniently rewritten in terms of a
dualized, asymmetric two-index $\beta$-space tensor $h^m_{\ q}$:
\begin{equation}
 h_{[kl], q} \equiv \eta_{klm}h^{m}_{\phantom{m}q}\qquad, \qquad h_{pq}= - \frac 12 \eta^{kl}_{ \ \ p} \,  h_{[kl], q}  \qquad.
\end{equation}
Notice that we use the  $\beta$-space Levi-Civita tensor,
with (because of  $\det (G_{ab}) = -2$)   $\eta_{klm} = \sqrt 2 \, \varepsilon_{klm}$ and $\eta^{klm} = -\frac1{\sqrt2} \, \varepsilon^{klm}$, with $\varepsilon_{123}=\varepsilon^{123}=1$. Moreover, we move indices by means of $G_{ab}$ and  $G^{ab}$. We also introduce the notation $h$ for the $G$-trace
of  $h_{pq}$, {\it i.e.} 
\begin{equation}h :=  G^{pq} \, h_{pq} \qquad .
\end{equation}

As we have chosen $\epsilon = +$, the considered ${{\Psi}}_{(3)}^+$ state belongs to ${\mathbb H}_{(3,0)} \oplus {\mathbb H}_{(1,2)}$. The operator ${\cal S}^{(0)}_+$ projects ${\mathbb H}_{(3,0)}$ on ${\mathbb H}_{(2,0)}$ and ${\mathbb H}_{(1,2)}$ on ${\mathbb H}_{(0,2)} \oplus {\mathbb H}_{(2,0)}$. As a consequence the constraint ${\cal S}^{(0)}_+ \, {{{\Psi}}_{(3)}^+}=0 $ leads to the two equations :
\begin{equation}
\label{Idf1}
\left( \frac i2 \, \partial_p + \alpha_p - \mu_p + \overset{(1)}{\rho}_p \right) f + h_{pk} \, \nu^k - \nu_p \, h = 0
\end{equation}
\begin{equation}
\label{Iroth1}
\frac i2 \, \partial_{[p} \, h_{q]}^{\ r} + \alpha_{[p} \, h_{q]}^{\ r} +  \overset{(1)}{\rho}_{[p} \, h_{q]}^{\ r} - 2 \, \rho_{[p\mid a}^{\phantom{[p\mid a}r} \, h_{q]}^{\ a} + \delta_{[p}^{\ r} \, \nu_{q]} \, f = 0 \qquad .
\end{equation}
On the other hand ${\cal S}^{(0)}_-$ projects ${\mathbb H}_{(3,0)}$ on the zero vector but ${\mathbb H}_{(1,2)}$ on ${\mathbb H}_{(1,1)}$.  Thus acting on ${{\Psi}}_{(3)}^+$  it leads to the  equation : 
\begin{equation}
\label{Idivh2}
\left( \frac i2 \, \partial_k + \alpha_k + \mu_k - \overset{(1)}{\rho}_k \right) h_p^{\ k} + 2 \, \rho_{kpl} \, h^{lk} = 0 \qquad .
\end{equation}
Acting with $\widetilde {\cal S}^{(0)}_+$, ${\mathbb H}_{(3,0)}$ is mapped onto ${\mathbb H}_{(2,2)}$ via the $\tilde D_+^{klm}$ term (and thus in the spinor $\widetilde {\cal S}^{(0)}_+{{\Psi}}_{(3)}^+$, the $f$ term only appears in conjunction with $\nu_k$), while ${\mathbb H}_{(1,2)}$ is projected on the same subspace via the action of all the terms of $\widetilde {\cal S}^{(0)}_+$, except the one proportional to $\nu_k$. The corresponding equation obtained from $\widetilde {\cal S}^{(0)}_+{{\Psi}}_{(3)}^+=0$ is :
\begin{equation}
\label{Iroth2}
\frac i2 \, \partial_{[p} \, h_{\ q]}^{r} - \alpha_{[p} \, h_{\ q]}^{r} + \overset{(1)}{\rho}_{[p} \, h_{\ q]}^{r} - 2 \, \rho_{[p\mid a}^{\phantom{[p\mid a} r} \, h_{\ q]}^{a} + \delta_{[p}^{\ r} \, \nu_{q]} \, f = 0 \qquad .
\end{equation}
Finally $\widetilde {\cal S}^{(0)}_-$ maps  ${\mathbb H}_{(3,0)}$  onto  ${\mathbb H}_{(3,1)}$, while it
maps   ${\mathbb H}_{(1,2)}$  both onto  ${\mathbb H}_{(3,1)}$ (coupling again $f$ and $h_{pq}$) and onto ${\mathbb H}_{(1,3)}$. The corresponding two equations are
\begin{equation}
\label{Idf2}
\left( \frac i2 \, \partial_p - \alpha_p - \mu_p + \overset{(1)}{\rho}_p \right) f + \nu_a \, h_{\ p}^a - \nu_p \, h = 0
\end{equation}
and
\begin{equation}
\label{Idivh1}
\left( \frac i2 \, \partial_k - \alpha_k + \mu_k - \overset{(1)}{\rho}_k \right) h_{\ p}^k + 2 \, \rho_{kpl} \, h^{kl} = 0 \qquad .
\end{equation}

We have thereby obtained a heavily overconstrained system of differential 
equations for the ten  unknowns $f, h_{pq}$. To make progress with this system, it is
useful to separate the asymmetric tensor  $h_{pq}$  into antisymmetric ($A$) and symmetric ($S$)  parts
(we do not subtract the trace $h = G^{pq} S_{pq}$ from the symmetric part): 
\begin{equation}\label{hSA}
 \quad A_{pq}  := h_{[p\,q]} \qquad , \quad S_{pq}  := h_{(p\,q)}  \qquad .
\end{equation}

Let us briefly indicate the results obtained by using such a decomposition in the previous equations, when
considering appropriate combinations of various equations. First, by
comparing  the derivatives of $f$ given by Eq. (\ref{Idf1}) with those given by Eq. (\ref{Idf2}) we obtain 
an {\it algebraic} relation between $A_{pq}$ and the scalar $f$
\begin{equation}
\nu^k \, A_{[kp]} = 2 \, \alpha_p \, f
\end{equation}
(whose compatibility is guaranteed by the relation  $\nu^p \, \alpha_p \equiv 0$). 

The latter constraint tells us that  the three components of the antisymmetric part $A_{[pq]}$ only depend
on one unknown function, say  $\lambda(\beta)$, and that we can replace $A_{pq}$ by the
following tensorial combination (with known coefficients) of the two scalar unknowns $f$ and $\lambda$:
\begin{equation}
\label{hpqA}
A_{[pq]} = \frac2{\nu^2} \, (\nu_p \, \alpha_q - \alpha_p \, \nu_q) \, f +  \eta_{pqr} \, \nu^r \, \lambda \, .
\end{equation}
Here,  $\nu_p $ and $ \alpha_p $ are explicitly known, and we denoted
\begin{equation}\label{nu2}
\nu^2  : = \nu^k \, \nu_k = \frac{xyz^2 + yzx^2 + zxy^2 - x^2 y^2 - y^2 z^2 - z^2 x^2}{8(x-z)^2 \, (y-z)^2} \qquad .
\end{equation}
The next step is to compare two different expressions for the gradient of the trace $h$: one expression is obtained
by taking the trace of Eq. (\ref{Iroth1}) and subtracting it from the divergence equation (\ref{Idivh2});  the second one
is obtained by doing the same operations for the Eqs. (\ref{Iroth2}) and Eq. (\ref{Idivh1}). Finally, by equating
the two different values of  $\frac i2 \, \partial_p \, h$ so obtained, we get an {\it algebraic} relation between $h$ and 
$A_{pq}$, namely
\begin{equation}2 \, \alpha_p \, h  = \left( \mu^k -\overset{(1)}{\rho}\strut^k + \overset{(2)}{\rho}\strut^k \right) A_{[kp]} 
\end{equation}
But,   from the definitions (\ref{muk}, \ref{rho1}, \ref{rho2}, \ref{nuk}), one finds that 
\begin{equation}
\mu^k - \overset{(1)}{\rho}\strut^k + \overset{(2)}{\rho}\strut^k  = \frac{(2z-x-y)}{(x-y)} \, \nu^k \qquad ,
\end{equation}
so that the previous relation $\nu^k \, A_{[kp]} = 2 \, \alpha_p \, f$ yields a simple proportionality between
$f$ and $h$:
\begin{equation}\label{fph}
f = \frac{(x-y)}{(2z-x-y)} \, h \qquad .
\end{equation}

In other words, at this stage we can eliminate the four functions  $f$ and $A_{pq}$ in terms of the two scalar functions
$\lambda$ and $h = G^{pq} S_{pq}$. The final problem is then to obtain differential equations for
the remaining unknowns, namely $\lambda$ and the six components of $S_{pq}$.

We can first obtain a differential equation for $\lambda$ (containing $S_{pq}$ in its lower-order coefficients)
in the following way.  The difference between Eqs. (\ref{Iroth1}) and (\ref{Iroth2})  yields a partial differential equation (PDE)
 for  $A_{pq}$ of the form
\begin{equation}\label{Arot}
\frac i2 \partial_{[p} \, A_{q]r} + \overset{(1)}{\rho}_{[p} \, A_{q]r} - 2 \, \rho_{[p \mid \ r\mid}^{ \ a} \, A_{q]a} + \alpha_{[p} \, S_{q]r} = 0 \qquad .
\end{equation}
Introducing in this equation the expression above of   $A_{pq}$ in terms of $\lambda$ and $f$, and 
projecting the  indices $pqr$ of this equation by a combination of the type 
$$
 \eta^{pqs} \nu_s \delta^r_t - \frac12 \eta^{pqr} \nu_t \, ,
 $$
 yields and equation for $\lambda$ of the form 
 \begin{equation} \label{lamgrad}
\frac i2\, e^{-\tilde\xi}\partial_p\left(e^{+\tilde\xi}\,\lambda\right)+\Lambda_p (S_{kl} , x,y,z)=0
\end{equation}
where
\begin{equation}
\tilde\xi = \ln \frac{(y-x)^{3/8} \, (xyz)^{1/4}}{  (z-x)^{5/8} \, (z-y)^{5/8}} \qquad ,
\end{equation}
and where $\Lambda_p (S_{kl} , x,y,z)$ denotes an expression linear in the $S_{pq}$ components, that
we do not explicitly write here.
In deriving this equation for $\lambda$, one must make use of an equation for the gradient of $f$ obtained
by summing Eqs. (\ref{Idf1}), (\ref{Idf2}); namely
\begin{equation}
e^{-\xi} \, \frac i2 \, \partial_p \, (e^{\xi} \, f) + \nu_k \, h^{(kp)} = 0
\end{equation}
with $\xi$ given by (from Eqs. (\ref{mukpot}, \ref{rho1}, \ref{rho2}))
\begin{eqnarray}
\xi &= &4 (r_1 - \mu) - 2 r_2 \nonumber \\
&= &\frac18 \ln\left\vert \frac{(z-x)^5 \, (z-y)^5}{(y-x)^5 \, x^2 \, y^2 \, z^2} \right\vert\qquad .
\end{eqnarray}

To close the system, we need a set of differential equations for $S_{pq}$.  Such a system is obtained from
Eqs.  (\ref{Idivh2}), (\ref{Idivh1}),  (\ref{Iroth1}) and (\ref{Iroth2}). It reads
\begin{equation}
\label{Srot}
\frac i2 \partial_{[p} \, S_{q]r} + \overset{(1)}{\rho}_{[p} \, S_{q]r} - 2 \, \rho_{[p \mid \ r\mid}^{ \ a} \, S_{q]a} + \alpha_{[p} \, A_{q]r} +G_{r[p}\nu_{q]}f= 0
\end{equation}
\begin{equation}
\label{Sdiv}
\frac i2 \, \partial_k \, S^{kp} + 2 \, \rho_{k \ l}^{\ p} \, S^{kl} + \left( \mu_k - \overset{(1)}{\rho}_k \right) S^{kp} - \alpha_k \, A^{kp} = 0
\end{equation}
In the r.h.s.'s one should replace $A_{pq}$ and $f$ in terms of $\lambda$ and  $h=G^{ab} S_{ab}$, using the
algebraic relations found above.

To summarize:  At level $N_F=3$, we have two independent sectors ($+$ or $-$) which are totally equivalent to
each other.
In each sector, the problem is reduced to the coupled dynamics of seven   unknown functions~:  the 
symmetric components $S_{pq}$ components  of the dual of the original $h_{[pq] , r}$ wave function,
and the scalar function $\lambda$ parametrizing part of the antisymmetric components   $A_{pq}$.
These seven unknown functions must satisfy twelve first-order partial differential  equations,
namely  (\ref{Srot}, \ref{Sdiv} and \ref{lamgrad}). We have checked the consistency of this system
(which satisfies Bianchi-like identities similar to the ones discussed at level $N_F=2$). 
Note that the two equations   (\ref{Srot}, \ref{Sdiv}) for $S_{pq}$ are of the ``curl" and ``div" type.
A new feature, however, is the coupling between $S_{pq}$ and the scalar degree of freedom $\lambda$ (
which had no analog at level 2).  A rough counting of the free data in the general solution (which would need
to be firmed up by a detailed analysis of the type we gave at level 2) is that the general solution in
each independent ($+$ or $-$) sector at level 3 depends on two (real) functions of two variables, to which
must be added an arbitrary constant entering the integration of the (gradient) equation for $\lambda$.


\bigskip
\setcounter{equation}{0}\section{Asymptotic plane-wave-type solutions at levels $N_F=2$ and $N_F=3$}

As explained in the previous sections, while there exist only discrete states at levels $N_F=0,1,5,6$,
at the intermediate levels $N_F=2,3,4$ there exists a mixture of discrete-states and continuous states (parametrized
by arbitrary functions).  We have proven the existence of the latter states by studying the Cauchy problem for
the PDE's satisfied by the wave function at level $N_F=2$ (arguing that the similar PDE systems at levels $N_F=3,4$
will feature similar solutions). However, it was evidently impossible to express these continuous states in closed form.
In the present section, we try to get some familiarity with the structure, and physical meaning, of these states by approximating them (in
some asymptotic regime) by  plane-wave type solutions. This could be done in the high-frequency, WKB approximation,
but, we shall actually study a regime where one can use a better approximation  than the usual WKB one.

We recall that a solution at some fixed fermionic level $N_F=N$, has the general structure
\begin{equation}
\Psi_{(N)} =   f^{\epsilon_1 \epsilon_2 \cdots \epsilon_N}_{a_1 a_2 \cdots a_N} (\beta) \widetilde b^{a_1}_{\epsilon_1} \cdots    \widetilde b^{a_N}_{\epsilon_N}  {\vert\, 0\,\rangle\hspace{-0.5mm}\strut_-}
\end{equation}
where the components $ f^{\epsilon_1 \epsilon_2 \cdots \epsilon_N}_{a_1 a_2 \cdots a_N} (\beta)$ of the wave function satisfy  a set of (Dirac-like) first-order partial differential equations  implied by the susy constraints (\ref{SAEq}). In addition,
they also satisfy a more familiar second-order Klein-Gordon-type spin-dependent WDW equation.

The WKB approximation would consist in looking for solutions where the tensorial wave function
$ f^{\epsilon_1 \epsilon_2 \cdots \epsilon_N}_{a_1 a_2 \cdots a_N} (\beta) $ would be the product of a slowly-varying
tensorial amplitude, and of a high-frequency scalar phase-factor $ e^{i S(\beta)/\varepsilon}$, with $\varepsilon \to 0$.
This high-frequency limit would mean that we consider the limit of large momenta 
$\pi_a \approx \partial_a S/\varepsilon \to \infty$.  Here, we shall instead consider a regime where the momenta are
not required to tend to infinity, so that we will be able to simultaneously retain effects linked to various
powers of the momenta. To do that, we consider the quantum analog of the classical BKL approximation, {\it i.e.} 
we take the formal ``far-wall'' limit where the various exponential potential walls entering either the susy
constraints, or the WDW equation become small. To be in such a regime, one needs all the $\beta$'s 
to be large and positive, keeping also large and positive some of their differences.  Geometrically, this
corresponds to being deep in the middle of a Weyl (or billiard) chamber, far from all its boundary walls.
For instance,  if we are within our canonical Weyl chamber $ \beta^1 \leq \beta^2 \leq \beta^3$, we need to have
$ \beta^1 \gg 1$, $\beta^2 - \beta^1 \gg 1$ and $\beta^3 - \beta^2 \gg 1$. Note that this implies 
$\beta^0 := \beta^1 + \beta^2 + \beta^3 \gg 1$.

In this limit, the susy constraint operators simplify to 
\begin{equation} {\widehat{\mathcal S}_A^{(0)}} = \frac i2  \Phi^a_A \partial_{\beta^a} + i \, \Phi \,  \Phi \, \Phi
\end{equation}
where the terms cubic in $\Phi$ have two origins:  the supergravity cubic terms $\widehat {\cal S}_A^{\rm cubic}$, Eq. (\ref{ScA}), and the (Weyl-chamber-dependent) far-wall limit of the symmetry-wall hyperbolic-cotangent  contribution  Eq. (\ref{SsA}).

Correspondingly, the far-wall limit of the Hamiltonian constraint has the structure
\begin{equation}2 \, \widehat H^{(0)}=  -G^{ab}   (\partial_a- \varpi_a )(\partial_b-  \varpi_b) +  \hat\mu^2 
\end{equation}
 where $\varpi_a$ is the  Weyl-chamber-dependent limit of $ A_a(\beta) = \partial_a \, \ln \, F  = F^{-1}   \partial_a  F $.
 We recall that
  \begin{equation} \label{F'}
 F(\beta)  = e^{\frac34 \, \beta^0} (\sinh \beta_{12} \, \sinh \beta_{23} \, \sinh \beta_{31} )^{-1/8} .
  \end{equation} 
 In our canonical Weyl chamber,  $ \beta^1 \leq \beta^2 \leq \beta^3$, we have
 \begin{equation} \label{varpi}
 (\varpi_a) = \left( 1, \frac34, \frac12  \right)
 \end{equation}
 
 We shall therefore be considering plane-wave-type solutions having wave functions of the form
\begin{equation}
 f^{\epsilon_1 \epsilon_2 \cdots \epsilon_N}_{a_1 a_2 \cdots a_N}(\beta)=  A^{\epsilon_1 \epsilon_2 \cdots \epsilon_N}_{a_1 a_2 \cdots a_N} \exp[i\,\pi_a\beta^a]\quad 
\end{equation}
where $A^{\epsilon_1 \epsilon_2 \cdots \epsilon_N}_{a_1 a_2 \cdots a_N}$ is some $\beta$-independent tensorial
amplitude.
We recall that it is convenient to rescale the wavefunction according to 
\begin{equation}
   \Psi'(\beta) = F(\beta)^{-1} \Psi(\beta) \sim e^{- \varpi_a \beta^a}  \Psi(\beta)
\end{equation}
This implies that the corresponding primed  plane-wave-type wavefunction
\begin{equation}
 f'^{\epsilon_1 \epsilon_2 \cdots \epsilon_N}_{a_1 a_2 \cdots a_N}(\beta)=  A^{\epsilon_1 \epsilon_2 \cdots \epsilon_N}_{a_1 a_2 \cdots a_N} \exp[i\,\pi'_a\beta^a]\quad 
\end{equation}
has the same tensorial amplitude $A$ but features a primed momentum $\pi'_a$ which differs from the momentum $\pi_a$
entering the original wave function
\begin{equation}\pi'_a = \pi_a + i \, \varpi_a
\end{equation}
 
 It was shown above that the equations satisfied by the wave function $ f^{\epsilon_1 \epsilon_2 \cdots \epsilon_N}_{a_1 a_2 \cdots a_N}(\beta)$ could be written in a purely real form.  When looking, as we do here, for plane-wave solutions
 it will be necessary to consider complex tensorial amplitudes $A$. We recall also that (as is clear from the expression
 above of the Hamiltonian constraint) it is the primed momentum  $\pi'_a$ , rather than  $\pi_a$ which has to satisfy
 the (real) mass-shell condition
 \begin{equation} G^{ab}  \pi'_a   \pi'_a  + \mu^2 =0
 \end{equation}
 In most of this section, we shall assume that we are interested in {\it real} solutions of this mass-shell condition,
 {\it i.e.}  real values of  the $\pi'_a$'s, corresponding to propagating waves.  
 This implies that the   original  $\pi_a$'s are complex.

\bigskip

\subsection {$N_F = 2$ asymptotic plane-wave solutions}

In the $N_F=2$ case, one can look for plane-wave solutions   in the ${\mathbb H}_{(1,1)_S}$ subspace, {\it i.e.} 
\begin{equation}{{\Psi}}^S_{(1,1)} =  \, k_{(pq)}(\beta)  \ \tilde b_+^p \, \tilde b_-^q \,  {\vert\, 0\,\rangle\hspace{-0.5mm}\strut_-} \qquad ,
\end{equation}
with
\begin{equation}\label{KpqAs}
k_{(pq)}(\beta) = K_{pq} \, e^{i\,\pi'_k \beta^k} \, e^{+\varpi_k \beta^k}
\end{equation}
and with a  primed momentum satisfying the $N_F=2$ (tachyonic) mass-shell condition
\begin{equation} \label{shell2}
G^{ab} \pi^{\prime}_a \, \pi^{\prime}_b  = \pi^{\prime a} \, \pi^{\prime}_a = + \frac38
\end{equation}
If, as we are mainly assuming,  the components $\pi'_a$ are real, the 3-vector  $\pi'^a$ must be spacelike. 
Note that $ k_{(pq)}(\beta)$ denotes the wave function of the original, unprimed,  state $\Psi$.

The tensorial amplitude $ K_{pq} $ has  to satisfy the two equations that results from
the plane-wave/far-wall limit of  Eqs. (\ref{divk}, \ref{rotk}), {\it i.e.} 
\begin{equation}
 \frac12 \, (\pi^{\prime k} -i\, \varpi^k) \, K_{kp} -  \bar \varphi^k  \, K_{kp} + 2 \bar \rho^{kl}_{\ p} \, K_{kl} = 0
\end{equation}
\begin{equation}
\frac12 \, (\pi' -i\, \varpi)_{[p}\, K_{q]r} + \bar \varphi_{[p} \, K_{q]r} -  \bar \mu_{[p} \, K_{q]r} - 2 \bar \rho_{[p\mid r}^{\ k} \, K_{q]k}= 0 \qquad .
\end{equation}
Here the overbar indicates that one must take the far-wall limit of the various coefficient functions $\varphi(\beta), \rho(\beta), \mu(\beta)$. The values of these limits generally depend on the considered Weyl chamber.  However, 
whatever be the Weyl chamber, the asymptotic values of $\alpha_k$  are always $\{0,0,0\}$. 
On the other hand,  the limit of $\mu_k$ is either  $i/2\{1,-1/2,1\}$ or $i/2\{-1/2,1,-1\}$ according to whether $y \geq x$ or $x \geq y$, irrespectively of the value of $z$. Another example is provided by the asymptotic behaviour of $\overset{(1)}\rho\strut_k$.  
In the canonical Weyl chamber $(a)$, where $z\geq y\geq x$,  it goes to $i/2\{-3,-5/2,0\}$, but in the Weyl chamber $(e)$, where $x\geq z\geq y$, its limit is $i/4\{-1,-3,-1/2\}$. This lack of obvious symmetry with respect to permutations of $x$, $y$ and $z$ is not a problem.  The equations will remains invariant only if an exchange between the $k$ indices is accompanied by a redefinition of the $\Phi_A^k$ matrices that represent  the Rarita--Schwinger field. Anyway,
their physical consequences will be the same in all Weyl chambers.

In the present case, we find that the linear system satisfied by the six tensorial amplitude $K_{pq}$
 is of rank five; its general solution therefore depends on only  one arbitrary constant, say  $C_{2}$.
 It can be written as :
\begin{equation}
\label{AsKpq}
K_{pq} = C_{2}\left(\pi'_p \, \pi'_q + L_{pq}^k \, \pi'_k + m_{pq}\right)
\end{equation}
where, after performing some linear algebra, and working in Weyl chamber $(a)$, the two $ 3 \times 3$
matrices $ L_{pq}(\pi') = L_{pq}^k \, \pi'_k$ and  $m_{pq}$ are given by
\begin{equation}
L_{pq}^k \, \pi_k^\prime = -i \begin{pmatrix}
3 \, \pi'_1 + \pi'_2 + \pi'_3 &\frac32 (\pi'_1 + \pi'_2) &\frac12 (\pi'_1 + 3 \pi'_3) \\
{ \ } \\
\frac32 (\pi'_1 + \pi'_2) &2 \pi'_2 + \pi'_3 &\frac12 (\pi'_2 + \pi'_3) \\
{ \ } \\
\frac12 (\pi'_1 + 3\pi'_3) &\frac12 (\pi'_2 + \pi'_3) &\pi'_3
\end{pmatrix}
\end{equation}
and
\begin{equation}
m_{pq} = -\frac14 \begin{pmatrix}
13 &9 &3 \\
9 &5 &1 \\
3 &1 &1
\end{pmatrix} \qquad .
\end{equation}
In this case, the expressions of the analog of the matrices $ L_{pq}(\pi') $ and  $m_{pq}$  in the other
 Weyl chambers are simply obtained from this one by the permutation of the indices corresponding to the ordering of the scale factors of the considered Weyl chamber, with respect to the reference one.
 
 Several comments on these plane-wave solutions are in order. First, the fact that they depend only on
 one (complex) amplitude (for each momentum direction), is the plane-wave transcription of our general finding that the continuous states
 at level $N_F=2$ depend on two (real) arbitrary functions of two variables. [In both cases, this represents one 
 scalar degree of freedom; corresponding to the general solution of a Klein-Gordon like equation.]
 Second,  if we consider real momenta  $\pi'_a$ with parametrically large components, the mass-shell condition
 (\ref{shell2}) reduces to the constraint that   $\pi'_a$ be approximately {\it null} : $\pi'^2 \approx 0$.
 In that (WKB) limit we recover the plane-wave analog of the classical cosmological-billiard dynamics
  \cite{Belinsky:1970ew,Misner:1969hg,Damour:2002et} : the Universe is represented by a massless
  particle moving along a straight line within a (Kac--Moody) billiard.  At the classical
  level, we know that when this particle will approach one of the potential walls defining the boundary
  of this billiard chamber, it will ``bounce'' on that wall and be reflected back within the central
  region of that chamber. At the quantum level, if we consider the full WDW equation, {\it i.e.}  the
  Hamiltonian constraint  (\ref{WDW}), with Hamiltonian   (\ref{H0}) or  (\ref{H'}), it is clear that, in the high-frequency
  WKB limit, the wave (or  wave-packet) (\ref{KpqAs}) will also bounce and reflect on the quantum analogs
  of the potential walls, if we decide to impose the boundary condition that the wave function must
  exponentially decay (rather than grow) under the potential walls. For an explicit proof of this
  (expected) behavior, see, e.g., Ref. \cite{Damour:2011yk} which considered the coupling
  of Bianchi Universes to  a spin-$\frac12$ field. [Though this case is technically simpler than
  the  spin-$\frac32$ we are now considering, it has many similarities with it.]
  We leave to future work a detailed study of how, within the present supergravity framework,
  the tensorial wave  (\ref{AsKpq}) reflects on a potential wall, and of the relation
  between the incident and outgoing ``polarization tensors'' $K_{pq}$.
  
  In the case where the   $\pi'_a$'s are parametrically large components, it is instructive to see how the restricted
  structure  (\ref{AsKpq})  of the plane wave solutions follows from the supersymmetry constraints. In that limit
  the susy constraints approximately reduce to 
  \begin{equation} {\widehat{\mathcal S}_A^{(0)}} \approx - \frac 12  \Phi^a_A \pi'_a
\end{equation}
which is simply the Fourier-space massless Dirac operator in $\beta$ space.  In this limit, the anticommutator 
identity (\ref{SSLSH}) simplifies to a usual supersymmetric quantum mechanical identity
\begin{equation} {\widehat{\mathcal S}_A^{(0)}} {\widehat{\mathcal S}_B^{(0)}} + {\widehat{\mathcal S}_B^{(0)}}  {\widehat{\mathcal S}_A^{(0)}} \approx \frac14 G^{ab} \pi'_a \pi'_b  \, \delta_{AB}
\end{equation}
 which clearly  exhibits the necessity of the approximate mass-shell condition $\pi'^2 \approx 0$. In this limit it is easy
 to find the general solution of the chiral-basis susy constraints
 \begin{equation} 
0 = - 2 {\cal S}^{(0)}_\epsilon   \vert {\bm \Psi}\rangle = \pi'_a  b^a_\epsilon   \vert {\bm \Psi}\rangle
\end{equation}
 \begin{equation} 
0 = - 2  \widetilde {\cal S}^{(0)}_\epsilon   \vert {\bm \Psi}\rangle = \pi'_a  \widetilde b^a_\epsilon   \vert {\bm \Psi}\rangle  
\end{equation}
Indeed, starting from the null vector  $\pi'_a$ in $\beta$-space, one can define a (real) null basis of $\beta$ space
made of two null vectors and a space-like one, say  $\pi'_a$,  $q_a$, and $r_a$, such that the only non-zero
$G$-scalar-products between these vectors are $ \pi' \cdot q = 1$ and  $r \cdot r =1$.  One can then replace the
original $\beta^a$-coordinate-based annihilation/creation operators $ b_{\epsilon}^a$ , $\widetilde b_{\epsilon}^{a}$
by their projections on this null basis, {\it i.e.}  $  b_{\epsilon}(\pi') := b_{\epsilon}^a \pi'_a$,  $  b_{\epsilon}(q):= b_{\epsilon}^a q_a$,  $  b_{\epsilon}(r) := b_{\epsilon}^a r_a$, etc.  Writing a general state at level 2 in terms of the 
corresponding null-basis creation operators, and using the basic anticommutation relations 
$ \left\{ b_{\epsilon}(u) , \tilde b_{\epsilon'}(v)  \right\} = 2 \,  u \cdot v \, \delta_{\epsilon\, \epsilon'}$, etc, it is easily found that the general solution
of the conditions $  b_{\epsilon}(\pi') \vert {\bm \Psi}\rangle= 0 =  \widetilde b_{\epsilon}(\pi') \vert {\bm \Psi}\rangle$ is
\begin{equation} \label{WKBNF2}
C_2  \,  \widetilde b_{+}(\pi')   \widetilde b_{-}(\pi')   {\vert\, 0\,\rangle\hspace{-0.5mm}\strut_-}
\end{equation}
which is equivalent to the leading-order term in the more general far-wall solution (\ref{AsKpq}) in the limit where
 the   $\pi'_a$'s are parametrically large. Let us note in passing that the approximate form  (\ref{WKBNF2}) can
 also be written (in the same approximation) as  $ \widetilde {\cal S}^{(0)}_+   \widetilde {\cal S}^{(0)}_-   {\vert\, 0\,\rangle\hspace{-0.5mm}\strut_-}$,
 which is reminiscent of an ansatz suggested by Csordas and Graham \cite{Csordas:1995kd}. However, we have
 shown that, within our framework, such an ansatz (saying that the general $N_F=2$ solution is obtained by
 acting on some $N_F=0$ scalar state $f(\beta)  {\vert\, 0\,\rangle\hspace{-0.5mm}\strut_-}$ by   $ \widetilde {\cal S}^{(0)}_+   \widetilde {\cal S}^{(0)}_-$)  is not correct
 beyond the high-frequency, plane-wave limit.

 Here, we focussed on asymptotic far-wall waves having a real (shifted) momentum $\pi'_a$, because
 this looks most natural in view of the formal hermiticity of the Hamiltonian operator $H'$, Eq. (\ref{H'}),
  corresponding to  the rescaled state  $\Psi'$,  Eq. (\ref{psi'}).  However, it might also be possible to
consider far-wall solutions where   the components $\pi'_a$ are complex, say  $\pi'_a= p_a - i q_a$, where the
two real 3-vectors $p_a$, $q_a$ would satisfy   $G^{ab} p_a q_b  =0$ and  $G^{ab} p_a p_b  - G^{ab} q_a q_b 
= +\frac38$.  The wave function of such waves would be of the type
\begin{equation}
k_{(pq)}(\beta) = K_{pq} \,  e^{i\, p_a \beta^a} \,  e^{+( q_a + \varpi_a) \beta^a}
\end{equation}
A particular case would be the situation where $\pi'_a$ is purely imaginary, {\it i.e.}  $p_a=0$ and $\pi'_a =  - i q_a$,
corresponding to real, exponentially-behaving (non-oscillating) plane waves
of the type
\begin{equation}
k_{(pq)}(\beta) = K_{pq} \,  e^{+( q_a + \varpi_a) \beta^a}
\end{equation}
In that case, the real vector $q_a$ must satisfy 
$G^{ab} q_a q_b = - \frac38$, and therefore it must be time-like.  

We have seen above that the covariant components $\varpi_a$ are given by Eq. (\ref{varpi}). The corresponding
contravariant components $G^{ab} \varpi_b$  read
\begin{equation}\varpi^a= \left( - \frac18, - \frac38, - \frac58 \right)
\end{equation}
If we conventionally define the ``future'' in $\beta$ space as the direction in which $\beta^0 = \beta^1 + \beta^2 + \beta^3$
increases (in other words the direction of decreasing volume of the Universe; {\it i.e.}   towards the cosmological singularity), the vector $\varpi^a$ is past-directed (and $  e^{\varpi_a \beta^a} $ increases toward the future).  Then, 
focussing on the case where $\pi'_a$ is purely imaginary, if the real time-like 3-vector $q^a$ is also
past-directed, the sum $  q^a + \varpi^a$ will be time-like and past-directed, so that the real factor 
$e^{+( q_a + \varpi_a) \beta^a}$ will increase towards the cosmological singularity. On the other hand, if we consider
a  time-like vector   $q^a$  which is future-directed, the sum  $  q^a + \varpi^a$ may have several different types
of $\beta$-space orientations.  Let us only note here the fact that the squared length of  $ \varpi^a$ is
\begin{equation} \varpi^2 = G_{ab}  \varpi^a \varpi^b= - \frac{23}{32}
 \end{equation}
 This is larger (in absolute value) than the squared magnitude of  $q$: $ q^2 =   G_{ab}  q^a  q^b = -\frac38$.
 Therefore, in the particular case where $q^a$ would be taken to be proportional to   $ \varpi^a $, the 
 sum   $  \pm q^a + \varpi^a$  would remain future-directed whatever be the sign $\pm$, {\it i.e.}  the direction
 of $q^a$. In the general case where we retain a non-zero real part $p_a$ in $\pi'_a$ there are even more
 possibilities. However, before considering more seriously all those possibilities involving complex values of
 the shifted momentum, one should study whether, when they impinge on one of the gravitational or symmetry walls,  
 they  can be matched to a reflected wave, modulo the presence of an exponentially decaying wave under the considered
 wall (as was shown to be the case for  real-$\pi'_a$ waves  coupled to a spin-$\frac12$  field \cite{Damour:2011yk}).

\bigskip

\subsection{$N_F = 3$ asymptotic plane-wave solutions}

The study of plane-wave solutions at level $N_F = 3$ leads to similar conclusions.  One finds that the general structure
describing such plane waves is either
\begin{equation}
{{\Psi}}_{(3)}^+
= \left[ \sqrt{2}\,  f^+  \tilde b_+^1 \,  \tilde b_+^2 \,  \tilde b_+^3 + S^{+}_{pq}  \left( \frac12 \, \eta^p_{\ \  kl} \, \tilde b_{-}^k \, \tilde b_{-}^l \, \tilde b_{+}^q \right) \right]  {\vert\, 0\,\rangle\hspace{-0.5mm}\strut_-} \qquad .
\end{equation}
or a similar ${{\Psi}}_{(3)}^-$ state.  Each such state is parametrized by a  {\it symmetric} tensorial wave function  
$S^{+}_{pq}$, or  $S^{-}_{pq}$. Indeed, the scalar  $f^+$ or $f^-$  is not independent from $S^{\epsilon}_{pq}$, but is
proportional to its trace: $f^{\epsilon} = \sigma \, G^{pq} S^{\epsilon}_{pq}$, where the factor $\sigma$ is equal
either to $0$, $+1$ or $-1$, depending on the considered Weyl chamber.

Indeed, by taking the plane-wave limit
of our general $N_F = 3$ analysis above, one finds that  the antisymmetric components   $A_{[pq]}$ 
 must vanish. For instance, when working in our canonical Weyl chamber (or, more generally
in any chamber where  $z = e^{2 \beta^3}$ is larger than $x$ or $y$), one first notices (from its definition in Eq. (\ref{nukpot}))
that all the components
of the vector  $\nu_k$ vanish.  Therefore, the second contribution to $A_{pq}$ in Eq. (\ref{hpqA}) (proportional to 
$\eta_{pqr} \nu^r \lambda$) vanish.  On the other hand, in the first contribution (proportional to $f  \nu_{[p} \, \alpha_{q]} / \nu^2$), one finds that the factor  $ \nu_{[p} \, \alpha_{q]} / \nu^2$ has a finite, non-zero limit (recall that  $\alpha_p \to 0$
far from the gravitational walls). However,  Eq. (\ref{fph}) 
shows that, in the case we are considering ($z$ dominant),  the scalar $f$
tends to zero with respect to $h = G^{pq} S_{pq}$. Finally, in this case ($z$ dominant), both $A_{pq}$ and $f$ vanish
(in the notation above, we have $\sigma=0$). If we are in a different Weyl chamber (with a subdominant $z$),  
neither $f$ nor the components of  $\nu_k$ will not vanish. Instead, they will have some finite limits.  First, 
Eq. (\ref{fph})  shows that $ f = \sigma \, h$, where $\sigma=+1$ if $y$ is dominant, and $\sigma=-1$ if $x$ is dominant.
Second, in such cases, the first contribution to  $A_{pq}$ in Eq. (\ref{hpqA})  will again vanish (now because 
 $ \nu_{[p} \, \alpha_{q]} / \nu^2 \to 0$).  As for the second contribution, it will again vanish, but now because $\lambda \to 0$
 in the considered cases. Indeed, the equation
 \begin{equation}
\label{Adiv}
\frac i2 \, \partial_k \, A^{kp} + \left( \mu_k - \overset{(3)}{\rho}_k \right) A^{kl} - \alpha_k \, S^{kl} = 0 \qquad ,
\end{equation}
which follows from the general $N_F=3$ equations, implies, asymptotically, the constraint
\begin{equation}
\lambda \left( -\frac12 \, \pi^{\prime k} + \mu^k - \overset{\ \ (3) \, k}{\rho} \right) \eta_{kpl} \, \nu^{\ell} = 0
\end{equation}
which can only be satisfied, for an arbitrary $\pi'^k$  on its mass shell, if $\lambda$ vanishes. 

 To derive the asymptotic structure of the symmetric part $S_{(pq)}$,  we have to deal (as we did in the $N_F=2$ case) with the div $+$ curl system satisfied by  $S_{(pq)}$: namely, the divergence equation (\ref{Sdiv}) (where the last term can be neglected) and the curl equation (\ref{Srot}) where one should use $ f= \sigma \,   G^{pq} S_{pq}$ in the last term.  
 The final result for the structure of  $S_{(pq)}$ depends on the considered Weyl chamber (both because of the different
 values of $\sigma$, and of the different far-wall limits of the coefficients entering the div $+$ curl system).
 
 For instance, in the Weyl chamber $(a)$, we obtain plane-wave amplitudes of the form
 \begin{equation}
S_{pq} = C_{3}\left(\pi'_p \, \pi'_q + L_{pq}^k \, \pi'_k + m_{pq}\right) \, ,  \qquad f  = 0\, ,\qquad h=C_3\left[ -\frac 12 +   \frac i 2 \, (\pi'_2 - \pi'_3)  \right]\qquad ,
\end{equation}
 with, now,
\begin{equation}\label{LkpqNF3CWa}
L_{pq}^k \, \pi_k^\prime = i \begin{pmatrix}
-\pi'_1 + \pi'_2 + \pi'_3 &-\frac12 \, \pi'_2 + \pi'_3 &-\frac12 (\pi'_1 - \pi'_3) \\
{ \ } \\
-\frac12 \, \pi'_2 + \pi'_3 &-\pi'_2 + \pi'_3 &-\frac12 \, \pi'_2  \\
{ \ } \\
-\frac12 (\pi'_1 - \pi'_3) &-\frac12 \, \pi'_2  &-\pi'_3
\end{pmatrix}
\end{equation}
and
\begin{equation}\label{mpqNF3CWa}
m_{pq} = +\frac14 \begin{pmatrix} 5&2&1 \\ 2&2&0 \\1&0&-1 \end{pmatrix} \qquad .
\end{equation}
By contrast, in, say,  Weyl chamber $(b)$,  the corresponding amplitudes are given by
 \begin{equation}
S_{pq} = C_{3}\left(\pi'_p \, \pi'_q + L_{pq}^k \, \pi'_k + m_{pq}\right) \, ,  \qquad f  = C_3  \left[  \frac14 +   \frac i 2 \, (\pi'_2 - \pi'_3)  \right]=h \qquad ,
\end{equation}
 with
\begin{equation}
L_{pq}^k \, \pi_k^\prime = i \begin{pmatrix}
-\pi'_1 + \pi'_2 + \pi'_3 &-\frac12 (\pi'_1 - \pi'_3) &\frac12 \, \pi'_2 \\
{ \ } \\
-\frac12 (\pi'_1 - \pi'_3) &-\pi'_2 &-\frac12 \, \pi'_2  \\
{ \ } \\
\frac12 \, \pi'_2 &-\frac12 \, \pi'_2  &\pi'_2 - \pi'_3
\end{pmatrix} \qquad ,
\end{equation}
\begin{equation}
m_{pq} = \frac14 \begin{pmatrix} 5&0&1 \\ 0&-1&-1 \\ 1&-1&2 \end{pmatrix}  \qquad .
\end{equation}

Contrary to what occurred at level $N_F=2$, the transformation rules of these amplitudes, when  swapping Weyl chambers, is far from obvious.

Most of the comments we made above in the $N_F=2$ case apply {\it mutatis mutandis}. In particular,  the fact
that the general plane-wave solution at level $N_F=3$  depends on only two complex constants $C_3^+, C_3^-$
is the plane-wave transcription of our finding above that there are, in each $\pm$ sector, two arbitrary (real) functions
of two variables.  In addition, it is also an instructive exercize to 
see how the special structure of the $N_F=3$ plane-wave solution emerges from the susy constraints in the
limit where the components $\pi'_a$ get large. In this limit,   $\pi'_a$ is approximately null ($ \pi'^2 \approx 0$),
and one can again conveniently introduce a null basis $\pi'_a,  q_a, r_a$, and corresponding projected
annihilation operators  $  b_{\epsilon}(\pi') := b_{\epsilon}^a \pi'_a$,  $  b_{\epsilon}(q):= b_{\epsilon}^a q_a$,  $  b_{\epsilon}(r) := b_{\epsilon}^a r_a$.  Using their anticommutation relations it is then easy to find the general solution,
at level $N_F=3$ of  the conditions $  b_{\epsilon}(\pi') \vert {\bm \Psi}\rangle= 0 =  \widetilde b_{\epsilon}(\pi') \vert {\bm \Psi}\rangle$.  One finds  that it is obtained by acting on the $N_F=2$ solution   $\widetilde b_{+}(\pi')   \widetilde b_{-}(\pi')   {\vert\, 0\,\rangle\hspace{-0.5mm}\strut_-} $ by
arbitrary combinations of the ``transverse'' creation operators   $  \widetilde b_{\epsilon}(r)$, {\it i.e.} 
\begin{equation} \label{WKBNF3}
\left( c_+  \widetilde b_{+}(r) + c_- \widetilde b_{-}(r) \right)  \widetilde b_{+}(\pi')   \widetilde b_{-}(\pi')   {\vert\, 0\,\rangle\hspace{-0.5mm}\strut_-} 
\end{equation}
It is easy to see that such a solution is equivalent to the above results in the limit where the components $\pi'_a$
get large.  The factorized form (\ref{WKBNF3}) suggests that 
one might obtain the general $N_F=3$ solutions by acting on the
general $N_F=2$ solution by some suitable raising operator. However, we have shown that this was not true
beyond the high-frequency plane-wave limit.

Finally, let us note that the $N_F=4$ plane-wave solutions can be easily obtained from the $N_F=2$ ones by 
the exchange $ {\vert\, 0\,\rangle\hspace{-0.5mm}\strut_-} \to {\vert\, 0\,\rangle\hspace{-0.5mm}\strut_+}$ and $b^a_{\pm} \to \widetilde  b^a_{\pm}$. [As the gravitational-wall terms are negligible in
the considered limit, one does not need to worry about the additional complex shift of the $\beta$'s.]

\smallskip


\setcounter{equation}{0}\section{Bouncing Universes and  boundary conditions in quantum cosmology}

\smallskip
Since the pioneering work of DeWitt \cite{DeWitt:1967yk},  the issue of boundary conditions (near big
bangs or big crunches) in quantum cosmology has been much discussed.  Several proposals have been
made. In particular,  DeWitt has suggested to impose the vanishing of the wave function of the Universe on
the singular ``zero-volume'' boundary of superspace, 
Vilenkin  \cite{Vilenkin:1982de,Vilenkin:1986cy}  suggested 
a boundary condition selecting a wave function tunnelling from ``nothing" into superspace, while
Hartle and Hawking \cite{Hartle:1983ai} have suggested determining a unique wave function for the
Universe by considering a path integral over compact Euclidean geometries.  See \cite{Vilenkin:1987kf}
for a comparison of the predictions from the latter two different choices within a restricted two-dimensional
minisuperspace model, and  see Refs. \cite{Furusawa:1985ef,Berger:1989jm,Graham:1990jd,Kirillov:1992kh,Benini:2006xu} for  studies of the wave function
of the (bosonic) Bianchi IX model.

Another context within which the issue of  boundary condition at a spacelike singularity is 
important is that of evaporating black holes. In particular, Horowitz and Maldacena \cite{Horowitz:2003he}
have suggested the need of imposing a ``final state boundary condition" at a black hole singularity
in order to make sure that no information is absorbed by an evaporating black hole. 

We wish to point out here that our finding  that supergravity predicts
the presence (in the major part of our Hilbert space) of a {\it tachyonic} ({\it i.e.}  negative) squared-mass $\mu^2$
in the WDW equation (\ref{WDW})  naturally leads to a kind of final-state boundary condition at the singularity
that might be relevant to the black hole information-loss problem.

Let us start by  explaining in simple terms the origin of the  squared-mass term $\mu^2$, and its   {\it a priori}   importance
near the singularity.  It is well-known that the supergravity Lagrangian density $ L$  (per unit proper spacetime volume) contains terms quartic in the fermions:  $ L_4 \sim \psi^4$.  Such terms will correspond to a proper energy density
$\rho_4 \sim \psi^4$. We have seen above that, when quantizing the spatial zero-modes of $\psi$ the variables $\Psi$
satisfying a Clifford algebra (with a numerically fixed r.h.s. of order unity in Planck units) are obtained by
rescaling $\psi$ according to $\psi = g^{- 1/4} \Psi$, Eq. (\ref{rescaledpsi}), where $g= (a \, b \, c)^2 $ denotes the
determinant of the spatial metric.  As a consequence the proper energy density linked to the quartic fermionic terms
scales with the proper spatial volume $\mathcal V_3 =  a \, b \, c$ as
\begin{equation} \label{rho4}
\rho_4 \sim g^{-1} =  \left( \mathcal V_3 \right)^{-2}=    \left(  a \, b \, c \right)^{-2} = \bar a ^{-6}
\end{equation}
where $\bar a := ( a b c)^{1/3}$ denotes the geometric average of the three scale factors.
As the volume $\bar a ^3$ of the Universe decreases,  the energy density $\rho_4$ increases faster than
the other well-known contributions to the energy density, such as the energy density 
$\rho \sim ( a b c )^{- ( 1 + w)} = \bar a^{-3 (1+w)}$ associated with  a fluid with the equation of state $ p = w \rho$, with $ w <1$.
Well-known examples are: (i) a cosmological constant ($w=-1$) with $\rho = Cst$. ; and (ii) thermal radiation ($w=\frac13$)
with   $\rho \sim \bar a^{-4}$.  The anisotropy energy associated with the Bianchi IX curvature,
namely $\rho_{\rm curv} \sim a^2/( b^2 c^2) + {\rm cyclic} $ plays initially a special role because of the
Kasner oscillations which can make, e.g., $ a \gg b , c$, thereby allowing (as proven in Ref. \cite{Belinsky:1970ew})
the anisotropic curvature energy 
$\rho_{\rm curv} $ to be more important than any ordinary fluid-type energy (having $ w <1$). However,  when averaging over the billiard motion
of $ \beta^1 = - \ln a$,   $ \beta^2 = - \ln b$,  $ \beta^3 = - \ln c$,  within some chamber, all the separate
scale factors $ a, b, c$ will eventually decrease and formally tend toward zero (though at different,
and chaotically changing speeds), so that the ratio $\rho_{\rm curv}/ \rho_4 \sim a^4 + b^4 + c^4 $
will eventually decrease and tend toward zero as   $\mathcal V_3 =  a \, b \, c  \to 0$.  This reasoning shows that,
when going toward the singularity,  the anisotropic potential $V_g(\beta) = \frac14 (a^4 + b^4 + c^4) - \frac12 (b^2 c^2 + c^2 a^2 + a^2 b^2)$ will initially dominate over usual energy densities (such as thermal energy, when included),
but will ultimately be dominated by the effect of  the squared-mass term $\mu^2$.
The latter conclusion does not depend on the sign of  $\mu^2$. We are, however, going to see that the sign of
$\mu^2$ has crucial consequences for the issue of boundary conditions at the singularity.  Let us also note that
the dependence of the fermionic energy density (\ref{rho4}) on the spatial volume formally
corresponds to a stiff equation of state $p = \rho$, with index $w=+1$ (as that corresponding
to a massless scalar field).

After these heuristic considerations, let us consider the technical aspects of the behavior of
the quantum wave function near the singularity.  As is well-known from the study of the classical Bianchi IX
dynamics \cite{Misner:1969hg,Chitre:1972} and its generalizations  \cite{Ivashchuk:1994da,Ivashchuk:1994fg,Damour:2002et},  the asymptotic behavior of the dynamics of the scale factors near the singularity is best exhibited by
replacing the flat Lorentzian coordinates $\beta^a$ by the corresponding (hyperbolic) polar coordinates $\rho, \gamma^a$:
\begin{equation}
\beta^a = \rho \, \gamma^a  \quad {\rm with} \  \rho = \sqrt{-G_{ab} \beta^a \beta^b} , \, G_{ab} \gamma^a \gamma^b = -1 \, .
\end{equation}
In other words, the variable $\rho$ (which should not be confused with the notation used above for
the proper energy density)  is the Lorentzian radius, while the corresponding Lorentzian ``angular coordinates"
are encoded in the
two independent components of  the vector $ \gamma^a$  running on
the unit hyperboloid, which is a realization of the Lobachevski plane. [The unit-hyperboloid vector  $ \gamma^{ a}$ 
should be distinguished from the  notation  $ \gamma^{\hat a}$ used for Dirac matrices.]
In terms of these ``polar" coordinates, the
metric in $\beta$ space becomes
\begin{equation}G_{ab} d \beta^a d \beta^b= - d \rho^2 + \rho^2  d \sigma^2
\end{equation}
where $d \sigma^2$ is the constant-curvature ($K=-1$) metric  on the unit $\gamma$ hyperboloid. The corresponding
d'Alembertian operator in $\beta$ space reads
\begin{equation}\Box_{\beta} = +  G^{ab} \partial_a \partial_b=   - G^{ab} \hat\pi_a  \hat\pi_b = - \frac{1}{\rho^{d-1} } \partial_{\rho} \left( \rho^{d-1} \partial_{\rho} \right) + \frac{1}{\rho^2}  \Delta_{\gamma}
\end{equation}
where, for more generality, we have provisionally considered the case of any $\beta$-space  dimension $d$ ($=$ the number of spatial dimensions). In our case, $d=3$ and we have
\begin{equation} \frac{1}{\rho^{2} } \partial_{\rho} \left( \rho^{2} \partial_{\rho} \right)= \partial_{\rho}^2 + \frac{2}{\rho}   \partial_{\rho}=  \frac{1}{\rho } \partial_{\rho}^2 \,  \rho
 \end{equation}
In terms of the rescaled wave function $\Psi'$ and of these polar coordinates, the WDW equation (\ref{H'}) reads
\begin{equation} \label{WDWrho}
\left(    \frac{1}{\rho } \partial_{\rho}^2 \,  \rho -   \frac{1}{\rho^2}  \Delta_{\gamma}  + \hat\mu^2 + \widehat W (\beta) \right) \Psi'(\rho, \gamma^a) =0
\end{equation}
Leaving to future work a study of the near-singularity limit of  the first-order susy constraints (\ref{constraints}), we shall 
only give here an approximate treatment of  the asymptotic behavior of the solutions of the second-order WDW equation.
When approaching the cosmological singularity we have $\rho \to + \infty$, and all the  potential
terms in  $ \widehat W (\beta) =  \widehat W (\rho \gamma) $ become very sharp functions of $\gamma$ on Lobachevski space (because of the factor $\rho$ multiplying the argument of $\widehat W$).  In the {\it interior} of the intersection
of a Weyl chamber of $AE_3$ on the unit $\gamma$ hyperboloid,
{\it i.e.}  when, say, $ 0 < \gamma^1 < \gamma^2  <  \gamma^3$,  the potential  $ \widehat W (\rho \gamma) $  will
tend toward zero as   $\rho \to + \infty$. On the other hand, when  one goes on the other side of the gravitational wall
({\it i.e.}  when, say,   $  \gamma^1 < 0 $)  the relevant bosonic gravitational-wall term $ \propto + e^{- 4 \rho \gamma^1}$
tends toward $ + \infty$, and dominates the spinorial $J$-dependent term  $ \propto  J_{11}  e^{- 2 \rho \gamma^1}$.
This suggests (as in the purely bosonic case) that we can replace the gravitational-wall terms by an infinite, sharp wall
located at $\gamma^1=0$.  The case of the symmetry walls is a    {\it a priori}    more subtle because they are purely
quantum (and spin dependent), and also because they are not exponential, but proportional to $1/ \sinh^2  \beta_{ab}$.
However, a local analysis of the regular solutions near these walls shows that the exact wave function $\Psi'$ (as  well
as $\Psi$)  vanish on the symmetry walls.  Finally, we can impose, in the asymptotic limit  $\rho \to + \infty$ that
the wave function  $\Psi'$ vanishes on all the boundaries of each Weyl chamber,  while, in the interior, it
satisfies the equation
\begin{equation} \label{qbilliard1}
\left(    \frac{1}{\rho } \partial_{\rho}^2 \,  \rho -   \frac{1}{\rho^2}  \Delta_{\gamma}  + \hat\mu^2  \right) \Psi'(\rho, \gamma^a) =0
\end{equation}
As $ \hat\mu^2$ is a $c$-number at each fermionic level, and as we have just seen that  $ \Psi'(\rho, \gamma^a) $ satisfies
Dirichlet boundary conditions on the $\gamma$-space walls $\gamma^1=0$,  $\gamma^1 = \gamma^2$  and 
$\gamma^2 = \gamma^3$, we can expand (at each level $N_F$) the general solution of  (\ref{qbilliard1}) in  a series
of separated modes of the form
\begin{equation} \Psi'(\rho, \gamma^a) = \sum_n  R_n(\rho) \, Y_n(\gamma^a)
 \end{equation}
Here, the ``angular factors" $Y_n(\gamma^a)$'s are eigenmodes, with Dirichlet boundary conditions,
 of the Laplace-Beltrami operator on, say, the triangular billiard chamber with boundaries  
 $\gamma^1=0$,  $\gamma^1 = \gamma^2$  and 
$\gamma^2  = \gamma^3$ on the unit hyperboloid, while  $R_n(\rho)$ is a corresponding radial factor.
The latter Dirichlet billiard is the quantum version of the so-called Artin billiard, whose domain is half the 
famous keyhole-shaped fundamental domain of the modular group $SL(2, {\mathbb Z}) $. The spectrum of
our quantum triangular Dirichlet billiard corresponds to the  spectrum of  odd cusp automorphic forms. See, e.g.,
\cite{Cartier:1979,Terras} for nice accounts of the theory of such Maass automorphic waveforms.  The eigenvalues
$\lambda_n$, with
\begin{equation} \Delta_{\gamma}  \,   Y_n(\gamma^a) =  - \lambda_n Y_n(\gamma^a) \, ,
 \end{equation}
 are often written as $\lambda_n = \frac14 + r_n^2$. The  fundamental Dirichlet eigenmode has
 $r_1= 9.5336952613536\dots $, which corresponds to the surprisingly large lowest eigenvalue
 $\lambda_1= 91.14134533635 \dots$ . 
 
 The differential equation that each radial factor $R_n(\rho) $ must satisfy reads
 \begin{equation} \label{qbilliard2}
\left(    \frac{1}{\rho } \partial_{\rho}^2 \,  \rho +   \frac{\lambda_n}{\rho^2}   + \mu^2  \right)  R_n(\rho) =0
\end{equation}
 Like for ordinary 3-dimensional quantum mechanical spherically symmetric problems we can consider
 the rescaled radial function $u_n(\rho) := \rho  R_n(\rho) $, which satisfies a one-dimensional Schr\"odinger
 equation. However, as $\rho$ is a timelike, rather than a spacelike, variable, we must reverse the sign of the
 analog one-dimensional potential. In other words,  one can think of $\rho$ as the position
 of a quantum particle moving, with {\it zero} energy,  in the potential (modulo a factor 2)
 \begin{equation} \label{Urho}
U(\rho)=   -   \frac{\lambda_n}{\rho^2}   - \mu^2 
 \end{equation}
 with a wavefunction satisfying
  \begin{equation} \label{qbilliard3}
\left(  -  \partial_{\rho}^2  +   U(\rho)  \right)  u_n(\rho) =0
\end{equation}
 
 The qualitative features of this quantum problem near the singularity ({\it i.e.}  as $\rho \to + \infty$)
 crucially depend on the sign of $\mu^2$, because  $\lim_{\rho \to + \infty} U(\rho) = - \mu^2$.
 [We are aware of the fact that all the solutions we are going to discuss can be written 
 in terms of (suitably modified) Bessel functions. However, it is more illuminating for our purpose
 to focus on the approximate analytic expressions that are relevant near the singularity.]
 
 If   $\mu^2$ is strictly positive (which happens only at fermionic level $N_F=3$ where $\mu^2 = \frac12$),
 $U(\rho)$ becomes negative  near the singularity  ($\rho \to + \infty$).  The general solution near the singularity
 will then be a superposition of incoming and outgoing waves
 \begin{equation}\rho \, R_n(\rho) \equiv  u_n(\rho) \approx a_n  e^{i \mu \rho} +   b_n  e^{-i \mu \rho}   \  , \   {\rm as} \  \rho \to + \infty
 \end{equation}
 The frequency of these waves only depend on $\mu = \sqrt{\mu^2}$ and not on the spatial eigenvalues $\lambda_n$.
 The possibility of such incoming or outgoing waves near the singularity signals a possible information-loss 
 (or information-gain) at the singularity.  At the classical level, the presence of such oscillating modes means
 that a positive $\mu^2$ ultimately quenches the BKL chaotic oscillations of the scale factors, and 
 (like would the presence of a massless scalar field) leads to a final, monotonic, power-law approach
 toward a zero-volume singularity. 
 
 In our supergravity context, $\mu^2$ never vanishes.  Let us, however,  to allow comparison of our results with  those obtained in previous works, where  $\mu^2=0$ was generally assumed, discuss what happens 
 when  $\mu^2=0$ in the quantum problem  (\ref{qbilliard3})  .  In that case, it is the subdominant term 
 $  -   \frac{\lambda_n}{\rho^2}  $ in the potential that matters.  The fact that it is negative leads again to
 a wavelike behavior near the singularity, with the presence of both positive and negative
 frequencies. However, in that case one should take as position variable $\lambda = \ln \rho$. 
 One easily finds that the general solution of  (\ref{qbilliard3}) then reads  \cite{Kleinschmidt:2009cv}
 \begin{equation} R_n(\rho) = a_n \rho^{-1/2} e^{ i r_n \ln \rho} +   b_n \rho^{-1/2} e^{- i r_n \ln \rho}
 \end{equation}
 where $ r_n = \sqrt{ \lambda_n - \frac14}$ is the eigenvalue parametrization introduced above.
 Again the simultaneous possibility  of such incoming or outgoing waves  signals a possible information-loss 
 (or information-gain) at the singularity.  At the classical level, the presence of such oscillating modes means
 that a vanishing $\mu^2$ leads to unending BKL chaotic oscillations of the scale factors,   toward a zero-volume singularity. 
 
 Let us now consider the case where $\mu^2$ is strictly negative (which happens  at all fermionic levels,
 apart from $N_F=3$).  In that case  $U(\rho)$ becomes {\it positive} ({\it i.e.}  repulsive) near the singularity so that the general
 solution is a superposition of exponentially decreasing or increasing solutions :
  \begin{equation}\rho \, R_n(\rho) \equiv  u_n(\rho) \approx a_n  e^{- |\mu| \rho} +   b_n  e^{+ |\mu| \rho}   \  , \   {\rm as} \  \rho \to + \infty
 \end{equation}
 where $ |\mu| := \sqrt{ - \mu^2}$. The presence of  possible solutions that are exponentially growing as  $\rho \to + \infty$
 suggests (similarly to the case of a quantum particle impinging on a repulsive potential wall) that we should
 impose as boundary condition at the singularity the absence such growing modes, {\it i.e.}  the vanishing of
 all the coefficients $b_n$.  In other words, it is natural to require that $ \rho \, R_n(\rho) \sim   e^{- |\mu| \rho} 
 \to 0$ as  $\rho \to + \infty$. At the classical level the absence of oscillating solutions near the singularity tells
 us that a negative $\mu^2$, {\it i.e.}  a negative fermionic energy density 
$ \rho_4 \sim -   |\mu|^2 \left( \mathcal V_3 \right)^{-2} \sim -    |\mu|^2  \left(  a \, b \, c \right)^{-2} $, has the
effect not only of stopping the chaotic BKL oscillations, but even of stopping the collapse of the Universe
toward small volumes,  and to naturally force the Universe to ``bounce''  toward large volumes.
It is interesting to see that supergravity naturally predicts (in most cases) that quartic-in-fermion terms
(linked to spatial zero-modes) lead to such a stopping, and reversal, of the collapse.  Though these negative
fermionic energy densities are of quantum origin, it seems consistent (within our fully quantum
framework) to take them into account and to conclude that they indeed allow for cosmological bounces.
In other words, our work realizes (within our minisuperspace context) a wish expressed by DeWitt   \cite{DeWitt:1967yk},
namely showing the dynamical consistency of imposing the vanishing of the wave function of the Universe at the
zero-volume boundary of superspace.

For completeness, let  us give the exact solution  of the separated quantum radial equation (\ref{qbilliard2}),
corresponding to one spatial mode $Y_n(\gamma)$. When imposing our suggested decaying
boundary condition at the singularity, it is of the form
\begin{equation}R_n(\rho) = \frac{a'_n}{  \rho^{1/2}}  \,  K_{ i r_n}(  |\mu| \rho)
\end{equation}
where  $ r_n = \sqrt{ \lambda_n - \frac14}$ and where $K_{i \nu}(z)$ is the $K$-Bessel function for
a pure imaginary order.  The latter  imaginary-order, real-argument  K-Bessel function
is real,  exponentially decaying for large argument, and real-oscillatory when $ |\mu| \rho \lesssim r_n$.
Viewed from a classical limit standpoint, the above radial wavefunction  describes a bounce of $\rho$ around the minimal value
$\rho_{\rm min} \approx r_n/  |\mu| $.

If we come back for a moment to the classical dynamics of (diagonal) Bianchi IX cosmological models,  it is interesting
to note that the effect of adding a negative $\mu^2$ to the ordinary (classical) bosonic potential 
$ W_g (\beta) = 2 \, V_g(\beta)$, with Eq. (\ref{Vg}),  has been studied in the literature.  Indeed, Refs.  \cite{Rugh:1990aq}
and  \cite{Christiansen:1994zq}  have considered a modification of the usual BKL dynamics equivalent to
adding a negative $\mu^2$, with the motivation that such a ``physically unacceptable" negative energy term had
been unwittingly included in some previous numerical studies of BKL chaos, thereby leading to unexpected, 
erratic oscillations of the three-volume.  In a follow-up paper   \cite{Christiansen:1994zq}, it was further noted that
the Bianchi IX dynamics modified by a negative $\mu^2$  contains numerous {\it closed orbits} in $\beta$ space,
{\it i.e.}  Universes that bounce {\it periodically}.  The existence of such classical cyclically bouncing Universes
(of the type of the old cycloid-based Friedman Universe, but with a regular minimum volume state) is intuitively
understandable in view of the  closed-Universe-recollapse property of classical Bianchi IX models. Let us recall
that Lin and Wald  \cite{Lin:1989tv} have proven  that vacuum Bianchi IX models cannot expand for an
infinite time, but must recollapse. [The later reference \cite{Lin:1990tq} has extended this result
to the non-vacuum case, under the condition that  the matter content satisfies the dominant energy condition, and 
that the average pressure is non-negative.  Strictly speaking, their results do not apply to our case, but,
as our negative energy (and pressure) fermionic term $\rho_4= p_4$  decreases very fast $\propto  \left(  a \, b \, c \right)^{-2} $ during the expansion,  we are considering here that the recollapse is actually induced by the large-volume
limit of the bosonic Bianchi IX potential $V_g(\beta)$ (which, modulo a rescaling by $ g = ( a b c)^2$ is the anisotropic equivalent of the well-known
Friedman curvature-potential term $\propto - k/ a^2$, with $k=+1$, responsible for the recollapse
of closed Friedman Universes).]

In Figure \ref{BouncingUniverses}  we sketch [in $\beta$ space, indicating the Lorentzian coordinates
$\xi^{\hat 0},  \xi^{\hat 1}, \xi^{\hat 2}$ defined in  Appendix  \ref{Varia}  below] two of the simplest  cyclically bouncing  Bianchi IX  models
(with an additional negative $\mu^2$) found in Ref.   \cite{Christiansen:1994zq}:  namely the
ones labelled (i) and (vii) in Table 1 there. [They refer to a diagonal
Bianchi IX model, without the symmetry walls present in our supergravity framework.]  The 
funnel-type  structure surrounding these periodic curves is a sketchy representation of the bosonic potential
$V_g(\beta)$; it indicates the locus of the $\beta$-space points where  $V_g(\beta)= + 1$.  For values of order
unity of the momenta $\pi_a = G_{ab} \dot \beta^b$,  the level set   $V_g(\beta)= + 1$ represents the
approximate location of the potential wall which confines  $\beta(\tau)$  motions oriented in spacelike or null
directions,
and leads to the usual billiard description. What is not represented (and must be mentally added by the reader) is the fact
that, both in the upper part (where the negative $\mu^2$ term dominates over  $V_g(\beta)$) and in the lower part
of the funnel (where the potential $V_g(\beta)$ becomes  deeply negative because, for large, nearly  isotropic spaces
$ a \approx b \approx c$  we have  $V_g(\beta) \approx - \frac34 a^4$) there are other potential walls that can
confine   $\beta(\tau)$  motions oriented in a time-like direction, and make it bounce backwards in  $\beta^0$ ``time".
Indeed, we have seen above that, for instance, the potential $U(\rho)$, Eq. (\ref{Urho}), describing the motion
in a $\beta$-space timelike direction (like the $\rho$ one) was the opposite of the usual potential $V_g(\beta)$ 
(augmented by the $\mu^2$ term), so that a negative  $V_g(\beta)$  wall approached in a timelike direction
is roughly equivalent to a positive   $V_g(\beta)$  wall approached in a spacelike direction.
In other words, we should think of the funnel represented in  Figure \ref{BouncingUniverses}  as being a kind
of closed ``bottle'' (in the sense of ``magnetic bottles")  within which a $\beta$-space motion is confined in all directions.

{\begin{figure}[h] 
\begin{center}
\includegraphics[height=75mm]{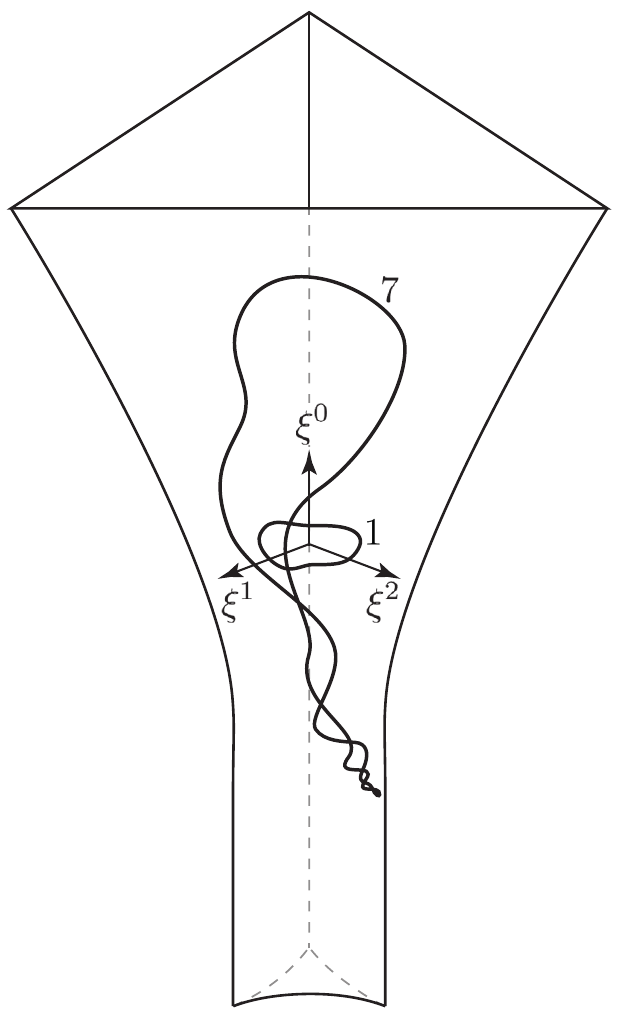} 
\end{center}
\caption{\label{BouncingUniverses}\footnotesize Two examples of periodically bouncing Universes
confined within the Lorentzian $\beta$-space ``bottle"  made by the bosonic Bianchi IX potential  $V_g(\beta)$ 
augmented by a negative $\mu^2$ term.}
\end{figure}}

Evidently, the periodic curves confined within such a $\beta$-space bottle (and sketched in Fig. \ref{BouncingUniverses} )
are just fine-tuned examples of generic classical orbits, which will chaotically  \cite{Rugh:1990aq,Christiansen:1994zq}
oscillate up and down (as well as sideways, as in  standard BKL chaos) in this bottle. We have explictly checked,
by numerically integrating the classical $\beta$-particle equations of motion, in presence of an additional negative
$\mu^2$ term,  that, indeed, generic classical orbits tend to  fill in a chaotic way the funnel represented
in  Fig. \ref{BouncingUniverses}. Note, however, that if one considers motions having,
at some point, very large momenta  $\pi_a = G_{ab} \dot \beta^b$ their confining funnel will be
correspondingly larger (though only logarithmically so).  However, we think that the existence of a (presumably
infinite) number of periodic $\beta$-space orbits is conceptually important for the following reason.
Studies of  the relation between classical chaos and quantum chaos (notably through the basic contributions
of Selberg \cite{Selberg:1956} and Gutwiller \cite{Gutzwiller:1971fy} ) have shown that there is an intimate link (embodied in some trace formula)
between the closed orbits of a classical system, and the eigenvalue spectrum of the corresponding
quantum dynamical system.  This classical/quantum correspondence suggests that the existence of a discrete
spectrum of periodic orbits in $\beta$-space (when $\mu^2 <0$)  signals the presence of a corresponding
discrete set of quantum states confined within the bottle of   Fig. \ref{BouncingUniverses}, {\it i.e.}  describing 
quantum bouncing Universes, satisfying the boundary condition that the wavefuntion $\Psi$ vanishes exponentially
{\it both} when $\beta^0 \to + \infty$ (small volumes) and  when  $\beta^0 \to - \infty$ (large volumes).
Our results above for fermionic levels $N_F= 0,$ and $1$ has rigorously established  (by explicit construction)
the existence of two such (square-integrable) discrete states (confined in all directions). [They might be considered as the
first two states in the expected tower of discrete states.]  On the other hand, our results on continuous\footnote{The discrete
states at level 2 were found to be non square-integrable.}  states at 
levels $N_F=2, 3 $ and $4$ has shown that one could construct continuous families (parametrized by several
arbitrary functions of two variables) of square-integrable states, that exponentially decayed when going
under the gravitational walls ({\it i.e.}  sideways in   Fig. \ref{BouncingUniverses}).  Among these, our emphasis on
the importance of having a negative $\mu^2$, suggests that we should restrict our attention to the cases
$N_F=2$ and $4$ for which  $\mu^2 = - \frac38$.
However, our construction did
not give us any freedom of imposing boundary conditions neither as  $\beta^0 \to + \infty$, nor as $\beta^0 \to - \infty$.
Our discussion above of  the behavior of quantum billiards (with $\mu^2 <0$)  at the singularity, suggests that
the imposition of the condition that   $\Psi$ vanishes exponentially when $\beta^0 \to + \infty$  will eliminate
half of the solution space (by setting all the $b_n$'s to zero). This roughly leaves a solution space
containing only one arbitrary function of two variables. [Indeed, each sequence $\{ a_n \}$ or  $\{ b_n \}$
parametrizes an arbitrary function of two variables, $\sum_n a_n Y_n(\gamma)$, satisfying Dirichlet conditions on our 
$\gamma$-space Weyl chamber.]  The imposition of a similar exponential decay of   $\Psi$ when 
$\beta^0 \to - \infty$ ({\it i.e.}  for large volumes) might further restrict  the arbitrariness described by the
sequence  $\{ a_n \}$ to leave only a much sparser discrete sequence of states, conceivably equivalent
to having, say, only one arbitrary function of {\it one} variable.  [Though, at this stage, we cannot discard
the possibility that this second restriction might eliminate all discrete states.] A toy model showing
the subtleties involved in such a reduction follows.

Let us consider a simple model of a Lorentzian dynamics within a potential that confines motions both
in spacelike and in timelike directions. Namely, a simple WDW-type equation
representing a two-dimensional Lorentzian harmonic oscillator :
\begin{equation}H \, \Psi(t,x) =0
\end{equation}
with
\begin{equation} \label{toy}
2 \, H = + \partial_t^2 -   \partial_x^2  - \omega_t^2\, t^2 +   \omega_x^2\, x^2 
\end{equation}
We have $ H = - H_t + H_x$ where both $H_t= \frac12 \left(- \partial_t^2 +  \omega_t^2\, t^2  \right)  $ and 
$H_x =  \frac12 \left(- \partial_x^2 +  \omega_x^2\, x^2  \right)  $ are usual, confining harmonic oscillators.
The harmonic frequency  for timelike motions (in the $t$ direction) is $ \omega_t >0$,
while the harmonic frequency for spacelike motions (in the $x$ direction) is $ \omega_x >0$ . The eigenvectors of $H$ can be looked for in a factorized
form $ \Psi(t,x) = f(t) \, g(x)$. As both $f(t)$ and $g(x)$ must be eigenfunctions of  confining-type harmonic oscillators,
they will be both restricted to a discrete spectrum if we impose that $\Psi(t,x)$ exponentially decays {\it both} in
timelike and spacelike directions (and for {\it both} signs of these two axes).  Under these conditions, we must have
$ \Psi(t,x) = h_m(t) \, h_n(x)$ (where $m,n$ denote natural integers, and $h_n$ are the usual Hermite eigenmodes), and the eigenvalues of the total $H$
( $H \Psi = E_{m n} \Psi$)
are restricted to the values
\begin{equation} E_{m n} = -  ( m + \frac12)  \omega_t  + ( n + \frac12)  \omega_x
 \end{equation}
 The WDW equation demands that we only consider states such that $E_{mn} =0$. As a consequence, we see that : 
 (i) if the ratio of the two frequencies $ \omega_t$, $ \omega_x$ is rational there will exist a restricted set of modes satisfying
 all our conditions [e.g., in the simple case where $ \omega_t= \omega_x$ we get the one-integer
 sequence  $\Psi_n(t,x) =h_n(t) \, h_n(x)$ of solutions]; (ii) on the other hand, if the ratio $ \omega_t/ \omega_x$ is irrational, there
 does not exist any solution satisfying our conditions [though, there exists modes of the type $f(t) h_n(x)$ that
 will decay in both spatial directions, as well as when $t \to - \infty$, but that blow up when  $t \to + \infty$.]
 In our case, we can hope that supersymmetry will relate the behavior in timelike and spacelike directions
 and allow for the existence of a final, sparse discrete set of solutions decaying in all directions. The fact that
 we have proven the existence of such states at levels $N_F=0$ and $N_F=1$ is a good indication in this sense.

 We initially hoped that the existence of classical bouncing solutions (as sketched in  Fig. \ref{BouncingUniverses})
 might entail the existence of corresponding quantum states. In particular, it is tempting to interpret the
 lowest classical periodic solution, labelled 1 in  Fig. \ref{BouncingUniverses},  as corresponding to the unique
quantum  ground-state at level $N_F=0$. [As sketched in the figure, the latter classical solution describes a Universe
which has a nearly constant Planck-size volume (nearly constant  $\xi^{\hat 0} =\frac{ \sqrt{6}}{2}  \beta^0$),
but whose ``shape'' oscillates.  This roughly fits with the wavefunction (\ref{Sol00}) of the $N_F=0$ state.]
However, the toy model (\ref{toy}) shows (when $a/b$ is irrational) that the existence of classical bouncing and confined 
solutions does not guarantee the existence of corresponding quantum states. 
 We leave to future work a study of the existence of quantum bouncing solutions at levels $N_F=2$ and $N_F=4$.
 
 In this work, we only considered the dynamics of pure supergravity, without extra matter content. Our aim was
 not to suggest a phenomenological description of early cosmology that might later turn into our observed Universe,
 but rather to investigate the conceptual role of supergravity in the dynamics close to a big bang or big crunch
 singularity. If, however, we contemplate an extension of our model containing, say, some type of inflationary
 sector (with an inflaton field $\phi$), we will have a modification of  the WDW equation (\ref{WDWrho}) consisting 
 (notably) of adding
 both a derivative term proportional to $-  \partial_\phi^2$ and an additional contribution to the potential term
 $W(\beta)$  proportional to  $ + (a  b c)^2 V(\phi)$, where $V(\phi)$ denotes the inflationary potential (which is
 chosen to be positive so as to be able to mimic a positive cosmological constant).
 When considering the dynamics of a time-like ({\it i.e.}  volume-like) gravitational
 degree of freedom (such as $\beta^0 = - \ln ( a bc)$)  the additional term 
 $W_\phi= + (a  b c)^2 V(\phi) = e^{- 2 \beta^0} V(\phi)$  must be considered (as explained above) as being
 a downfalling cliff rather than a repulsive wall. Therefore, from the point of view of the quantum dynamics of
 $\beta^0$  the confining (near isotropic  metrics) wall   $W_g(\beta)$ that led to the above recollapse
 at large volumes will be eventually counteracted (on the large volume side) by the deconfining, attractive 
 effect of    $W_\phi$.  In other words, we have here a situation where the wavefunction for  $\beta^0$ can
 tunnel through the potential barrier linked to  $W_g(\beta)$, to emerge on the inflationary side where
 it can lead to an exponentially expanding space.  [In picturesque terms the ``bottle'' of Fig. \ref{BouncingUniverses}
 should be thought of as leaking, by a quantum tunnel effect, on its bottom side, corresponding to large volumes.]
 Such models have been often considered in the
 literature, see, e.g., \cite{Vilenkin:1982de,Vilenkin:1986cy,Vilenkin:1987kf,Hartle:1983ai,Graham:1990jd}.
 The new aspect that our work might provide is a specific proposal for the ``initial wave function of the Universe'',
describing a sort of quantum storage ring within the upper part of the bottle of  Fig. \ref{BouncingUniverses}, corresponding to Planckian-size Universes.

\smallskip

\setcounter{equation}{0}\section{Summary and conclusions}

Let us summarize our main results:

\begin{enumerate}

\item  We have studied the  dynamics of a triaxially squashed 3-sphere (a.k.a. Bianchi IX model) in $D=4$, $\mathcal N=1$ supergravity by means of a new approach that gauge-fixes, from the start, the six degrees of freedom describing
possible local Lorentz rotations of the tetrad. In our approach, the only constraints to consider are:  the four susy constraints,
$ {\cal S}_A \approx 0$, the Hamiltonian constraint $H \approx 0$ and the three diffeomorphism constraints $H_i \approx 0$.

\item The quantization of this constrained Hamiltonian system has been done by first canonically quantizing the six bosonic
($g_{ij}(t)$), and twelve fermionic ($\psi^A_{\hat a}$) gauge-fixed degrees of freedom. The six metric degrees of freedom
are parametrized by means of three logarithmic scale factors $\beta^1 = -\log a$, $\beta^2 = -\log b$, $\beta^3 = -\log c$ measuring the squashing of the three-geometry, and by three Euler angles   $\varphi^1, \varphi^2, \varphi^3$ parametrizing
the orientation of the quadratic form $g_{ij}$ w.r.t. the Cartan-Killing metric $k_{ij}$  associated with the $SU(2)$ 
homogeneity symmetry of the squashed 3-sphere.  The canonical quantization of the gravitino leads (similarly
to the Ramond string) to a Spin(8,4) Clifford algebra for a suitably rescaled, and linearly transformed,  gravitino
zero-mode $\widehat \Phi^a_A$ ($a=1,2,3; A=1,2,3,4$), see Eq. (\ref{ACRPhi}).  This implies that the wavefunction of the Universe is a 64-dimensional spinor depending on six bosonic variables $\Psi_\sigma(\beta^a, \varphi^a)$, $(\sigma=1,\cdots,64)$.

\item The constraints are then imposed \`a la Dirac as restrictions on the state: $\widehat {\cal S}_A \, \vert{\bm\Psi}\rangle = 0 \qquad , \quad \widehat H \, \vert{\bm\Psi}\rangle = 0 \qquad , \quad \widehat H_i \, \vert{\bm\Psi}\rangle = 0 $.  Because
of our choice of parametrization of the Euler angles (connecting $g_{ij}(t)$ to the Cartan-Killing metric $k_{ij}$ associated
with the Bianchi IX structure constants), one finds that the diffeomorphism constraints are equivalent to requiring that
the wavefunction $\Psi_\sigma(\beta^a, \varphi^a)$ does not depend on the three Euler angles $ \varphi^a$. 
The remaining constraints are uniquely ordered by requiring that they be hermitian, and are found to consistently close,
see Eq. (\ref{SSLSH}).

\item The (rotationally reduced) susy constraints $\widehat {\cal S}_A \, \vert{\bm\Psi}\rangle = 0$ yields four simultaneous
Dirac-like equations, $\hat{\mathcal S}_A \Psi = \left( + \frac{i}2 \, \Phi_A^a \partial_{\beta^a} + \ldots \right) \Psi(\beta) = 0$ (where the $\Phi_A^a$'s are four separate triplets of $64 \times 64$ gamma matrices) describing the propagation of the
64-component spinorial wavefunction $\Psi(\beta)$  in the  three-dimensional space of the
logarithmic scale factors  $\beta^1 = -\log a$, $\beta^2 = -\log b$, $\beta^3 = -\log c$. The latter $\beta$-space is endowed
with the Lorentzian-signature metric $G_{ab}$,  Eq. (\ref{Gab}), induced by the kinetic terms of the Einstein-Hilbert action.
Each one of the Dirac-like equations  $\hat{\mathcal S}_A \Psi(\beta) =0$
forms a first-order symmetric hyperbolic system.  In addition, $ \Psi(\beta)$ satisfies initial-value-type constraints
in $\beta$-space, and a second-order Klein-Gordon-type Wheeler-DeWitt equation, $\hat H \Psi = \left(-\frac12 \, G^{ab} \partial_a \partial_b + \ldots \right) \Psi = 0$,  which is a consequence of the susy constraints.

\item The operatorial content of the   $\hat{\mathcal S}_A $'s and of $\widehat H$ reveals a hidden hyperbolic Kac--Moody
structure which confirms (and extends at the fully quantum level) previous conjectures about a correspondence
between supergravity and the dynamics of a spinning particle on an infinite-dimensional coset space (
$AE_3/K(AE_3)$ in our present context).  {\em The newest aspect of  this hidden Kac--Moody structure is the fact
that all the terms in $\widehat H$ that are quartic in fermions give rise to a ``squared-mass term" $\widehat \mu^2$ in
the Wheeler-DeWitt equation which commutes with all the operators   $\hat S_{12}, \hat S_{23} , \hat S_{31}, \hat J_{11} , \hat J_{22} , \hat J_{33}$ that are the building blocks of the quantum Hamiltonian $\widehat H$ }(and which are
second-quantized versions of the generators of
the Lie algebra $K(AE_3)$, {\it i.e.}   the maximally compact subalgebra of the hyperbolic Kac--Moody algebra $AE_3$).
In addition, the operator   $\widehat \mu^2$ is found to be expressible in terms of the square of  a certain (centered) 
fermion number $\widehat N_F -3 \equiv \widehat C_F := \frac12 \, G_{ab} \overline{\widehat  \Phi}^a \, \gamma^{\hat 1\hat 2\hat 3} \, \widehat\Phi^b $, which also commutes with all the operators  $\hat S_{12}, \hat S_{23} , \hat S_{31}, \hat J_{11} , \hat J_{22} , \hat J_{33}$.

\item  Representing the Clifford gravitino generators $\widehat \Phi^a_A$ in terms of two sets of annihilation and creation
fermionic operators $b_+^a $,  $ b_-^a$,  $\widetilde b_+^a  $,   $\widetilde b_-^a$ (where $\widetilde b \equiv b^{\dagger}$) allows one to decompose the
fermionic Hilbert space into various fermion-number levels, ${\mathbb H}_{(N^F_+ , N^F_-)}$. These correspond to
constructing the 64 states of ${\rm Spin} (8,4)$ by acting with a certain number of $b_{\pm}^{a\dagger}$ operators on 
 the empty state $\vert 0 \rangle_-$ (annihilated by the six $b_{\epsilon}^a$'s).  Actually, $\widehat N_F = \widehat C_F + 3$ counts the total number $N^F_+  + N^F_- $ of $b^{\dagger}_{\pm}$ operators.
The use of the ``chiral"  operators  $b_+^a $,  $ b_-^a$,  $\widetilde b_+^a$,   $\widetilde b_-^a$ allows one to write explicitly the susy constraints in a
convenient form, see Eq. (\ref{Seps}). One of the main new results of our approach is that we succeeded in
describing in detail the complete solution space, say   ${\mathcal V}^{(N_F)}$, of  the susy constraints  $\widehat {\cal S}_A \, \Psi(\beta) = 0$, at fermionic level  $N_F= N^F_+  + N^F_- $. 
It is a mixture of discrete-spectrum states (parametrized by a few constant parameters, and known in explicit form)
 and of continuous-spectrum states 
(parametrized by arbitrary functions entering some initial-value problem): ${\mathcal V}^{(0)} = V_1^{(0)}$ is one-dimensional; ${\mathcal V}^{(1)} = V_2^{(1)}$ is two-dimensional; ${\mathcal V}^{(2)} = V_3^{(2)} \oplus V_{1,\infty^2}^{(2)}$ is the direct sum of a three-dimensional space $V_3^{(2)}$ and of an infinite-dimensional space $V_{1,\infty^2}^{(2)}$ parametrized by one constant and two (real) functions of two (real) variables (together with an additional arbitrary constant); ${\mathcal V}^{(3)} = V_{2,\infty^2}^{(3)} \oplus V_{2,\infty^2}^{(3)}$ is the direct sum of two infinite-dimensional spaces, each one of which involves as free data two parameters and two functions of two variables. Moreover, when $4 \leq N_F \leq 6$, there is a duality under which ${\mathcal V}^{(N_F)}$ is one-to-one mapped to ${\mathcal V}^{(6-N_F)}$. 
Our results significantly differ from the conclusions of previous works. 

\item  At fermionic levels $2 \leq N_F \leq 4$, where there are  continuous-spectrum states,  we have explicitly described
the kind of plane-wave states they give rise to in the asymptotic far-wall limit where the various exponential potential
terms in the susy constraints are small.  In this regime, the wavefunction of the Universe looks like a spinorial
plane-wave that bounces between well-separated spin-dependent potential walls, probably leading
to a spinorial arithmetic chaos linked to the Weyl group of $AE_3$.

\item  A surprising result is that {\em supergravity predicts that the squared-mass term $\widehat \mu^2$ entering the Wheeler-DeWitt
equation is  negative over most of the fermionic Hilbert space}. This is a quantum effect (quartic in the fermions)
which has important implications for the dynamics of the geometry near the big bang, or big crunch,
(small-volume) singularity.  Indeed,  the corresponding contribution to the
energy density,  $\rho_4 \sim \mu^2  \left( \mathcal V_3 \right)^{-2}=   \mu^2  \left(  a \, b \, c \right)^{-2}$, dominates
the other contributions when the spatial volume  $ \mathcal V_3 = a b c$ tends toward zero. When considered at the classical level, such a negative $\rho_4$ necessarily leads to a halting of the collapse of the Universe, and makes
its volume bounce back toward larger volumes. We suggest, at the quantum level, to  require that the wavefunction
$\Psi(\beta)$ satisfy the corresponding quantum boundary condition to vanish for small volumes. When considering
a big crunch, this boundary condition is a kind of final-state boundary condition, that might be important for the
resolution of the information-loss problem in black hole evaporation. We also suggest that this quantum avoidance
of  zero-volume singularities would lead to a ``bottle effect'' between small-volume-Universes and large-volume
ones, and to a corresponding  storage-structure made of a discrete spectrum of  quantum states (starting with
the Planckian-size Universes described by the discrete susy states at levels $N_F=0$ and 1).

\end{enumerate}

\smallskip

Our results  open new perspectives that we hope to discuss in future work.  Among them, let us mention:
\begin{itemize}

\item[ (i)]  studying  the { \it quantum fermionic billiard} defined by the reflection of the plane-wave states discussed
above on the various potential walls; 

\item[(ii) ]  discussing the existence of a discrete set of quantum states confined
within the Lorentzian ``bottle"  associated with a negative eigenvalue of $\widehat \mu^2$, and their eventual link with
the classical periodic orbits in $\beta$ space; 

\item[(iii)] defining a norm on the solutions of the susy constraints;

\item[(iv)]  discussing the matching of our early Bianchi IX dynamics to a later inflationary era;

\item[(v)]  generalizing the $\mathcal N =1$, $D=4$ case considered here to more supersymmetric cases, and in particular to
 the   $\mathcal N =8$, $D=4$ case, or,  the  $\mathcal N =1$, $D=11$  one, where the relevant Kac--Moody algebra should be $E_{10}$; 
 
 \item[(vi)]   including the
effect of inhomogeneous modes on the dynamics of the spatial zero-modes considered above; 

\end{itemize}

\smallskip\noindent\emph{Acknowledgments.~} We thank  G. Bossard, P. Cartier,  P. Deligne, V. Kac, A. Kleinschmidt, V. Moncrief, and H. Nicolai  for informative discussions. Ph. S. reiterate its thanks to IHES for its kind hospitality; his work has been partially supported by ``Communaut\'e fran\c caise de Belgique -- Actions de Recherche concert\'ees'' and by IISN-Belgium (convention 4.4511.06).

\bigskip

\appendix

\setcounter{equation}{0}\section{ Summary of notation} \label{Varia}

\smallskip
To facilitate the reading let us recap below the definitions of the different variables parametrizing the metric degrees of freedom.

Scale factors of the metric:
\begin{equation} a = e^{-\beta^1} , \,  b = e^{-\beta^2}, \,  c = e^{-\beta^3},  \, a \, b \, c = e^{- (\beta^1 + \beta^2 + \beta^3)} \equiv e^{- \beta^0} \, .
 \end{equation}
 Note that the billiard limit  $\beta^0 \equiv \beta^1 + \beta^2 + \beta^3 \to + \infty$ corresponds to the small three-volume
 limit  $ a \, b \, c \to 0$.

 Diagonal metric components
 \begin{equation}
e^{-2\beta^1} \equiv a^2 \equiv \frac1x \, , \quad e^{-2\beta^2}  \equiv b^2 \equiv \frac1y\, , \quad e^{-2\beta^3} \equiv c^2 \equiv \frac1z \, .
\end{equation}

Let us note also the definitions
 \begin{equation} \beta_{1\,2}=\beta^1-\beta^2=-u\ ,\ \beta_{2\,3}=\beta^2-\beta^3\ ,\ \beta_{3\,1}=\beta^3-\beta^1=-v\\
\end{equation}
\begin{equation}\beta^0=\beta^1+ \beta^2+ \beta^3  \, \\ 
\end{equation}
\begin{equation}\label{Apxivar}
\xi^{\hat 0}=\frac {\sqrt{6}}2\,\beta^0\ ,\ \xi^{\hat 1}=\frac {\sqrt{2}}2\,\beta_{2\,3}\ ,\ \xi^{\hat 2}=\frac {\sqrt{6}}6(\beta_{1\,2}-\beta_{3\,1})\\
\end{equation}
\begin{equation}T=e^{\beta^0}\ ,\ X=e^{\beta_{2\,3}}\ ,\ Y=e^{\beta_{1\,2}-\beta_{3\,1}}\\
\end{equation}
\begin{equation}{\mathcal U}=-\frac 12 \coth \beta_{1\,2}=\frac 12 \coth u\ ,\ 
{\cal V}=-\frac 12 \coth \beta_{3\,1}=\frac 12 \coth v\\
\end{equation}

Let us also recall
\begin{eqnarray}  
&\theta=\det(\theta^{\hat \alpha}_\mu)=N\,\sqrt{g} \ , \   g=\det[(g_{i\,j})]=\det[(h^{\hat a}_i)]^2  = ( a \, b \, c )^2&\\
&\psi_{\hat \alpha}=g^{-\frac 14}\,\Psi_{\hat \alpha} \quad  , \quad \overline \Psi=i\,  \Psi^T   \gamma_{\hat 0}&\\
&\Phi^k=\gamma^{\hat k}\Psi^k\quad ,\quad \overline \Phi^k=-\overline\Psi^k\gamma^{\hat k}&
\end{eqnarray}

\begin{eqnarray}
 \gamma^{\hat 1}=
\left(
\begin{array}{cccc}
0  &   1&0&0   \\
1  &  0 & 0  &0\\
 0 & 0  & 0  &1\\
 0 & 0  &1   & 0
\end{array}
\right)\quad&,&\quad
\gamma^{\hat 2}=
\left(
\begin{array}{cccc}
-1  &  0&0&0   \\
0  &  1 & 0  &0\\
 0 & 0  & -1  &0\\
 0 & 0  &0  & 1
\end{array}
\right)\quad, \\
 \gamma^{\hat 3}=
\left(
\begin{array}{cccc}
0  &  0&0&-1   \\
0  & 0 & 1  &0\\
 0 &1  & 0  &0\\
 -1 & 0  &0  & 0
\end{array}
\right)\quad&,&\quad\gamma^{\hat 0}=
\left(
\begin{array}{cccc}
0  &  0&0&1   \\
0  & 0 &-1  &0\\
 0 &1  & 0  &0\\
 -1 & 0  &0  & 0
\end{array}
\right)\quad,\\
\gamma^5 =\gamma^{\Hat 0}\gamma^{\Hat 1}\gamma^{\Hat 2}\gamma^{\Hat 3}&=&
\left(
\begin{array}{cccc}
0  &  -1&0&0   \\
1  & 0 &0  &0\\
 0 &0  & 0  &1\\
 0& 0  &-1  & 0
\end{array}
\right) \qquad .
\end{eqnarray}

\setcounter{equation}{0}
\section{Characteristics of fermionic subspaces}\label{fermionicspaces}
The  following two tables summarize the dimensions and  eigenvalues of  the quadratic fermionic
 operators $\sim \bar \psi \psi$ that play a basic  r\^ole in underlying the quantum dynamics discussed in the text.

The first table displays the decomposition of the 64-dimensional Spin(8,4) spinorial space into the     
irreducible subspaces defined in Sec. 10, as well as into eigensubspaces of the fermion number operators $\widehat N_F$ and $\widehat N^F_+$. It provides: the dimensions of these subspaces,  the eigenvalues of 
$\widehat N_F$, of its centered version $\widehat C_F=\widehat N_F-3$,  of the squared-mass operator $\widehat \mu^2= 1/2- (7/8) \widehat C_F^2$ [see Eqs (\ref{mu2} --\ref{CFNF})], and of the partial $\widetilde b_+$ number operator $\widehat N^F_+$.

{\small
\begin{center} 
\begin{tabulary}{1000pt}{|C|C|C|C|C|}
\hline 
&&&& \\
dim & $N_F$ &$C_F $&$\mu^2$ &$N^F_+$  \\ 
&&&& \\ \hline \hline
&&&& \\
1 &0 &$-3$ &$-\frac{59}8$ &0  \\ 
&&&& \\ \hline
&&&& \\
$3 \oplus 3$ &1 &$-2$ &$-3$ &$1 \oplus 0$  \\ 
&&&& \\ \hline
&&&& \\
$6 \oplus 3 \oplus 3 \oplus 3$ &2 &$-1$ &$-\frac38$ &$1 \oplus2 \oplus 1 \oplus 0$  \\ 
&&&& \\ \hline
&&&& \\
$10 \oplus 10$ &3 &0 &$\frac12$ &$\left(2\vert_{_9} \oplus 0\right)\oplus \left(1\vert_{_9}\oplus 3\right)$  \\ 
&&&& \\ \hline
&&&& \\
$3 \oplus 3 \oplus 3 \oplus 6$ &4 &1 &$-\frac38$ &$3 \oplus 1 \oplus 2  \oplus 2$  \\ 
&&&& \\ \hline
&&&& \\
$3 \oplus 3$ &5 &2 &$-3$ &$3 \oplus 2$  \\ 
&&&& \\ \hline
&&&& \\
1 &6 &3 &$-\frac{59}8$ &3  \\
&&&& \\
\hline
\end{tabulary}
\end{center}
\label{table1}}


The second table  provides the eigenvalues of the Kac--Moody-related operators $\widehat J_{11}$, $\widehat S_{12}$ and $\widehat S^2_{12}$ that are the building blocks of the susy constraints and of the Hamiltonian. It displays how
these eigenvalues are split   along the irreducible     subspaces of the total  64-dimensional fermionic space
defined in Sec. 10. Let us notice that the operators $\widehat J_{11}$, $\widehat J_{22}$, $\widehat J_{33}$ and $\widehat S_{12}$ commute with $\widehat N_F$ and $\widehat N^F_+$, while only the squares of the other spin operators $\widehat S_{23}$ and $\widehat S_{31}$ commute with $\widehat N_F$. The finer subspace decompositions of the 15-dimensional spaces $N_F=2$ or 4 in 6 + 3+ 3+ 3-dimensional subspaces provide invariant subspaces only for  $\widehat J_{11}$, $\widehat J_{22}$, $\widehat J_{33}$ and  $\widehat S_{12}^2$.
{ \small
\begin{center} 
\hglue-2cm
\begin{tabulary}{1000pt}{|C|C|C|C|C|}
\hline 
&&&& \\
dim & $N_F$ &$  J_{11}$ &$  S_{12}$ &$S_{12}^2$ \\ 
&&&& \\ \hline \hline
&&&& \\
1 &0 &$\left(-\frac12\right)$ &$(0)$ &$(0 )$ \\ 
&&&& \\ \hline
&&&& \\
$3 \oplus 3$ &1 &$(1,-1 \vert_2) \oplus (1,-1 \vert_2)$ &$\left( -\left.\frac12\right \vert_2 , \frac32 ,\right) \oplus \left(-\frac32,\left.\frac12\right \vert_2  \right)$ &$\left( \frac94 , \frac12 \bigl\vert_{_2} \right) \oplus \left( \frac94 , \frac12 \bigl\vert_{_2} \right)$ \\ 
&&&& \\ \hline
&&&& \\
$6 \oplus 3 \oplus 3 \oplus 3$ &2 &$\left( -\frac32 \bigl\vert_{_3} , \frac12 \bigl\vert_{_2}, \frac52  \right) \oplus \left( -\frac32 , \frac12 \bigl\vert_{_2} \right)   $ &$\left( -2 \vert_2 , -1 \vert_1 , 1 \vert_2 , 0 \vert_5 , 2 \vert_2 \right)$ &$(0 \vert_4,4 \vert_2 ) \oplus(1 \vert_3)$ \\ 
&&&& \\
&&$\oplus \left( -\frac32 , \frac12 \bigl\vert_{_2} \right) \oplus \left( -\frac32 , \frac12 \bigl\vert_{_2} \right)$ &$\oplus (1,-1 \vert_2)$ &$\oplus (0,4 \vert_2 )\oplus (1 \vert_3)$ \\
&&&& \\ \hline
&&&& \\
$10 $ &3 &$\left((-2 \vert_2 , 0 \vert_5 , 2 \vert_2)\oplus 0\right) $ &$\left(\left( -\frac52 , -\frac12 \bigl\vert_{_4}  , \frac32 \bigl\vert_{_4}\right)\oplus  -\frac12 \right )$ &$\left(\left( \frac14 \bigl\vert_{_4}, \frac94 \bigl\vert_{_4} , \frac{25}4  \right) \oplus\frac 14\right) $ \\ 
&&&& \\
$\oplus 10$&&$\oplus\left (-2 \vert_2 , 0 \vert_5 , 2 \vert_2)\oplus 0\right)$&$\oplus\left( \left( -\frac32 \bigl\vert_{_4} , \frac12 \bigl\vert_{_4} , \frac52 \right)\oplus \frac 12\right)$ &$\oplus\left(\left( \frac14 \bigl\vert_{_4}, \frac94 \bigl\vert_{_4} , \frac{25}4  \right) \oplus \frac 14\right)$\\
&&&& \\ \hline
&&&& \\
$3 \oplus 3 \oplus 3 \oplus 6$ &4 &$\left(  -\frac12 \bigl\vert_{_2},\frac32  \right)\oplus \left( -\frac12 \bigl\vert_{_2},\frac32  \right)  $ &$(-1 ,  1 \vert_2)$ &$(1 \vert_3)\oplus (1 \vert_3) $ \\ 
&&&& \\
&&$\oplus \left( -\frac12 \bigl\vert_{_2}, \frac32  \right)\oplus \left( -\frac52 , \frac32 \bigl\vert_{_3} , \frac12 \bigl\vert_{_2} \right)$ &$\oplus (-2 \vert_2 , -1 \vert_2 , 1 \vert_1 , 0 \vert_5 , 2 \vert_2)$ &$\oplus (4 \vert_2 , 0) (4 \vert_2 ,0 \vert_4)$ \\
&&&& \\ \hline
&&&& \\
$3 \oplus 3$ &5 &$\left(-1 , \frac12 \bigl\vert_{_2} \right) \oplus \left( -1 , \frac12 \bigl\vert_{_2} \right)$ &$\left( - \frac12 \bigl\vert_{_2}  , \frac32 \right)\oplus \left( -\frac32 , \frac12\bigl\vert_{_2} \right)$ &$\left( \frac94 , \frac14 \bigl\vert_{_2} \right) \oplus \left( \frac94 , \frac14 \bigl\vert_{_2} \right)$ \\ 
&&&& \\ \hline
&&&& \\
1 &6 &$(+\frac12)$ &$(0)$ &$(0)$ \\
&&&& \\
\hline
\end{tabulary}
\end{center}
\label{table2}}


\setcounter{equation}{0}
\section{$L^C_{AB}$ operator components}\label{AppLform}

When working in the chiral basis ({\it i.e.}  replacing the original Majorana indices $A,B,C = 1,2,3,4$ by a pair of indices
$\epsilon, \widetilde \epsilon$ referring, on the model of  (\ref{bbd}) to  the combinations $+=1 + i \, 2$, $-=3 -  i \, 4$,
$\widetilde +=1 - i \, 2$, $\widetilde -=3 +  i \, 4$) 
 the operators $L^C_{AB}$ occurring in our basic anticommutation relations Eq.(\ref{SSLSH}) read 
 (with $\epsilon, \sigma, \rho = \pm$) : 

\begin{eqnarray}
  L_{\epsilon\sigma}^\rho &= &-i\left[  \mu_k \, b_{\epsilon}^{k} \, \delta_{\epsilon}^{\sigma} \, \delta_{\epsilon}^{\rho} + \frac 12 \left( \overset{(1)}{\rho}_k - \overset{(2)}{\rho}_k \right) \left( b_{-\epsilon}^k \, \delta_{\epsilon}^{-\sigma} \, \delta_{\epsilon}^{\rho} + b_{\epsilon}^k \, \delta_{\epsilon}^{-\sigma} \, \delta_{\epsilon}^{-\rho} \right) + \, \nu_k \, b_{-\epsilon}^k \, \delta_{\epsilon}^{\sigma} \, \delta_{\epsilon}^{-\rho}\right] \\
 L_{\epsilon\sigma}^{\tilde \rho} &= &0 \\
  L_{\epsilon\tilde\sigma}^\rho &= &\frac i 2\left[\mu_k \, \tilde b_{\sigma}^k \, \delta_{\epsilon}^{\sigma} \, \delta_{\epsilon}^{\rho} + \left( \overset{(3)}{\rho}_k - \overset{(2)}{\rho}_k \right) \tilde b_{-\sigma}^k \, \delta_{\epsilon}^{\sigma} \, \delta_{\epsilon}^{-\rho}  
 +\nu_k \, \tilde b_{\sigma}^k \, \delta_{\epsilon}^{-\sigma} \, \delta_{\epsilon}^{-\rho} + \left( \overset{(1)}{\rho}_k - \overset{(3)}{\rho}_k \right) \tilde b_{-\sigma}^k \, \delta_{\epsilon}^{-\sigma} \, \delta_{\epsilon}^{\rho}\right] \qquad .
\end{eqnarray}

\setcounter{equation}{0}\section{Explicit tensor components}\label{Apptauform}

The components of the completely symmetric, traceless object $\tau_{abc}$ introduced in Eq. (\ref{rhoklm}) are given by :
$$ 
  \tau_{111}=\frac{i (x (y+z)-2 y z)}{4 (x-y) (x-z)}\ ,\ 
  \tau_{222}=-\frac{i
   (x (y-2 z)+y z)}{4 (x-y) (y-z)}\ ,\ 
    \tau_{333}=-\frac{i (y z+x (z-2 y))}{4 (x-z) (z-y)}\ ,$$
    $$\tau_{112}=-\frac{i (x
   (y-4 z)+3 y z)}{20 (x-y) (x-z)}\ ,\ \tau_{113}=\frac{i (4 x y-3 z y-x
   z)}{20 (x-y) (x-z)}\ ,  $$
   $$
   \tau_{122}=\frac{i (x (y+3 z)-4 y z)}{20 (x-y)
   (y-z)}\ ,\  \tau_{223}=-\frac{i (4 x y-z y-3 x
   z)}{20 (x-y) (y-z)}\ ,$$
   $$\tau_{133}=\frac{i (x (3 y+z)-4
   y z)}{20 (x-z) (z-y)}\ ,\ 
  \tau_{233}= \frac{i (3 x y+z y-4 x z)}{20 (x-z)
   (z-y)}\ ,$$
$$\tau_{123}=-\frac{i}{20}$$

\setcounter{equation}{0}\section{$\mathbb H_{(1,1)_S}$ space : solving the constraint equations}  \label{solvingconstraints}

\smallskip

Let us show why the general solution of the constraint equations  (\ref{constraints}) (arising from the $2+1$ decomposition of the Maxwell-like equations (\ref{divk}), (\ref{rotk})
for the symmetric tensor $k_{pq}$ arising at level $N_F=2$) can be parametrized by {\it two} arbitrary functions of two variables (together with an additional constant). 

A preliminary useful observation is that, when re-expressed in the Lorentzian coordinates (\ref{Lor0}), (\ref{Lor1}), (\ref{Lor2}),
some components of $\tau_{klm}$ vanish, namely~: 
\begin{equation}
\tau_{\hat p \hat 0 \hat 1} = 0 = \tau_{\hat p \hat 0 \hat 2} \qquad . 
\end{equation}
As a consequence the constraint $\mathcal C_{\hat 0}$ only involves $k_{\hat 0 \hat 1}$ and $k_{\hat 0 \hat 2}$
\begin{equation}
\partial_{\hat 1} \, k_{\hat 0 \hat 2} - \partial_{\hat 2} \, k_{\hat 0 \hat 1} = (\partial_{\hat 1}  \gamma) \, k_{\hat 0 \hat 2} - (\partial_{\hat 2}  \gamma) \, k_{\hat 0 \hat 1} 
\end{equation}
where $\gamma = 2i (\rho^{(1)} +  \mu - \alpha)$ [see Eqs (\ref{alphak}) -- (\ref{rhoklm})]. 

Therefore, the general solution for  $k_{\hat 0 \hat 1}$, $k_{\hat 0 \hat 2}$ (considered at some given 
initial ``time'' $\xi^{\hat 0}$)  can be parametrized as :
\begin{equation}\label{kzerok}
k_{ \hat 0 \hat p} = e^{\gamma} \, \partial_{\hat p} \, K [\xi^{\hat 1} , \xi^{\hat 2}] \qquad ,
\end{equation}
where $K [\xi^{\hat 1} , \xi^{\hat 2}]$ is a first arbitrary function of two variables.

In this section, it will be often useful to give special names to the following exponential form of the
Lorentzian coordinates  $\xi^{\hat 0} , \xi^{\hat 1} , \xi^{\hat 2}$:
\begin{eqnarray}
T &= &e^{\frac{\sqrt 6 \, \xi^{\hat 0}}{3}} = e^{(\beta^1 + \beta^2 + \beta^3)} \nonumber \qquad ,\\
X &= &e^{\sqrt 2 \, \xi^{\hat 1}} = e^{(\beta^2 - \beta^3)} \nonumber \qquad ,\\
Y &= &e^{\sqrt 6 \, \xi^{\hat 2}} = e^{(2\beta^1 - \beta^2 - \beta^3)}\qquad . \nonumber
\end{eqnarray}
In terms of these exponentiated Lorentzian coordinates, the explicit expression of the integrating factor $ e^{\gamma}$
entering  the parametrization (\ref{kzerok}) of   $k_{\hat 0 \hat 1}$, $k_{\hat 0 \hat 2}$ reads
\begin{equation}
\label{egam}
e^{\gamma} : =\frac{ (X-Y)^{3/8}(1-X\,Y)^{3/8}(1-X^2)^{3/8}}{ X^{3/4}\,Y^{3/8} } e^{-\left(\frac1{2\,T\,Y^{2/3}}+\frac{(1+X^2)Y^{1/3}}{2\,T\,X}\right)} \qquad .
\end{equation}

Let us now consider the remaining constraints $\mathcal C_{\hat p}$, and   $\mathcal C'_{\hat p}$,  $\hat p = 1,2$.  Because of the vanishing, indicated above,
of several relevant components of the tensor  $\tau_{klm}$,  one finds that the two constraints   $\mathcal C_{\hat p}$
do not involve  $k_{\hat 0 \hat 0}$. [Because of the identity (\ref{identityconstr}) between the constraints, 
the two constraints $\mathcal C'_{\hat p}$ will provide a way to consistently determine  $k_{\hat 0 \hat 0}$ once
we will have determined the ``spatial'' components $k_{\hat p \hat q}$ (with $\hat p, \hat q = 1,2$) of the tensor $k_{\hat a \hat b}$ (see below).] 
 The essential issue is then to parametrize the general solution of the two constraints 
 $\mathcal C_{\hat p}$,  $\hat p = 1,2$, viewed as equations for the three unknowns $ k_{\hat 1 \hat 1} ,  k_{\hat 1 \hat 2} ,  k_{\hat 2 \hat 2} $. There are many ways of doing so.  By assuming that some linear combination of these three
 components is known, the constraints  $\mathcal C_{\hat p}=0$ will give two equations for two other (linearly-independent)
 combinations of the three  $k_{\hat p \hat q}$. Surprisingly, we found that the so-obtained system of
 two equations for two unknowns can be elliptic or hyperbolic, depending on the choice of combination
 that is assumed to be known. Among possible choices, we found one which has nice properties. [We shall
 see below that these special properties are linked to a corresponding special arrangement of the characteristic
 lines entering the initial-constraints system with respect to the symmetry walls.]
 It consists in taking as second arbitrary function parametrizing the solution of the constraints the particular
 combination
 \begin{equation}\label{arbH}
  k_{\hat 2 \hat 2} - 3  k_{\hat 1 \hat 1} =  H (\xi^{\hat 1} , \xi^{\hat 2})
\end{equation}
 Using this combination to eliminate $ k_{\hat 2 \hat 2}$, the two constraints $\mathcal C_{\hat p}=0$ then give
 a linear system of equations for  $ k_{\hat 1 \hat 1} $ and $ k_{\hat 1 \hat 2} $, with source terms depending
 on the given functions  $ H (\xi^{\hat 1} , \xi^{\hat 2})$ and  $ K (\xi^{\hat 1} , \xi^{\hat 2})$ 
 (that enter the $k_{ \hat 0 \hat p}$'s).  It is convenient to rewrite this system in terms of suitably rescaled
 versions of all the $k_{\hat a \hat b}$'s.  Namely, we set
 \begin{equation}\label{kelkt}
k_{\hat a \hat b} \equiv  e^{\lambda} \, \tilde k_{\hat a \hat b} \qquad \hat a , \hat b = \hat 0, \hat 1 , \hat 2
\end{equation}
 where the rescaling factor $ e^{\lambda}$ is defined as
 \begin{equation}  \label{elambda}
 e^{\lambda} := e^{\gamma} \, \frac{X^{1/2} (X-Y)^{1/2} \, (1-X\,Y)^{1/2}}{(1-X^2) \, Y^{1/3}}
 \end{equation}
 where $ e^{\gamma}$ is the integrating factor (\ref{egam}) introduced above.
 
 In terms of such rescaled versions of the  $k_{\hat a \hat b}$'s (and a correspondingly rescaled version of $H$),
 one gets the following system of two equations for  $ \tilde k_{\hat 1 \hat 2} $, and $ \tilde k_{\hat 1 \hat 1} $:
 \begin{eqnarray}
\label{DEPD1}
\partial_{\hat 1} \, \tilde k_{\hat 1 \hat 2} &-& \partial_{\hat 2} \, \tilde k_{\hat 1 \hat 1} =  s_1 \\
\label{DEPD2}
- \left(\partial_{\hat 2} +  \partial_{\hat 2} (\mu + \lambda) \right)  \, \tilde k_{\hat 1 \hat 2} &+& 3\, \left( \partial_{\hat 1} +  \partial_{\hat 1} (\mu + \lambda) \right) \, \tilde k_{\hat 1 \hat 1} =  s_2 
\end{eqnarray}
Here, the new function $\mu$ is defined as
\begin{equation} \label{emu}
e^{\mu} = e^{-\gamma} \, \frac{(X-Y)^{1/2} \, (1-XY)^{1/2}}{X^{1/2} \, Y^{2/3}} \ ,
\end{equation}
while the (known) source terms appearing on the r.h.s.'s are given by
\begin{equation}s_1=  - \frac{\sqrt 6}3 \, \tilde k_{\hat 0 \hat 2} + c \, \tilde H
\end{equation}
\begin{equation}s_2=   \frac{\sqrt 6}3 \, \tilde k_{\hat 0 \hat 1} + (r \, \tilde H - \partial_{\hat 1} \, \tilde H)
\end{equation}
where
\begin{equation}\label{fctc}
c = \frac{Y(1+X^2) + 2X (1-2 Y^2)}{2 \sqrt 6 \, (X-Y) (1-XY)} = \frac{\sqrt 6}{12} \, \frac{x(y+z) + 2y z - 4x^2}{(x-y)(x-z)}
\end{equation}
and
\begin{eqnarray}\label{fctr}
r &= &- \frac{4X(1+X^2)(1+Y^2) + Y(1-18X^2-X^4)}{2\sqrt 2 \, (1-X^2) (X-Y) (1-XY)} \nonumber \\
&= &-\sqrt 2 \left( \frac{4 (x^2 + yz) (y+z) + x(y^2 + z^2) - 18 \, xyz}{4(z-y)(z-x)(y-x)}\right) \qquad .
\end{eqnarray}

The system (\ref{DEPD1}),  (\ref{DEPD2}), can be viewed as a Dirac equation for the ``spinor''  
$\psi = ( \tilde k_{\hat 1 \hat 2},  \sqrt{3} \, \tilde k_{\hat 1 \hat 1})^T $, with source $s=(s_1, - \frac{1}{\sqrt{3}} s_2)^T$; namely
\begin{equation} \label{Dirac}
D_{\mu+\lambda} \, \psi = s
\end{equation}
with a Dirac operator (coupled to a ``connection'' $\omega_{\hat p} = \partial_{\hat p} \omega$ given by the gradient of  a function $\omega$) of the general form 
\begin{equation}
\label{Domega}
D_\omega=  \begin{pmatrix} \partial_{\hat 1}  &- \frac{1}{\sqrt{3}} \partial_{\hat 2}  \\ + \frac{1}{\sqrt{3}} \left(  \partial_{\hat 2} +  \omega_{\hat 2} \right) &  - \left( \partial_{\hat 1} +  \omega_{\hat 1}  \right)   \end{pmatrix}
\end{equation}
with the function $\omega$ given (in the  case of our specific Eq. (\ref{Dirac})) by the sum $\omega =  \mu+\lambda$.

It happens that, in our case, the Dirac-like equation (\ref{Dirac})) has special properties that allows one to control its
solutions, and even to explicitly compute its relevant Green's function. Let us start by noting that it is a Dirac equation
of the hyperbolic (rather than elliptic) type.  Indeed, if we absorb the factors $\frac{1}{\sqrt{3}}$ in a rescaling
of  $\xi^{\hat 2}$  (say $\xi^{2'} := \sqrt 3 \, \xi^{\hat 2}$)   the derivative terms in our Dirac equation take the form
$\gamma^{\hat 1}   \partial_{\hat 1} +  \gamma^{ 2'}   \partial_{2'}$, where the $ 2 \times 2$ matrices $\gamma^{p'}$
are given in terms of the standard Pauli matrices $\sigma_i^{\rm Pauli}$ by
\begin{equation}\gamma^1 \equiv \gamma^{ 1'}=  \sigma_3^{\rm Pauli} \ ,  \  \gamma^{2'}  = - i \,   \sigma_2^{\rm Pauli}
\end{equation}
This shows that these two gamma matrices define a Clifford algebra of Lorentzian signature:
$\gamma^{p'}  \gamma^{q'} +  \gamma^{q'}  \gamma^{p'} = 2 \, \eta^{p' q'}$, with $\eta^{p' q'}=  {\rm diag} (+1, -1)$.
The Dirac equation (\ref{Dirac}))
is therefore (as any Lorentzian Dirac equation) equivalent to a symmetric-hyperbolic first-order system for the unknowns
$\psi = ( \tilde k_{\hat 1 \hat 2},  \sqrt{3} \, \tilde k_{\hat 1 \hat 1})^T $, with known sources 
$s=(s_1, - \frac{1}{\sqrt{3}} s_2)^T$.  In addition, as the gamma matrices $\gamma^{p'}$ are real, we are discussing
here a real Dirac equation. From the point of view of looking for solutions of the
constraints  (\ref{DEPD1}),  (\ref{DEPD2}) , we can think of the 2-plane 
$\xi^{\hat 1},  \xi^{\hat 2}$ as being a 2-dimensional Lorentzian spacetime [though, with respect to the $G_{ab}$ metric
in $\beta$ space, it is a spacelike hypersurface, which we are using as initial Cauchy slice.]
These remarks suffice to prove that, locally, a general solution of the constraints [{\it i.e.}  of Eq.   (\ref{Dirac}))] is determined
by the two arbitrary functions of two variables $H, K$ (which enter the source term $s$), modulo some ``initial conditions"
in the auxiliary 2-dimensional Lorentzian space $\xi^{\hat 1},  \xi^{\hat 2}$ (which might involve arbitrary functions
of {\it one} variable, but no other arbitrary functions of two variables).

Surprisingly, it is possible to be more precise, and to solve {\it globally} the Dirac equation (\ref{Dirac})) when incorporating
boundary conditions that are natural for our problem. This arises because two remarkable facts happen to be true:
(i)  The first-order system (\ref{Dirac}))  is directly related to the well-known second-order Euler-Poisson-Darboux (EPD) equation; and (ii)  the characteristics lines, as well as the singular line, of this auxiliary EPD equation coincide with the
trace of the symmetry walls $\beta^a=\beta^b$ on the 2-plane $\xi^{\hat 1},  \xi^{\hat 2}$. Let us briefly explain these facts,
and how they allow one to solve Eq. (\ref{Dirac})). 

Let us start by exhibiting the connection of the characteristic lines of our Dirac equation (\ref{Dirac})) to the symmetry walls.
This follows simply from the fact that we have seen above that   
$\xi^{1'} :=  \xi^{\hat 1}$  and $\xi^{2'} := \sqrt 3 \, \xi^{\hat 2}$ were Lorentzian coordinates, so that the corresponding
null coordinates read
\begin{equation}\label{uvvar}
u = \frac{\xi^{\hat 1} - \sqrt 3 \, \xi^{\hat 2}}{\sqrt 2}=\beta^2-\beta^1 \qquad , \qquad v = \frac{\xi^{\hat 1} + \sqrt 3 \, \xi^{\hat 2}}{\sqrt 2}=\beta^1-\beta^3
\end{equation}
This result shows that the two symmetry walls   $\beta^1=\beta^2$  and  $\beta^1=\beta^3$ are characteristic for
the Dirac equation. As for the third symmetry wall,  $\beta^2=\beta^3$,  it enters our Dirac equation through a singularity
of the connection terms   $\omega_{\hat p} = \partial_{\hat p} \omega =   \partial_{\hat p} (\mu + \lambda)$. 
Indeed, by inserting the (several) changes of variables introduced above, one finds that
\begin{equation}\label{keyrelUV}
e^{- \omega} = e^{-(\mu + \lambda)} = \frac12 \left( \coth \left[    u\right] + \coth \left[  v\right]\right) \qquad .
\end{equation}
This formula shows that the gradients of $\omega$ have (pole-like) singularities not only when either $u$ or $v$ vanish,
but also along the line where $u=-v$, {\it i.e.} , in view of the definitions (\ref{uvvar}) of $u$ and $v$,  along the line where $\beta^2=\beta^3$. Summarizing, we have the following correspondences between  the symmetry walls 
[which are singular lines for the Dirac equation (\ref{Dirac}))] and some special lines in the Lorentzian 2-plane 
 $\xi^{\hat 1},  \xi^{\hat 2}$ [coordinatized by the null coordinates (\ref{uvvar})]
\begin{equation}(u=0) \leftrightarrow \left(   \beta^1=\beta^2 \right)  , \, (v=0) \leftrightarrow \left(   \beta^1=\beta^3 \right)  , \, (u+v=0) \leftrightarrow \left(   \beta^2=\beta^3 \right) \, .
\end{equation}
What is remarkable in these simple correspondences is not that the symmetry walls are singular lines for our Dirac 
equation (indeed, they were singular planes already in the original susy constraints), but that their traces
on the  Lorentzian 2-plane   $\xi^{\hat 1},  \xi^{\hat 2}$ have a special orientation with respect to the null
coordinates  (\ref{uvvar}).  Let us henceforth consider that we work within our canonical chamber $a$, 
{\it i.e.}  $\beta^1 \leq \beta^2 \leq \beta^3$.  This chamber has two boundaries: the null boundary $u=0$ ($  \beta^1=\beta^2$),
and the timelike boundary $u+v=0$ ($ \beta^2=\beta^3 $).  In rescaled Lorentzian coordinates 
$\xi^{1'} :=  \xi^{\hat 1}$, $\xi^{2'} := \sqrt 3 \, \xi^{\hat 2}$, these two boundaries are, respectively, the diagonal
$\xi^{1'} = \xi^{2'} $ , and the vertical axis $ \xi^{1'} =0$.  If we give ourselves some boundary conditions for $\psi$
on these boundaries, and if we can construct a Green's function $\mathcal G$ (satisfying these boundary conditions) for our
Dirac equation, we can conclude that  the convolution $ \mathcal G  \star s$ of the Green's function with the sources $s$
will define the (unique) solution $\psi$ satisfying the boundary conditions.  [We are assuming here for simplicity that
the data $H, K$ have a compact support, away from the boundaries, so that the source $s$ is regular and
compact-supported.]

Natural boundary conditions for $\psi$ are obtained as follows.  A local analysis, near a symmetry wall
 $\beta_{ab} \equiv \beta^a - \beta^b$  of the  solutions of  the 
$\mathbb H_{(1,1)_S}$-sector susy constraints shows that the general solution is a superposition of two types
of solutions: a  {\it regular} solution where the symmetric tensor $k_{pq}$  behaves like $\beta_{ab}^{+ 3/8}$ as $\beta_{ab} \to 0$, and a {\it singular} solution where   $k_{pq}$  behaves like $\beta_{ab}^{- 5/8}$. As was already mentioned above, 
the fact that there exist conserved Dirac-like currents that are bilinear in the wavefunction ({\it i.e.}  bilinear in  $k_{pq}$ for
the present case) suggests that we should impose that  $k_{pq}$ is square integrable when integrated over a spacelike
section in $\beta$-space (say $\int    d \xi^{\hat 1}  d  \xi^{\hat 2} \sim \int du \, dv$). 
 Imposing such a square-integrability requirement leads us to keeping, at {\it each} symmetry wall, only the solutions
 where  $k_{pq}$  behaves like  $\beta_{ab}^{+ 3/8}$.  We shall use this restriction in solving our Dirac-like system, and,
 in particular, in  constructing a Green's function incorporating these boundary conditions.
 
We succeeded in  constructing a Green's function $\mathcal G$ for our Dirac-like system, 
incorporating such boundary conditions, in the following way.  As the source $s$ has two independent components,
we can separately consider the two problems where one of the two components of $s$ vanish (and, when looking for
a Green's function, where the remaining component is a $\delta$ function). Let us first consider the case where $s_1=0$.
In that case, the explicit form of  the first equation of our system, namely Eq. (\ref{DEPD1}), says that there exists a scalar
field $\Phi$ such that 
\begin{equation}\label{fgPhi}
 \tilde k_{\hat 1 \hat 2}  = \partial_{\hat 2} \, \Phi \qquad \mbox{and} \qquad  \tilde k_{\hat 1 \hat 1}  = \partial_{\hat 1} \, \Phi \qquad .
\end{equation}
Inserting this form in the second equation of our system,  namely Eq. (\ref{DEPD2}),  leads to a second-order equation
for the potential $\Phi$.  This second-order equation remarkably happens to be equivalent to an EPD equation. This equivalence
occurs because the function $e^{- \omega} = e^{- \mu + \lambda}$ happens to enjoy the following special separation property:  
\begin{equation}\label{keyrel}
e^{- \omega} = e^{-(\mu + \lambda)} = \frac12 \left( \coth \left[    u\right] + \coth \left[  v\right]\right) \equiv  {\mathcal U}(u) +{\mathcal V}(v)  \qquad .
\end{equation}
In the last equation, we have introduced the new null coordinates  $\cal U$ and $\cal V$, defined as :
\begin{equation} \label{UV}
{\mathcal U} :=\frac12 \coth \left[  u \right] = \frac12 \left( \frac{X+Y}{X-Y} \right) \quad , \quad
{\mathcal V} :=\frac12 \coth \left[ v \right] = \frac12 \left( \frac{XY + 1}{XY - 1} \right) \qquad ,
\end{equation}
In terms of these  transformed null coordinates, the equation for the potential $\Phi$ becomes
 \begin{equation}
\partial_{\cal U} \frac 1{({\cal U}+{\cal V})}\,\partial_{\cal V} \Phi+\partial_{\cal V}\frac 1{ ({\cal U}+{\cal V})}\,\partial_{\cal U} \Phi=\frac{4\,s_2}{3\, (1-4\,{\cal U}^2)(1-4\,{\cal V}^2)}
\end{equation}
Let us recall that the general form of the homogeneous EPD equation is
\begin{equation}
\left[ \, \partial_{\mathcal U} \, \partial_{\mathcal V} + \frac{m}{\mathcal U+\mathcal V} \, \partial_{\mathcal U} + \frac{n}{\mathcal U+\mathcal V} \, \partial_{\mathcal V}  \right] \, F = 0
\end{equation}
It is easily seen that the differential operator appearing in the equation for $\Phi$ is of the EPD type with $m=n= -\frac12$.
As we are in the case where $m=n$, one can explicitly compute the Green's fuction for this differential operator. 
This is best seen by rescaling the potential $\Phi$  by a factor $ ({\mathcal U}+{\mathcal V})^{1/2} $. Namely, if we set
\begin{equation}
\Phi = ({\mathcal U}+{\mathcal V})^{1/2} \, {\widetilde \Phi} \qquad
\end{equation}
we find that the differential operator acting on $\widetilde \Phi$ reads
\begin{equation}\left( \partial_{\mathcal U} \, \partial_{\mathcal V} -  \frac34   \frac1{({\mathcal U}+{\mathcal V})^2} \right) {\widetilde \Phi} \qquad .\
\end{equation}
In terms of the new null coordinates  ${\mathcal U}, {\mathcal V}$  the two boundaries where we can impose boundary conditions are
\begin{equation} \left(   \beta^1=\beta^2 \right)   \leftrightarrow  ( {\mathcal U} = \infty) , \,   \left(   \beta^2=\beta^3 \right)  \leftrightarrow  ( {\mathcal U} + {\mathcal V} =0 )
 \end{equation}
 In the auxiliary 2-dimensional Minkowski space spanned by the null coordinates  ${\mathcal U}, {\mathcal V}$, we
 can think of the first boundary $ {\mathcal U} = \infty$ as being past null infinity ( $ \mathcal I^-$), while the second
 boundary $ {\mathcal U} + {\mathcal V} =0 $ would be the spatial origin. [In terms of  auxiliary ``time'' and ``radius''
 coordinates ${\cal T}$ and ${\cal R} $, with $ {\mathcal U} = {\cal R}  - {\cal T}$, and $ {\mathcal V} = {\cal R}  + {\cal T}$,  this interpretation would, respectively, correspond
 to the boundaries $ {\cal T} \to - \infty$ with ${\cal T}+{\cal R} =Cst$, and ${\cal R} =0$.]  By going through the various redefinitions of  independent
 and dependent variables, it is straightforward to relate the boundary conditions (at the two relevant
 symmetry walls $\beta_{12}$, $\beta_{23}$) on $k_{pq}$ discussed above to corresponding boundary conditions on
 $ {\widetilde \Phi}$ at the corresponding boundaries   $ {\mathcal U} = \infty$  ( $ \mathcal I^-$), and  $ {\mathcal U} + {\mathcal V} =2\, {\cal R}  = 0 $. More precisely, a local analysis of the equation for   $ {\widetilde \Phi}$ at these boundaries  yields the following.
 First, near   $ \mathcal I^-$ an incoming-radiation behavior for  $ {\widetilde \Phi}$, {\it i.e.}  $ {\widetilde \Phi}({\mathcal U} , {\mathcal V} )
 \sim \phi_{\rm in}( {\mathcal V} ) + O({\mathcal U}^{-1}) $ would correspond to a singular solution $k_{pq} \sim  \beta_{12}^{- 5/8}$.  Therefore, in terms of  $ {\widetilde \Phi}$ we should impose a no-incoming-radiation condition at   $ \mathcal I^-$
 (in the    ${\mathcal U}, {\mathcal V}$ plane). Second,  a local Fuchs-type analysis of the equation for   $ {\widetilde \Phi}$ 
 at the regular singular point  $ {\mathcal U} + {\mathcal V} =2\, {\cal R}  = 0 $ leads to an indicial equation for the exponents $s$ in
  $ {\widetilde \Phi} \sim ({\mathcal U} + {\mathcal V} )^s  \sim {\cal R} ^s$ of the form $s(s-1)= \frac34$. The solutions of this indicial
  equation are $s=\frac32$ and $s=- \frac12$. As the difference between these two exponents is an integer, the 
  (more regular) solution  built around $s=\frac32$ will be unambiguously defined, while the  (more singular)  solution built around
  $s=- \frac12$ will contain logarithmic terms (and an arbitrary constant).  Similarly to what happened at the other boundary,
  one finds that the logarithmic-free, more regular solution around ${\cal R}  \sim \beta_{23}=0$ corresponds to a
   square-integrable  solution  $k_{pq} \sim  \beta_{23}^{+ 3/8}$, while the more singular solution (containing
   logarithms) corresponds to a non square-integrable   $k_{pq} \sim  \beta_{12}^{- 5/8}$. Summarizing, our boundary conditions lead us to select solutions (and, in particular, a Green's function) for  $ {\widetilde \Phi}$ which satisfy the two
   conditions: (i) absence of incoming radiation on  $ \mathcal I^-$, and (ii) vanishing of  $ {\widetilde \Phi}$  at the spatial origin
   ${\cal R} =0$ according to   $ {\widetilde \Phi} \sim {\cal R} ^{3/2}$.  These conditions uniquely select a Green's function for the 
    $ {\widetilde \Phi}$ equation of the reflected-retarded form
\begin{equation}   \label{Gretref}
G_{-\frac34}[{\cal U}_P,{\cal V}_P;{\cal U},{\cal V}]
=\theta[{\cal U}_P+{\cal V}]\,\theta[{\cal U} -{\cal U}_P]\,\theta[{\cal V}_P-{\cal V}]\, R _{-\frac34}[{\cal U}_P,{\cal V}_P;{\cal U},{\cal V}]
\end{equation} 
Here, the field point is denoted ${\cal U}_P,{\cal V}_P$; ${\cal U},{\cal V}$ denotes the source point on which one
will integrate after the inclusion of the source term, and $\theta$ denotes Heaviside's step function. In addition, the ``Riemann'' function  $R_{-\frac34}$ is explicitly given by a Legendre function  of index $+ \frac12$.  More generally,  we have
\begin{equation}\label{LPundemi}
R_{-\frac14 \mp \frac12} \, [{\mathcal U}_P , {\mathcal V}_P ; {\mathcal U} , {\mathcal V}] = P_{\pm\frac12} \left[ 1-2 \, \frac{({\mathcal U}_P - {\mathcal U})({\mathcal V} - {\mathcal V}_P)}{({\mathcal U}_P + {\mathcal V}_P)({\mathcal U} + {\mathcal V})} \right]\qquad ,
\end{equation}
where the upper-sign case corresponds to  $R_{-\frac34}$, while the lower-sign (defining   $R_{+\frac14}$) will
correspond to the other EPD equation considered below, and where we adopt the following definition of  Legendre functions :
$$
P_{\nu}  [z] = \strut _2F_1 \left[ -\nu , 1+\nu ; 1 ; \frac{1-z}2 \right] \qquad .
$$

These two Green's functions satisfy
\begin{eqnarray}
\left (\partial_{{\cal U}_P}\partial_{{\cal V}_P}-\frac 3{4\,({\cal U}_P+{\cal V}_P)^2}\right)G_{-3/4}[{\cal U}_P,{\cal V}_P;{\cal U},{\cal V}]&=&-\delta[{\cal U}_P-{\cal U}]\,\delta[{\cal V}_P-{\cal V}]\nonumber\\
&&\\
\left (\partial_{{\cal U}_P}\partial_{{\cal V}_P}+\frac 1{4\,({\cal U}_P+{\cal V}_P)^2}\right)G_{1/4}[{\cal U}_P,{\cal V}_P;{\cal U},{\cal V}]&=&-\delta[{\cal U}_P-{\cal U}]\,\delta[{\cal V}_P-{\cal V}]\nonumber\\
&&
\end{eqnarray}
These  reflected-retarded Green's functions include three Heaviside step functions $\theta$. The two step functions 
$\theta[{\cal U} -{\cal U}_P]\,\theta[{\cal V}_P-{\cal V}]$ are the usual step functions defining a {\it retarded} Green's function,
having a support (w.r.t. the source point, for a given field point $P$)  in the {\it past} light cone of  ${\cal U}_P,{\cal V}_P$.
The additional step function $ \theta[{\cal U}_P+{\cal V}] $  geometrically corresponds to restricting the support of the
Green's function to what would be the image in the ${\cal T} , {\cal R} $ plane of a past light cone in a, say, four-dimensional
Minkowski spacetime ${\cal T, X, Y, Z}$ (with ${\cal R} = \sqrt{\mathcal X^2+ \mathcal Y^2 +\mathcal  Z^2}$). Indeed, the line  ${\cal U}_P+{\cal V} =0$
is easily seen to be the image in the ${\cal T} , {\cal R} $-plane of the continuation of the radial null geodesic emitted (toward the past)
by the field point  ${\cal U}_P,{\cal V}_P$ and reflected toward positive values of ${\cal R} $ after it encounters the origin ${\cal R} =0$. 
Such reflected-retarded Green's functions (solutions of the EDP equation) are also uniquely selected when considering
the  ${\mathcal T} , {\mathcal R} $-plane  Green's function for the radial equation  describing the propagation of   massless scalar
waves having a fixed multipolarity $ Y_{l m}( \theta, \varphi)$. In the latter case, it has been found that the (retarded, multipolar)  Green's function 
in the  ${\mathcal T} , {\mathcal R} $-plane was given by  Eq.(\ref{Gretref}) with  a Riemann function $R_l$ given by  Eq.(\ref{LPundemi}) with
a Legendre function $P_l$ instead of  $P_{\pm\frac12} $  (see Appendix D  in Ref. \cite{Blanchet:1985sp}). [In both cases,
the regularity condition at the radial origin ${\mathcal R} =0$ selects a $P_\nu$ solution instead of a $Q_\nu$ one (which would
contain logarithms).]

Let us briefly discuss the case where it is the second component of the source $s$ which vanishes, {\it i.e.}   $s_2=0$.
In that case, the second equation of our system, namely Eq. (\ref{DEPD2}), says that there exists a scalar
field $\Psi$ such that 
\begin{equation}\label{fgPsi}
e^{(\mu + \lambda)}  \tilde k_{\hat 1 \hat 2}  = 3 \, \partial_{\hat 1} \, \Psi \qquad \mbox{and} \qquad e^{(\mu + \lambda)} \,  \tilde k_{\hat 1 \hat 1}  = \partial_{\hat 2} \, \Psi \qquad .
\end{equation}
Inserting this form in the first equation of our system,  namely Eq. (\ref{DEPD1}),  leads to a second-order equation
for the potential $\Psi$.  This second-order equation again happens to be equivalent to an EPD equation.
When using the transformed null coordinates (\ref{UV}), it is an EPD equation with $m=n= +\frac12$, of the form
\begin{equation}
\partial_{\mathcal U} ({\mathcal U}+{\cal V})\,\partial_{\cal V} \Psi+\partial_{\cal V} ({\cal U}+{\cal V})\,\partial_{\cal U} \Psi=\frac{4\, s_1}{3\,(1- 4\,{\cal U}^2 )( 1-4\,{\cal V}^2 )}
\end{equation}
Using the rescaled potential
\begin{equation}
\Psi = ({\mathcal U}+{\mathcal V})^{- 1/2} \, {\widetilde \Psi} \qquad
\end{equation}
we now get the differential operator
\begin{equation}
\label{Heq}
\left( \partial_{\mathcal U} \, \partial_{\mathcal V} +\frac 14 \frac1{({\mathcal U}+{\mathcal V})^2} \right) {\widetilde \Psi} \qquad .
\end{equation}
The discussion of the boundary conditions for this equation is entirely analogous, {\it mutatis mutandis}, to the one above.
The two exponents at  $ {\mathcal U}+{\mathcal V} = 2 {\cal R}  =0$ are now $\frac12, \frac12+0$ (where the $+0$ indicates
a logarithmic correction $ {\cal R} ^{1/2} \ln {\cal R} $).  Again, one finds that one must select as regular solution the solutions 
of the $\widetilde \Psi$ equation that contain no incoming radiation on  $ \mathcal I^-$, and which are regular on the axis ${\cal R} =0$
(this excludes the solution containing a logarithm).  At the end of the day, this selects again a   reflected-retarded 
Green's function, which is now of the form
\begin{equation}  
G_{+\frac14}[{\cal U}_P,{\cal V}_P;{\cal U},{\cal V}]
=\theta[{\cal U}_P+{\cal V}]\,\theta[{\cal U} -{\cal U}_P]\,\theta[{\cal V}_P-{\cal V}]\, R_{+\frac14}[{\cal U}_P,{\cal V}_P;{\cal U},{\cal V}]
\end{equation} 
 with a Riemann function  $R_{+\frac14}$ given by the lower-sign case of Eq.  (\ref{LPundemi}), {\it i.e.}  given by
 a Legendre function of index $-\frac12$.
 
 The matricial Green's function for the original Dirac equation (\ref{Dirac}) can finally be read off from the following explicit solution
 for $\psi$ in terms of the two components of the source $s$, {\it i.e.}  the solution of the system (\ref{DEPD1}-\ref{DEPD2}):
 \begin{equation}
\left(
\begin{array}{c}
  k_{{\hat 1}{\hat 1}} \\
  k_{{\hat 1}{\hat 2}}   
\end{array}
\right)=\left(
\begin{array}{cc}
 3\,e^{-\mu}\partial_{\hat 1} \,e^{(\lambda+\mu)/2}& e^{\lambda}\partial_{\hat 2} \,e^{-(\lambda+\mu)/2} \\
  e^{-\mu}\partial_{\hat 2}\,e^{(\lambda+\mu)/2}&  e^{\lambda}\partial_{\hat 1} \,e^{-(\lambda+\mu)/2} \\
\end{array}
\right)\left(
\begin{array}{c}
 G_{1/4}\star\sigma_1 \\
  G_{-3/4}\star\sigma_2  
\end{array}
\right)
\end{equation}
where the $\star$ denotes an integration over $\mathcal U, \, \mathcal V$.   Here, the
new source terms $\sigma_1$ and $\sigma_2$ (which differ from $s_1, s_2$ by factors related to our
redefinitions above, and notably by a Jacobian linked to $du/d{\cal U}=1/(2\,{\cal U}^2-1/2))$, etc.) are given by
$$\sigma_1 : =\frac{2\,s_1}{3\,(1-4\,{\cal U}^2 )(1-4\,{\cal V}^2)\,\sqrt{{\cal U}+{\cal V}}}$$
$$\sigma_2 : =\frac{2\,\sqrt{{\cal U}+{\cal V}}\,s_1}{3\,(1-4\,{\cal U}^2)(1-4\,{\cal V}^2)}$$ 
Note that the presence of derivatives acting on the scalar Green's functions $ G_{1/4}$, $ G_{- 3/4}$ (which contain
step functions) means that the matricial Green's function for our Dirac system  contains $\delta$-functions
having their support on the (reflected) past light cone, in addition to step functions with support within the
interior of the latter light cone.

Finally, having (uniquely) obtained   $k_{\hat 1\hat 1}$, $k_{\hat 1 \hat 2}$ and $k_{\hat 2 \hat 2}$ in terms of the two arbitrary functions
$H$ and $K$, it only remains to determine   $k_{\hat 0 \hat 0}$ (on our chosen initial Cauchy slice  $\xi^{\hat 0}=$ Cst).
This is done by integrating the two constraints $\mathcal C'_{\hat p}$.  We already mentioned that this system is
integrable. It therefore determines  $k_{\hat 0 \hat 0}$ by a line integral, modulo an 
arbitrary solution of the homogeneous system that involves one free constant, $C_4$, namely 
\begin{equation}
k_{00}^{({\rm hom})} = C_4 \, \frac{(1-X^2)^{3/4} \, (X-Y)^{3/8} \, (1-XY)^{3/8}}{X^{3/2}} \ e^{-\frac{(1+X^2) Y^{1/3}}{2TX}}
\end{equation}


\end{document}